\documentclass[twocolumn]{aastex631}
\usepackage{amsthm,amssymb,mathtools}
\usepackage{natbib}
\usepackage{multirow}
\usepackage{physics}
\usepackage{gensymb}
\usepackage{xcolor}
\usepackage{graphics}
\usepackage[cp1251]{inputenc}
\usepackage{graphicx} 
\usepackage{subfigure}

\begin{document}
	
	\title{Photometric Analysis of the OGLE Heartbeat Stars}
	\author[0000-0002-3051-274X]{Marcin Wrona}
	\accepted{February 18, 2022}
	\published{April 4, 2022}
	\submitjournal{ApJ}
	\affil{Astronomical Observatory, University of Warsaw, Al. Ujazdowskie 4, 00-478 Warszawa, Poland}
	\author[0000-0003-2244-1512]{Piotr A. Ko\l{}aczek-Szyma\'nski}
	\affil{Astronomical Institute, University of Wroc\l{}aw, Kopernika 11, 51-622 Wroc\l{}aw, Poland}
	\author[0000-0002-3218-2684]{Milena Ratajczak}
	\affil{Astronomical Observatory, University of Warsaw, Al. Ujazdowskie 4, 00-478 Warszawa, Poland}
	\author[0000-0003-4084-880X]{Szymon Koz{\l}owski}
	\affil{Astronomical Observatory, University of Warsaw, Al. Ujazdowskie 4, 00-478 Warszawa, Poland}
	
	\correspondingauthor{Marcin Wrona}
	\email{mwrona@astrouw.edu.pl}

	\begin{abstract}
		We present an analysis of 991 heartbeat stars (HBSs) from the OGLE Collection of Variable Stars. The sample consists of 512 objects located toward the Galactic bulge, 439 in the Large Magellanic Cloud, and 40 in the Small Magellanic Cloud. We model the $I$-band OGLE light curves using an analytical model of flux variations reflecting tidal deformations between stars. We present distributions of the model parameters that include the eccentricity, orbital inclination, and argument of the periastron but also the period--amplitude diagrams. On the Hertzsprung--Russell diagram, our HBS sample forms two separate groups of different evolutionary status. The first group, including about 90 systems with short orbital periods ($P\lesssim50$~days), consists of an early-type primary star lying on (or close to) the main sequence. The second group, including about 900 systems with long orbital periods ($P\gtrsim100$~days), contains a red giant (RG). The position of the RG HBSs on the period--luminosity (PL) diagram strongly indicates their binary nature. They appear to be a natural extension of confirmed binary systems that include the OGLE ellipsoidal and long secondary period variables. We also present a time-series analysis leading to detection of tidally excited oscillations (TEOs). We identify such pulsations in about 5\% of stars in the sample with a total of 78 different modes. This first relatively large homogeneous sample of TEOs allowed us to construct a diagram revealing the correlation between the TEO's orbital harmonic number and the eccentricity of the host binary system.
		
	\end{abstract}

	\section{Introduction}
	Heartbeat stars (HBSs) are a subclass of ellipsoidal variable stars with eccentric orbits. The ellipsoidal variable stars are close binary systems whose components are deformed due to gravitational tidal forces. For a ``classical" ellipsoidal variable, which typically has a circular orbit, the light curve is characterized by a sinusoidal shape with two maxima and two minima per orbital period. This is a result of observing different sides of the stellar surface, which is distorted into a rotational ellipsoid (\citealt{1985ApJ...295..143M}). 
	
	If the orbit of the system is eccentric, the deformation strength depends on the orbital phase, and the strongest effect appears during the periastron passage. This variation of the tidal force affects the shape of the light curve, which starts to deviate from the typical sinusoidal shape. The response of the stellar surface and volume to the presence of varying tidal potential can be expressed as a sum of two effects \citep{1975A&A....41..329Z}.
	
	The first one is an equilibrium tide, which refers to the instantaneous deformation of a star due to tides. It is the only component of the tidal response provided that the system is a circular and synchronized binary. The equilibrium tide is responsible for the presence of the ``heartbeat" feature in the light curves of HBSs. This prominent characteristic resembles a single electrocardiogram pulse, hence the name of this group of variables.
	
	The second type of stellar tidal response is the dynamical tide, which may manifest itself as tidally excited oscillations (TEOs; e.g., \citealt{1970A&A.....4..452Z}, \citealt{1995ApJ...449..294K}, \citealt{2017MNRAS.472.1538F}). The periodically changing tidal potential may act as a driving force and induce even naturally damped pulsations, which would not be visible in the absence of a nearby companion. Most of the TEOs observed in main-sequence (MS) stars are high-radial order gravity modes \citep{10.3389/fspas.2021.663026}. 
	
	\cite{1995ApJ...449..294K} provided a convenient analytical model of the flux variations driven by tidal interactions (we will refer to it as the K95 model). The K95 model allows for determination of the orbital parameters of the system based only on the shape of the single-passband light curve. For instance, this model was successfully applied to derive orbital parameters for a sample of HBSs discovered by \cite{2012ApJ...753...86T}.
	The K95 model accounts for the ellipsoidal variability only. It neglects other proximity effects, such as the irradiation/reflection effect and Doppler beaming/boosting. While the latter should not be pronounced in our sample of HBSs because of their long orbital periods and hence low radial velocities, the former effect may play a potentially significant role (see, e.g., \citealt{2011ApJS..197....4W}, their Figure~7).
	
	The prototype of the entire HBS class is KOI-54 (HD~187091), studied in detail by many authors, e.g., \cite{2011ApJS..197....4W}, \cite{2012MNRAS.421..983B}, and \cite{2012MNRAS.420.3126F}. For the first time, the HBS was reported as a separate class of variable stars in the work of \cite{2012ApJ...753...86T}, but such objects were known in the literature much earlier (e.g., \citealt{2002MNRAS.333..262H}, \citealt{2009A&A...508.1375M}). A group of more than 100 ellipsoidal variables with an eccentric orbit was highlighted in the work of \cite{2004AcA....54..347S}. A subsample of them were analyzed later by \cite{2010MNRAS.405.1770N}, \cite{2012MNRAS.421.2616N}, and \cite{2017ApJ...835..209N}. Their studies showed that HBSs containing a red giant (RG) typically have longer periods than classical ellipsoidal variables. The vast majority of HBSs in our sample belong to this group; therefore, we could verify that HBSs are indeed an extension of the classical ellipsoidal systems to longer periods for a given brightness.
	
	The majority of papers on HBSs published so far have focused on individual objects, e.g., KOI-54 (\citealt{2011ApJS..197....4W}), KIC~3749404 (\citealt{2016MNRAS.463.1199H}), KIC~8164262 (\citealt{2018MNRAS.473.5165H}), and MACHO 80.7443.1717 (OGLE-LMC-HB-0254; \citealt{2019MNRAS.489.4705J}, \citeyear{2021MNRAS.506.4083J}, \citealt{2022A&A...659A..47K}). Due to the limited number of known HBSs, a detailed quantitative analysis has been challenging to carry out. A sketchy analysis of HBSs containing RG stars was introduced in the work of \cite{2004AcA....54..347S}, where the authors explained the reason for the unusual shape of the light curves of the ellipsoidal variables. Later, a subset of those systems were analyzed by \cite{2012MNRAS.421.2616N} and then by \cite{2014AJ....148..118N}. In both works, the authors measured changes in radial velocities confirming the binary nature of those stars and showing that the strongest brightness changes occur near the periastron passage. Afterward, \cite{2017ApJ...835..209N} presented an analysis of 81~ellipsoidal RG stars, including 22 HBSs. They mainly focused on the evolutionary status of those stars and the primary/secondary mass distributions. In turn, \cite{2014A&A...564A..36B} studied the asteroseismic properties of 18 RG HBSs found in the data obtained by NASA's Kepler Space Telescope (\citealt{2010Sci...327..977B}).
	
	Another set of HBSs, but consisting of stars located on or close to the MS, was found in the Kepler database (17 objects were the subject of an analysis by \citealt{2012ApJ...753...86T}). The catalog, including HBSs, among other variable stars, was released by \cite{2016AJ....151...68K}. Those HBSs are mainly low- and intermediate-mass A--F-type stars. They are characterized by a short orbital period (days to tens of days) and very small amplitudes of brightness variations (a few millimagnitudes) contrary, to the HBSs with an RG star, which have much longer periods (hundreds of days) and an order of magnitude higher amplitudes.
	
	Recently, using the 9th Catalogue of Spectroscopic Binary Orbits \citep{2004A&A...424..727P}, \cite{2021A&A...647A..12K} selected HBS candidates and examined their light curves delivered by the Transiting Exoplanet Survey Satellite (TESS) mission from sectors 1\,--\,16 (mostly the southern ecliptic hemisphere). The authors discovered 20 massive and intermediate-mass HBSs, seven of which exhibit several TEOs lying at low harmonics of the orbital frequency.
	
	In this work, we conduct a general analysis of 991 HBSs cataloged in the OGLE Collection of Variable Stars (OCVS; \citealt{2022ApJS..259...16W}, hereafter Paper I). In parallel to the analysis of the heartbeat phenomenon itself, we search for TEOs in the presented collection of OGLE HBSs. The derived sample of binaries exhibiting TEOs may be a valuable test bed for future studies on the influence of dynamical tides on the orbital evolution in binary systems, including those with evolved companions. 
	
	The structure of this paper is as follows. In Section~\ref{sec:data}, we describe the origin of the photometric data of the HBSs and the way we prepare them for the examination. In Section~\ref{sec:model}, we describe the modeling process of the light curves using the K95 model. In Section~\ref{sec:irr_ref}, we study the impact of irradiation and the reflection effect on the light curve of an HBS. In Section~\ref{sec:teo}, we introduce the method used to search for TEOs and present the results of the search. The core of our work is presented in Section~\ref{sec:results}, where we discuss the results of the analysis. In Section~\ref{sec:conc}, we summarize and conclude our work.
	
	\section{Photometric Data} \label{sec:data}
	In this work, we analyze the sample of HBSs found in the OGLE project database. The detailed specifications of the data are presented in Section~2 of Paper I.
	
	The time-series data used in the analysis were obtained using the 1.3 m Warsaw Telescope located at Las Campanas Observatory, Chile. The majority of the data come from the fourth phase of the OGLE project, in operation since 2010 (\citealt{2015AcA....65....1U}). The observations were conducted using Cousins $I$-band and Johnson $V$-band filters. The photometry was obtained using differential image analysis (DIA; e.g., \citealt{1998ApJ...503..325A}, \citealt{2000AcA....50..421W}). All of the data for each filter were calibrated separately to the standard photometric system using the scheme presented by \cite{2015AcA....65....1U}. We also corrected the photometric uncertainties based on the work of \cite{2016AcA....66....1S}.
	
	In the analysis and modeling of the light curves, we used mainly the $I$-band data. The $V$-band photometry was used to determine the $(V-I)$ color information for all HBSs from our sample. To prepare the data for modeling and analysis, we have taken the following steps.
	
	First, we cleared the light curves from outliers by rejecting all of the data that lay outside the $3\sigma$ level from the average flux, where $\sigma$ is the standard deviation of the flux in the entire range of time. This step was taken separately for all of the available data obtained by different OGLE phases (OGLE-II, OGLE-III, and OGLE-IV) because in some cases, there was a significant shift in the mean flux between those data. During this step, we also removed sets of data consisting of clearly improperly determined photometry, which could be caused, for instance, by bad weather conditions during the observing night or some failure in the DIA pipeline (this is a very common case for sources with high proper motions).
	
	The second step was to remove trends in the data sets. This procedure was done separately for data obtained during OGLE-II, OGLE-III, and OGLE-IV. To each part of the light curve, we fitted cubic splines. Then, the obtained sets of splines were subtracted from the data.
	
	In the third step, we shifted the detrended data in the fluxes from OGLE-II and OGLE-III to OGLE-IV (or to OGLE-III if there were no data from OGLE-IV). For this purpose, we calculated the mean flux for each phase and shifted the data to the latest phase by the difference between these averages.
	
	Finally, we used additional cleaning procedures. Using the combined data from all phases, we prepared light curves phase-folded with the orbital period and divided them into bins, the
	width of which was set on $0.1$ of the orbital phase. Then we calculated a standard deviation in each bin and removed points lying more than $A\cdot\sigma$ from the mean magnitude. The $A$ parameter was chosen individually for each star and mainly depended on the number and positions of outlying points. The $A$ parameter was usually about 3. 
	
	To assess photometric temperatures for the HBSs containing a hot primary, we used $UBV$ photometry obtained by \cite{2002ApJS..141...81M}. Data were taken with the Curtis Schmidt telescope at Cerro Tololo Inter-American Observatory (CTIO), Chile, using a Tektronix $2048\times2048$ CCD with a $2.32''/$pixel scale. Observations were made at the beginning of 1999 and in 2001 March/April. 
	
	For some objects, the aforementioned $UBV$ data were unavailable; therefore we decided to use $UBV$ photometry obtained by \cite{2004AJ....128.1606Z}. The authors used the 1 m Swope Telescope, located at Las Campanas Observatory, right next to the Warsaw Telescope. Observations were taken with the Great Circle Camera with a 2K CCD with a $0.7''$ pixel scale between 1995 October and 1999 December, with additional observations in 2001 December.
	
	In this work, we also present, among others, period--luminosity (PL) diagrams using the $W_{JK}$ Wesenheit index. To calculate this quantity, we used $JHK_{\rm s}$ photometry collected by \cite{2007PASJ...59..615K} using the InfraRed Survey Facility (IRSF) $1.4$ m telescope at Sutherland, the South African Astronomical Observatory, with the SIRIUS camera, which is equipped with three 1024 $\times$ 1024 HgCdTe arrays. The pixel scale for this camera is $0.45''$.
	
	As complementary data, we utilized photometry collected during the Two Micron All Sky Survey (2MASS; \citealt{2006AJ....131.1163S}, \citeyear{https://doi.org/10.26131/irsa2}). The 2MASS used two $1.3$ m telescopes located at Mount Hopkins, Arizona, and at the CTIO, Chile. The telescopes were equipped with three NICMOS3 $256\times256$ HgCdTe arrays with a pixel scale of $2''$. Observations were taken between 1997 June and 2001 February.
	
	\section{Modeling} \label{sec:model}
	\subsection{The K95 Model of a Light Curve} \label{sec:kumar_model}
	In this work, we decided to use the analytic model of tidally induced stellar deformation presented in \cite{1995ApJ...449..294K}, where the normalized flux variations are described by their Equation (44). The K95 model was derived under certain assumptions and simplifications, such as spin-orbit alignment, inclusion of dominant modes only (spherical harmonics with $l=2$, $m=0,\pm 2$), and alignment of the tidal bulge with the line connecting the mass centers of stars and without the irradiation and Doppler beaming effects. Therefore, even a model perfectly fitted to the light curve only gives an assessment of the orbital parameters. The slightly modified version of the K95 formula was used by many authors (e.g., \citealt{2012ApJ...753...86T}; \citealt{2019MNRAS.489.4705J}; \citealt{2021A&A...647A..12K}) to assess the basic orbital parameters of the system based on the light curve. According to \cite{2012ApJ...753...86T}, the relative change of the flux caused by tidal deformation as a function of time, $t$, can be expressed as 
	\begin{equation}
		\frac{\delta F}{F}(t)=S\cdot\frac{1-3\sin^2i\sin^2(\varphi(t)-\omega)}{(R(t)/a)^3}+C,
		\label{eq:kumar}
	\end{equation}
	where $S$ is the scaling factor of the amplitude, $C$ is the zero-point offset, $i$ is the inclination of the orbit, $\omega$ is the argument of the periastron, $\varphi(t)$ represents the true anomaly as a function of time, $R(t)$ describes the distance between the components of the system as a function of time, and $a$ is the semimajor axis.
	
	Both time-dependent variables $R(t)$ and $\varphi(t)$ can be rewritten as functions of the eccentricity, $e$, and eccentric anomaly, $E$:
	\begin{equation}
		\frac{R(t)}{a}=1-e\cos E(t),
		\label{eq:intro1}
	\end{equation}
	\begin{equation}
		\varphi(t)=2\arctan\left(\sqrt{\frac{1+e}{1-e}}\cdot\tan(\frac{E(t)}{2})\right).
	\end{equation}
	The quantities $E$ and $t$ are connected by Kepler's equation,
	\begin{equation}
		\frac{2\pi(t-T_0)}{P}=E-e\sin E,
	\end{equation}
	where $T_0$ is the time of periastron passage, and $P$ is the orbital period. 
	
	\begin{figure}
		\includegraphics[width=0.48\textwidth]{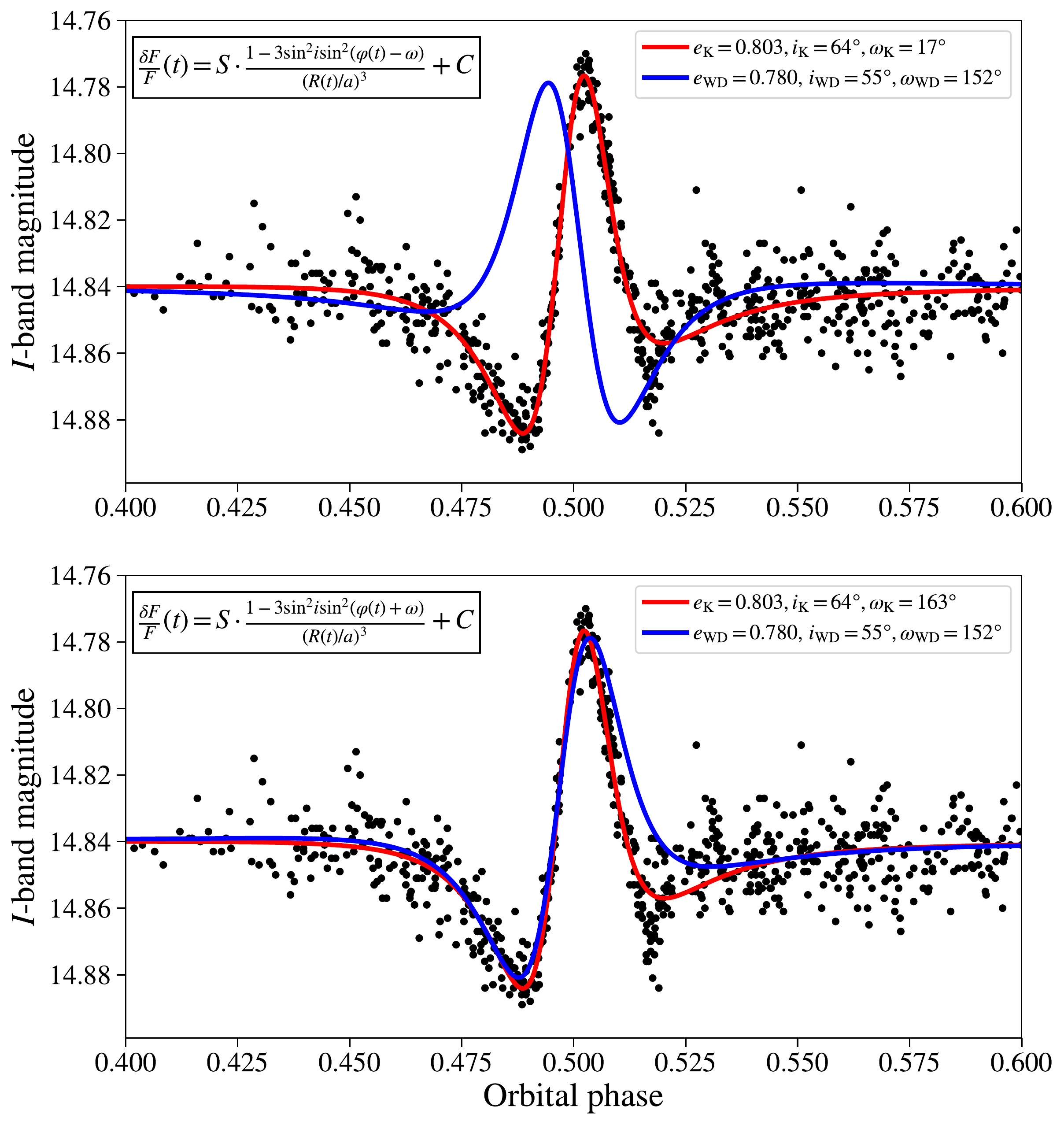}
		\caption{Phase-folded light curve of OGLE-BLG-HB-0081 (black dots). In the top panel, we plot lines based on the K95 model described by Equation (\ref{eq:kumar}), while in the bottom panel, we used the corrected version shown in Equation (\ref{eq:kumar2}). We obtained the orbital parameters of the system using two independent methods: the K95 model (red line, index K) and the PHOEBE Legacy program, which is based on the WD code (blue line, index WD).}	
		
		\label{fig:kumar_problem}
	\end{figure}
	
	However, we believe that there is a mistake in Equation (\ref{eq:kumar}). In the top panel of Figure~\ref{fig:kumar_problem}, we present the light curve of the HBS OGLE-BLG-HB-0081. In the modeling process, we used Equation (\ref{eq:kumar}) and the Markov Chain Monte Carlo (MCMC) fitting procedure (described in more detail in Section~\ref{sec:MCMC}). We obtained the following orbital parameters: $e_{\rm K}=0.803$, $i_{\rm K}=64\degree$, and $\omega_{\rm K}=17\degree$. The fitted model is represented by the red line in Figure~\ref{fig:kumar_problem}. We also performed modeling of that light curve using the PHysics Of Eclipsing BinariEs (PHOEBE) Legacy program\footnote{\url{http://phoebe-project.org/}} (\citealt{2005ApJ...628..426P}), which is based on the Wilson--Devinney (WD) code (\citealt{1971ApJ...166..605W}), and thus independent of the K95 model. The resulting values of $e_{\rm WD}=0.78$ and $i_{\rm WD}=55\degree$ are similar to the ones obtained with the K95 model (within 3$\sigma$), but the argument of periastron is far from that: $\omega_{\rm WD}=152\degree$. Setting these values to Equation (\ref{eq:kumar}) and plotting the results in Figure~\ref{fig:kumar_problem} we got the blue line. It is clearly seen that this model is far from being proper, but one can notice that the blue line is a symmetric reflection of the red one about the vertical axis.
	
	We also compared light curves generated on the basis of Equation (\ref{eq:kumar}) with synthetic light curves presented in Figure 5 of \cite{2012ApJ...753...86T}. The results turned out to be very similar to the ones described above. Our light curves were symmetric reflections of the ones presented by \cite{2012ApJ...753...86T}. We are certain that the fault is connected with the sign of the $\omega$ in Equation (\ref{eq:kumar}) because by inverting this sign, we get proper models. We were not able to track down the source of this discrepancy.
	Nevertheless, the version of Equation (\ref{eq:kumar}) with the plus sign before $\omega$ makes the model fit properly to the data and is in agreement with the synthetic light curves generated with alternative methods; thus, we decided to use the following equation instead of Equation (\ref{eq:kumar}):
	\begin{equation}
		\frac{\delta F}{F}(t)=S\cdot\frac{1-3\sin^2i\sin^2(\varphi(t)+\omega)}{(R(t)/a)^3}+C.
		\label{eq:kumar2}
	\end{equation}
	
	The default unit of a star's brightness in our catalog is magnitude, while during the modeling, we operated with a fractional change of the flux. We can express the change of the magnitude using Pogson's equation,
	\begin{align}
		\delta m = m - m_0 &= -2.5\log\left(\frac{F+\delta F}{F}\right)= \nonumber\\ 
		&=-2.5\log\left(1+\frac{\delta F}{F}\right),
	\end{align}
	where we used the median magnitude as a zero-point, $m_0$. Thus, we can calculate the fractional change of the flux using following formula:
	\begin{align}
		\frac{\delta F}{F} &= 10^{-0.4(m-m_0)} - 1, \\
		\sigma_f &= \frac{2.5(\delta F / F)\sigma_m}{\ln 10},
	\end{align}
	where $\sigma_f$ is an uncertainty of the flux change with a given uncertainty of the magnitude change, $\sigma_m$.

	\subsection{Fitting the K95 Model to the Light Curve}
	\label{sec:MCMC}
	In the fitting procedure, we decided to use the MCMC method, first to search for a proper model and investigate plausible degeneracies and then to estimate the uncertainties of the model's parameters. We used Python's \texttt{emcee} v3.0.2 package, described in detail by \cite{2013PASP..125..306F}.
	
	The time span of the OGLE HBS light curves often exceeds a dozen years; therefore, in the case of a high apsidal motion rate, the shape of the light curve can change significantly (e.g., \citealt{2016MNRAS.463.1199H}). That may degrade the quality of the K95 model fit. To estimate the role of apsidal motion, we visually compared phase-folded light curves from the first and last three observational seasons. After the eye inspection, we detected clearly visible changes in the shape of the light curve in only six objects: OGLE-BLG-HB-0240, OGLE-BLG-HB-0317, OGLE-BLG-HB-0358, OGLE-LMC-HB-0072, OGLE-LMC-HB-0146, and OGLE-SMC-HB-0001. A light curve of the last object and a brief description of the possible causes of the light curve's shape changes are presented in Section 4.3.4 in Paper I. We did not detect any significant changes in the rest of the HBSs; thus, we decided to neglect the apsidal motion in the fitting procedure.
	
	\subsubsection{Uniqueness of the Solution Found Using the K95 Model}
	\label{sec:mcmc_test}
	To assess the risk of degeneracy of the K95 model, we created 20,000 synthetic light curves using the K95 model with randomly selected parameters from the entire hyperspace. We also took into account the noise of the brightness and the uneven time sampling for the typical OGLE light curve. We assumed the normal distribution of the noise with the standard deviation depending on the mean magnitude, which was also randomly selected. The time sampling included the mid-season gaps and typical cadence of the observations for the central regions of the Magellanic Clouds (MCs) during the OGLE-IV project. 
	
	In the fitting process, we used the MCMC method. In the MCMC run, we used the K95 model, according to the formulae introduced in Section~\ref{sec:kumar_model}. We adopted the flat prior distribution for $e$, $i$, and $\omega$. We limited their final values to the physically reasonable range but with a small margin for angular variables to avoid sharp cuts at the final distribution: $0<e<1$, $0\degree<i<92\degree$, and $-10\degree<\omega<190\degree$. The range of the argument of periastron is only a half of the full angle because the stellar distortion due to tidal deformation is symmetric along the elongation axis. For the remaining parameters, we used the normal prior distribution. The range limits for $P$ and $T_0$ were connected to their estimated initial values $P_{\rm e}$, $T_{0, \rm e}$: $0.95P_{\rm e}<P<1.05P_{\rm e}$,  $T_{0, \rm e} - 0.5P_{\rm e}<T_0<T_{0, \rm e} + 0.5P_{\rm e}$. In the modeling process, we used 50 walkers and 20,000 steps for a single chain. As a result, we used parameters from the model with the maximum likelihood.
	
	We found that systems with low inclination angles ($i\lesssim20\degree$) are difficult to model, and the obtained orbital parameters are usually unreliable. This behavior is caused by the $\sin^2i$ term in the K95 model (Equation~\ref{eq:kumar2}). In our sample, however, there is a very low number of HBSs with such a low inclination angle (see Figure~\ref{fig:params_hist}); thus, to assess the level of the degeneracy of the K95 model, we will focus on the systems with $i>20\degree$.

	For 90\,\% of simulated light curves with $i>20\degree$ ($\approx16,000$ systems), the MCMC fitting procedure returned the correct values of the parameters within the $3\sigma$ region. For the remaining 10\,\% of the simulated light curves, we did not recover the initial parameters of the K95 model. In one of the most common cases (4\,\% of the sample), the program found two equivalent solutions for $\omega$, one about $0\degree$ and the second about $180\degree$. These two solutions are mathematically indistinguishable due to the $180\degree$ ambiguity, which is the result of the periodicity of the $\sin^2(\varphi(t)+\omega)$ term of the K95 model. We also noticed an excess in the number of systems (about 4.5\,\% of the sample) for which the program found $i\approx90\degree$, despite the true value of inclination often being far from $90\degree$. The rest of the parameters obtained for these systems (except for $P$ and $C$) often differ significantly from the original values. We found that such behavior is exhibited by systems in which the mean scatter of the light curve is comparable with the amplitude of the heartbeat. The third most common case (1.5\,\% of the sample) was observed for systems with low inclination angles ($i\lesssim30\degree$) and the argument of the periastron about $0\degree$ or $180\degree$. If we take a smaller $i$ of about $10\degree$ and a larger $e$ of about 0.3 for $e\approx0.2$ or about 0.1 for $e\approx0.7$ (the value of the correction is inversely proportional to the eccentricity), and if we change $\omega$ to the right angle and lower amplitude scaling factor by a factor $~3.5$, the obtained model will be similar to the initial one.

	\subsubsection{Modeling of the OGLE HBS Light Curves}
	
	We conducted fitting with the MCMC method for all 991 HBSs from our sample. We used light curves cleaned from outliers, detrended, and shifted in average flux to the latest OGLE phase, as described in Section~\ref{sec:data}. Stars with additional brightness variations, such as eclipses or spots, were manually cleaned by removing the affected part of the light curve if it could be easily separated from the heartbeat modulation. Nevertheless, in some cases, this procedure could affect the heartbeat shape; therefore, the final parameters may not be reliable. Each OGLE HBS for which we did not find a proper K95 model or the fitted model is unreliable has been appropriately flagged (see the description of Table~A2 in Paper I). 
	
	\begin{figure*}[h]
		\centering
		\includegraphics[width=0.75\textwidth]{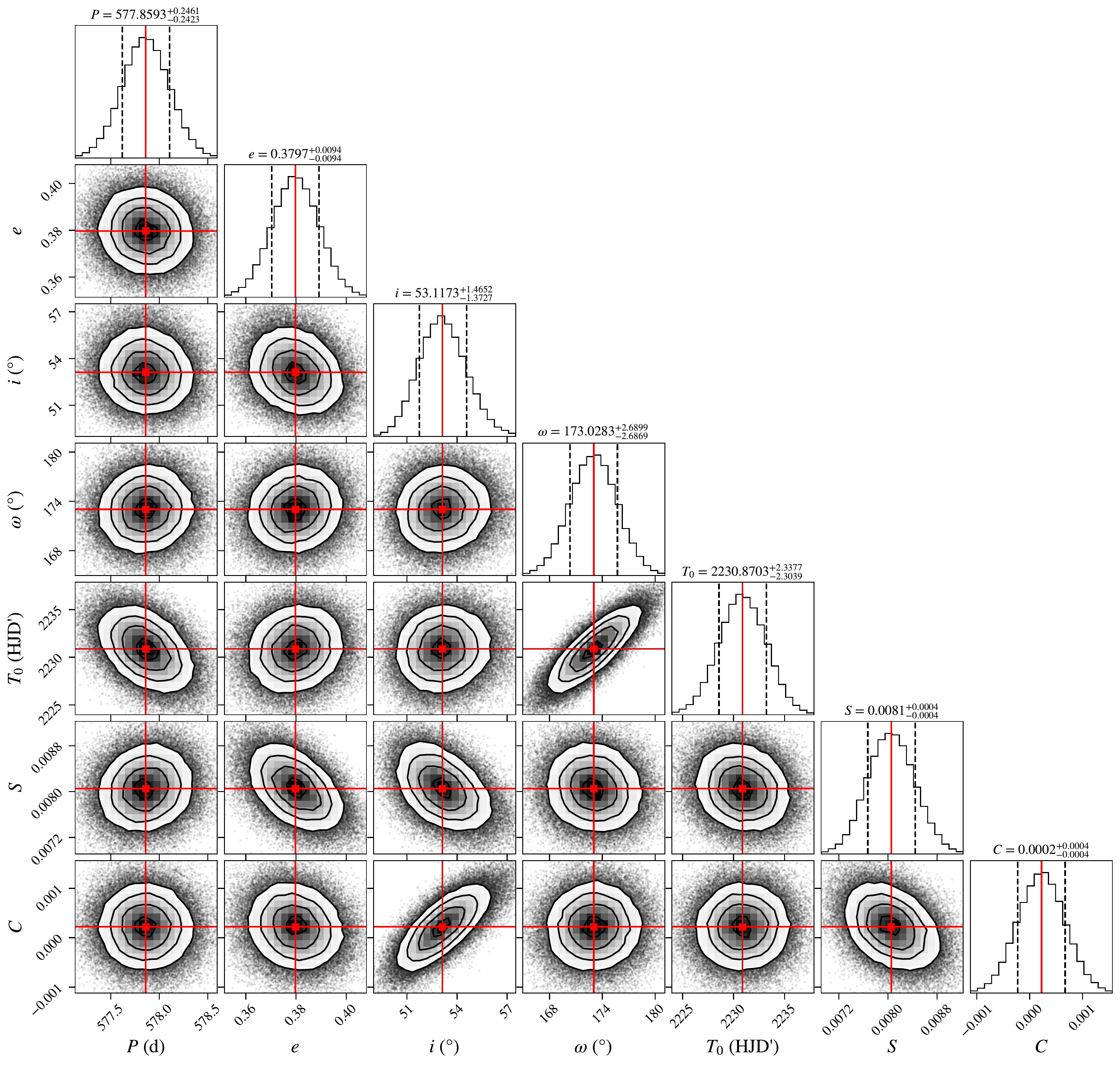}
		\includegraphics[width=0.80\textwidth]{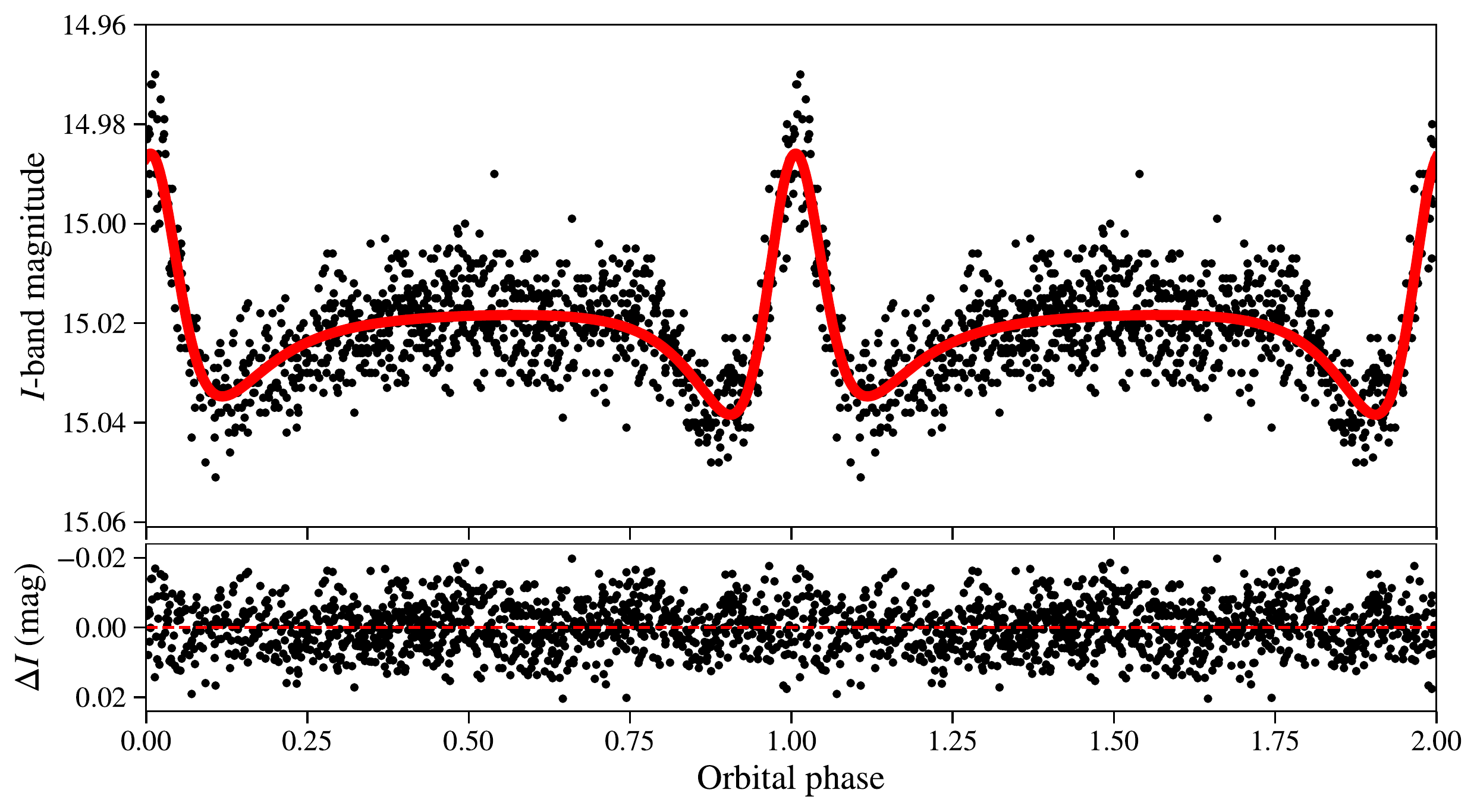}
		\caption{In the top panel, we present a corner plot with the results of the MCMC fitting procedure for OGLE-LMC-HB-0018. Red lines indicate the median values of the presented histograms for each parameter. Black dashed lines denote the positions of the 16th and 84th quantiles. The $T_0$ unit HJD' = HJD$-$2,450,000. In the middle panel, we plot the K95 model (solid red line) fitted to the phase-folded light curve of OGLE-LMC-HB-0018 (black dots). The bottom panel shows residuals from the fit. The dashed red line denotes a zero-point.}
		\label{fig:corner_lc}
	\end{figure*}
	
	We ran the fitting process with similar prior distributions of parameters and their limits as described in Section~\ref{sec:mcmc_test}. We calculated initial orbital periods, $P_{\rm e}$, using the FNPEAKS\footnote{\url{http://helas.astro.uni.wroc.pl/deliverables.php?active=fnpeaks}} program (created by Z. Ko\l{}aczkowski, W. Hebisch, and G. Kopacki), which is based on Fourier frequency spectra. The initial time of periastron passage, $T_{0,\rm e}$, was estimated based on the local extrema in the light curves.
	
	In the MCMC fitting process, for each star, we applied 50 walkers with 20,000 steps in a single chain. Only about 6\% of the HBSs showed signs of degeneracies. In most cases, the problems were $\omega$ bouncing between $0\degree$ and $180\degree$ or the program finding a different solution for the period. For those problematic stars, we used another MCMC run with a larger number of walkers and steps and narrower prior distributions. This approach led to proper models for most of the troublesome light curves. We could not find the satisfactory model for only about 2\% of our HBSs.
	
	Finally, we once more used MCMC fitting to derive the uncertainties of the K95 model's parameters. The priors were randomly selected using a normal distribution centered on the value from the previous step, with the standard deviation being one-hundredth of this value (for $T_0$, we used $P/100$, and for $C$, we used $C/1000$). We used 100 walkers and 25,000 steps in a single chain.
	
	As a final value of the given parameter, we used the maximum value of the distribution, but we also considered using the median value. In most cases, these two approaches give similar values, but for the non-Gaussian distribution (e.g., when the $i$ value is about $90\degree$), we found that the first one results in better-fitted models. As uncertainties $\sigma_{-}$ and $\sigma_{+}$, we assigned distances between the 50th and the 16th and 84th percentiles of the distribution, respectively. In Figure~\ref{fig:corner_lc}, we present a typical corner plot (top panel) and the corresponding phase-folded light curve including the model (bottom panel). Both corner plots and light curves with the fitted model for each OGLE HBS are available on the OGLE websites\footnote{\url{https://www.astrouw.edu.pl/ogle/ogle4/OCVS/blg/hb/} for the GB sample of HBSs (instead of ``blg'', use ``lmc'' or ``smc'' for the Large or Small Magellanic Cloud sample of HBSs, respectively).}. In most cases, there are no clearly visible correlations between parameters, except for the following pairs: $i$--$S$, $i$--$C$, $\omega$--$T_0$, and less often $P$--$T_0$.
	
	The final values of the K95 model parameters for each star with their uncertainties are presented in Paper I (their Table A3). 
	
	\section{The Impact of the Irradiation/Reflection Effect} \label{sec:irr_ref}
	In order to quantitatively estimate the impact of the irradiation/reflection effect on the shape of the OGLE $I$-band light curves of HBSs located in the Large Magellanic Cloud (LMC), we ran a set of dedicated simulations. We used PHOEBE\,2 modeling software (version 2.3; \citealt{2016ApJS..227...29P}; \citealt{2018ApJS..237...26H}; \citealt{2020ApJS..247...63J}; \citealt{2020ApJS..250...34C}) for this purpose and performed two groups of simulations. The first one refers to HBSs with the primary component being an RG. The majority of our HBSs fall into this group. In the second group, we considered HBSs with a massive primary component located on the MS or passing through the Hertzsprung gap. We will refer to these two groups of simulations as S-RG and S-MS, respectively.
	
	\begin{figure*}[h]
		\centering
		\includegraphics[width=0.95\textwidth]{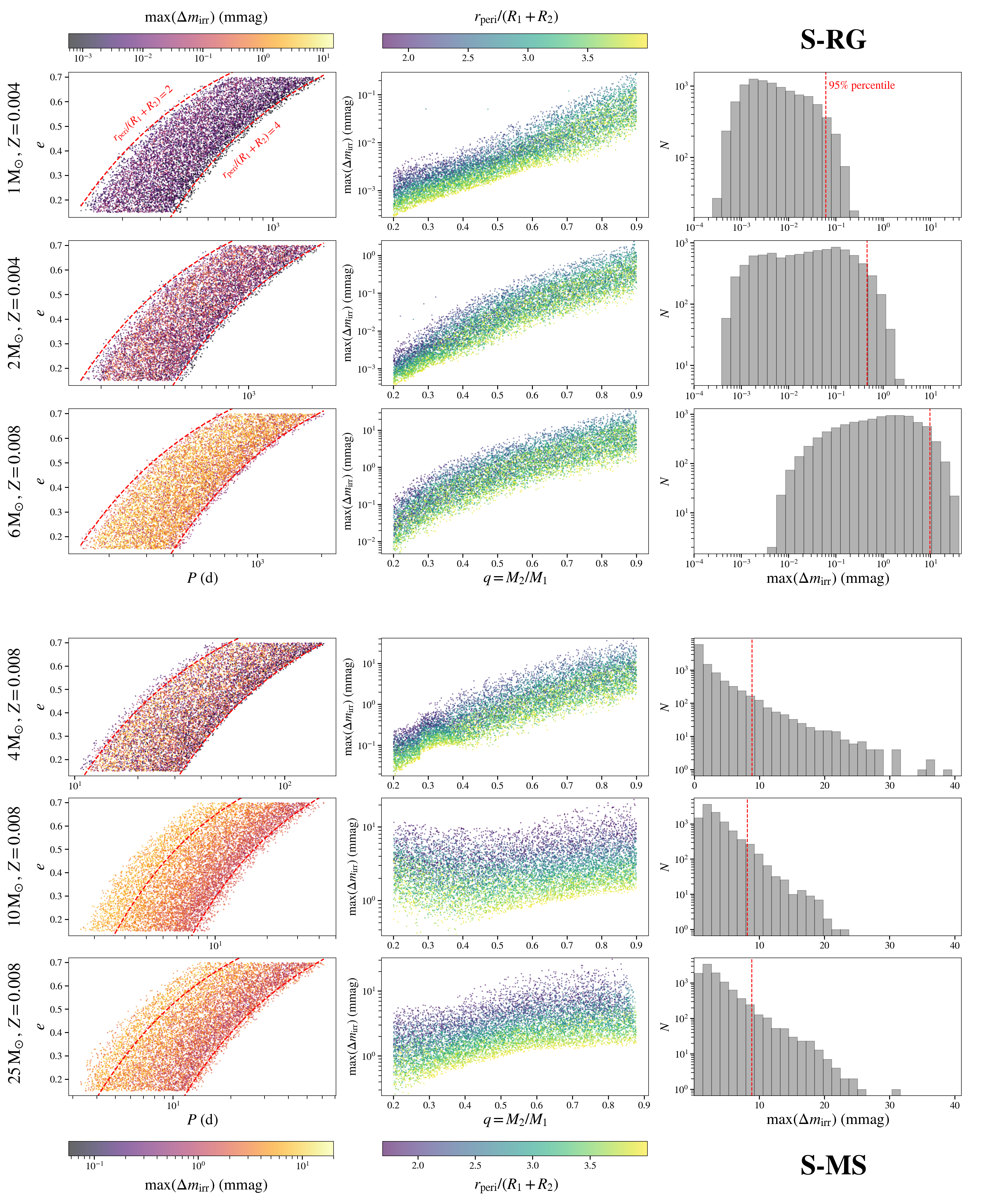}
		\caption{Results of the S-RG (upper three rows) and S-MS (lower three rows) simulations described in Section~\ref{sec:irr_ref} for different masses and evolutionary statuses of the primary component. Left column: orbital period--eccentricity distributions of simulated binaries. The maximum contribution of the irradiation/reflection effect in the $I_C$ passband, $\max(\Delta m_{\rm irr})$ is color-coded separately for the S-RG and S-MS simulations. The dashed red lines denote systems with $r_{\rm peri}/(R_1+R_2)=2$ or 4, assuming that $q=0.5$. Middle column: distributions of simulated binaries on the $q$--$\max(\Delta m_{\rm irr})$ plane. The periastron distance scaled by the sum of the radii of the components is color-coded. Right column: histograms of simulated $\max(\Delta m_{\rm irr})$ values. The vertical red dashed lines indicate the position of the 95th percentile.}
		\label{fig:irr-effect}
	\end{figure*}

	\begin{figure*}[h]
		\centering
		\includegraphics[width=0.90\textwidth]{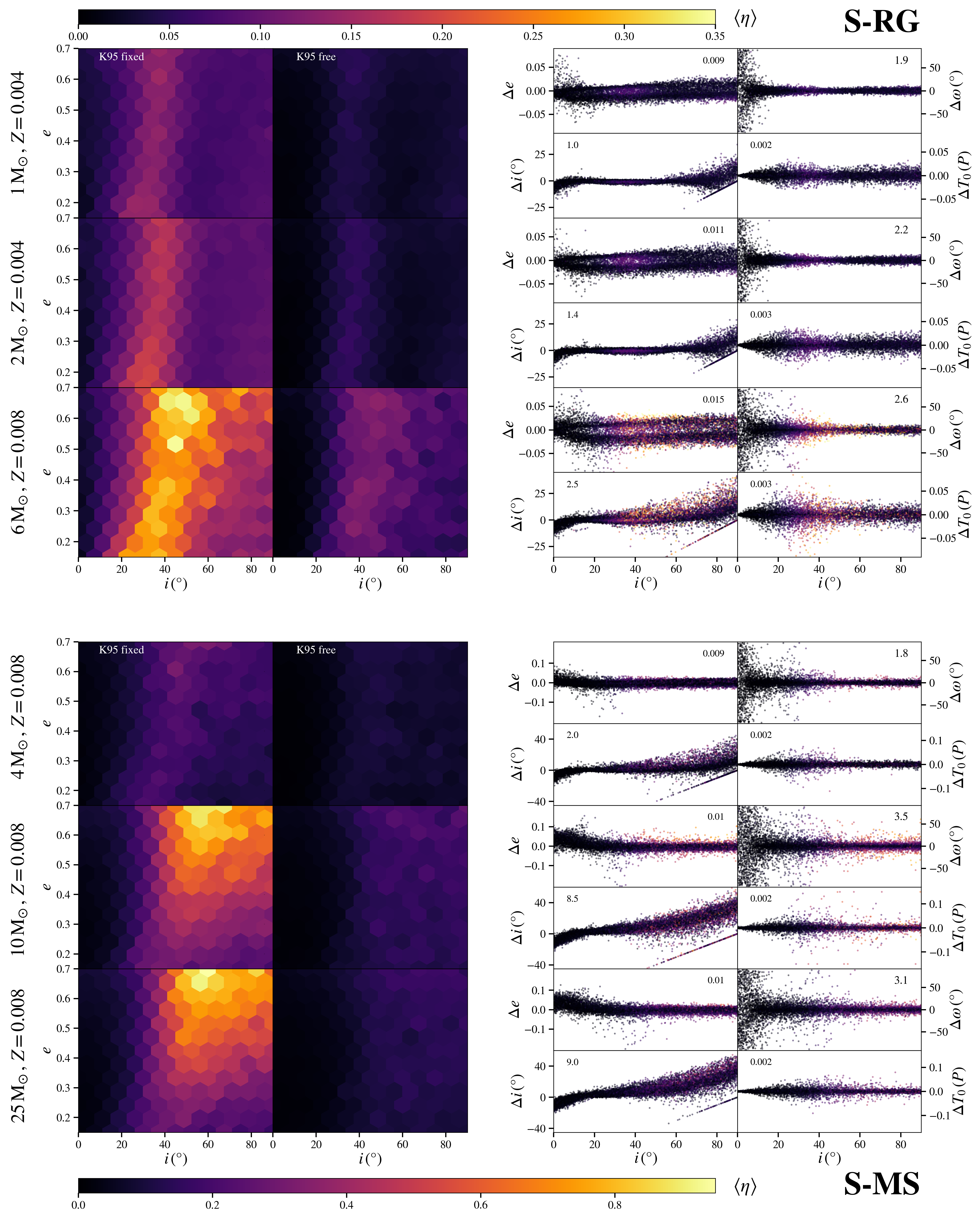}
		\caption{Results of the S-RG (upper part) and S-MS (lower part) simulations described in Section~\ref{sect:irr-effect-kumar}. Each row corresponds to a different mass and metallicity of the primary component, as indicated by the labels placed in the leftmost part of the figure. The series of hex-binned panels on the left shows the distributions of the average value of $\eta$, $\langle \eta \rangle$ (color-coded), for simulated orbits with different eccentricities and inclinations. The column of panels denoted as ``K95 fixed" presents the results for the K95 model with fixed orbital parameters during the fitting procedure. In turn, the column denoted as ``K95 free" corresponds to the fits with orbital parameters being free during the optimization. On the right, we present the absolute errors of the orbital parameters, $\Delta \mathcal{X}$, obtained from the ``K95 free" simulations. The value presented in the corner of each plot is equal to the median of the $|\Delta \mathcal{X}|$. See the discussion in Section~\ref{sect:irr-effect-kumar} for more details. Note the different range of $\langle \eta \rangle$ for S-RG and S-MS.}
		\label{fig:kumar-accuracy}
	\end{figure*}
	
	The S-RG simulations were done as follows. We considered three masses of RGs being a primary component, 1, 2, and 6$\,M_\odot$ with corresponding radii of 50, 75, and 100$\,R_\odot$ (see right panel of Figure~\ref{fig:HR_diagram}). For each mass of the primary component, we generated 10,000 binaries with mass ratios $q$, $e$, $i$, and $\omega$, and periastron distances relative to the sum of the radii of components $r_{\rm peri}/(R_1+R_2)$ drawn from uniform distributions $\mathcal{U}_{[\alpha,\beta]}$ on the interval $[\alpha,\beta]$. We comment on them below.
	\begin{itemize}
		\item $q\sim\mathcal{U}_{[0.2,0.9]}$ -- We did not consider $q>0.9$ because we realistically assumed that the secondary companions are MS stars. Here $q\approx 1$ would suggest that the secondary is also an RG, which in turn would result in very long orbital periods, unobserved by us. On the other hand, the overall strength of the tides is proportional to $q$; therefore, we omitted systems with $q<0.2$. The typical peak-to-peak amplitude of the heartbeat in our sample of RGs is relatively high, $\sim$\,0.05~mag; hence, it rather excludes the possibility of low-$q$ companions.
		\item $e\sim\mathcal{U}_{[0.15,0.70]}$ -- We adopted a representative range of eccentricities observed in the analyzed collection of OGLE HBSs.
		\item $i\sim\mathcal{U}_{[0,\pi/2)}$, $\omega\sim\mathcal{U}_{[0,2\pi)}$ -- Any possible values of $i$ and $\omega$ were allowed in the simulation.
		\item $r_{\rm peri}/(R_1+R_2)\sim\mathcal{U}_{[1,4]}$ -- The strength of the tidal forces at periastron falls off rapidly as $r_{\rm peri}^{-3}$. Therefore, we set a maximum limit of $r_{\rm peri}/(R_1+R_2)$ to 4 in order to simulate only the orbits with a chance to reproduce detectable heartbeat signals.
	\end{itemize}
	The values of $q,\,i,\,\omega$, and $r_{\rm peri}/(R_1+R_2)$ were generated independently from each other. The drawn values of $r_{\rm peri}/(R_1+R_2)$ were subsequently transformed to $P$, using Kepler's third law. During the simulations, only those systems were accepted that did not overflow at the periastron; i.e., the detached geometry was always preserved. The remaining physical parameters of the components, namely, their effective temperatures and the radii of the secondaries, were taken from the MIST isochrones of the age compatible with the parameters of the primary RG star. We assumed [Fe/H]\,$=-0.7$ for systems with 1 and 2$\,M_\odot$ primaries, while for 6$\,M_\odot$ we adopted a higher value of [Fe/H]\,$=-0.4$ in order to account for the metallicity gradient observed in the LMC. The bolometric albedos were set to 0.6 and 1.0 for components with convective and radiative envelopes, respectively. The surfaces of both components were simulated within PHOEBE\,2 with 4000\,--\,6000 triangular elements, depending on the drawn orbital parameters. The radiative properties of stars with effective temperatures above 4000\,K were obtained from ATLAS9 model atmospheres \citep{2003IAUS..210P.A20C}. Otherwise, the PHOENIX models (\citealt{1997ApJ...483..390H}, \citealt{2013A&A...553A...6H}) were used for cooler components. Both grids of model atmospheres are incorporated into PHOEBE\,2 with the accompanying limb-darkening tables. Finally, for each system, two $I_C$-band light curves were calculated in the full range of orbital phases, $\phi$; one with the irradiation/reflection effect being switched off, $F_{\rm irr-off}(\phi)$, and the second one with the aforementioned effect treated in the formalism developed by \cite{1990ApJ...356..613W}, $F_{\rm irr-on}(\phi)$. Since our calculations in PHOEBE\,2 were performed in the absolute scale, the following condition was satisfied for every $\phi$, $F_{\rm irr-on}(\phi)>F_{\rm irr-off}(\phi)$. This is due to the fact that the irradiation/reflection effect can only add flux to the beam. Next, we were searching for the largest difference between these two synthetic light curves, which we denote as $\max(\Delta m_{\rm irr})$. We defined this quantity in the following way:
	\begin{equation}
			\max(\Delta m_{\rm irr}) \equiv \max\limits_{\phi \in (0,1]} \left\{2.5\log\left[\frac{F_{\rm irr-on}(\phi)}{F_{\rm irr-off}(\phi)}\right]\right\}.
	\end{equation} 
	
	The upper part of Figure~\ref{fig:irr-effect} summarizes the effects of the S-RG simulations. The first thing that draws attention is the clear correlation between $q$, $r_{\rm peri}$, and $\max(\Delta m_{\rm irr})$ (second column in Figure~\ref{fig:irr-effect}). The larger the $q$ and lower the $r_{\rm peri}$, the more pronounced the impact of the irradiation/reflection effect. Nevertheless, for systems with the primary's mass $\lesssim 2\,M_\odot$, the $\max(\Delta m_{\rm irr})\lesssim 1$\,mmag. This fact allows us to state that the K95 model should return trustworthy orbital parameters for the majority of the OGLE HBSs containing an RG. This is because most of them are low-mass stars, as indicated by their positions on the Hertzsprung--Russell (H-R) diagram (Figure~\ref{fig:HR_diagram}). The situation is different for massive RGs, companions of which may have high effective temperatures and luminosities; hence, the irradiation/reflection effect is significant. In such a scenario, the 95th percentile of $\max(\Delta m_{\rm irr})$ is equal to circa 10\,mmag, but systems with $\max(\Delta m_{\rm irr})$ up to 40\, mmag can occur. Depending on the amplitude of the heartbeat observed in HBSs that contain the massive RG, one should be aware that the irradiation/reflection effect may become comparable to the ellipsoidal variability. Therefore, the orbital parameters obtained for such systems via fitting the K95 model to their light curves should be treated as estimates (see Section~\ref{sect:irr-effect-kumar}).
	
	The S-MS simulations were performed analogously to the S-RG, but here we considered different masses of the primary components, 4, 10, and 25\,$M_\odot$ (see left panel of Figure~\ref{fig:HR_diagram}). To realistically reflect the evolutionary phase of these components visible in Figure~\ref{fig:HR_diagram}, we assumed the 4\,$M_\odot$ primary to be in the middle of its Hertzsprung gap, while the 10 and 25\,$M_\odot$ primaries were located halfway between the zero-age MS (ZAMS) and terminal-age MS (TAMS). All models were assumed to have [Fe/H]\,$=-0.4$. The rest of the parameters and methods necessary to run S-MS simulations were identical to the S-RG setup described above.
	
	Similarly to the upper part of Figure~\ref{fig:irr-effect}, the lower part shows the results of the S-MS simulations. We found that for massive HBSs that are still on the MS or near it, the irradiation/reflection effect in the $I_C$ band is generally nonnegligible. The 95th percentile of $\max(\Delta m_{\rm irr})$ is equal to around 9\,mmag for all variants in these simulations. However, still numerous systems can have an amplitude of the irradiation/reflection effect in the range of 10\,--\,30\,mmag. Considering these facts, we would like to emphasize that orbital parameters derived by us for massive MS HBSs may deviate from actual values and should be treated with caution (see Section~\ref{sect:irr-effect-kumar}).
	
	\subsection{K95 Model Fitted to the Synthetic Light Curves}\label{sect:irr-effect-kumar}
	
	In the previous subsection, we were interested in the overall amplitudes of the irradiation/reflection effect among the HBSs from our sample. However, how this phenomenon will affect the derived orbital parameters is another question. Recalling that the K95 model neglects the aforementioned effect, one can expect that some systematic biases in the derived parameters might occur. Moreover, in some circumstances, the morphology of the model may not be able to effectively reproduce the actual brightness changes. To explore both this issue and the properties of the model, we performed two types of tests, which were done on the synthetic irradiated $I_C$-band light curves from the S-RG and S-MS simulations.
	
	First, we examined if the K95 model can successfully fit the synthetic light curves provided that the orbital parameters are fixed on the values injected into PHOEBE\,2. In other words, the only free parameters during the fitting procedure were $S$ and $C$ (see Equation~\ref{eq:kumar2}). In order to quantitatively describe the global effectiveness of the best fit in the entire range of orbital phases, we constructed the following metric:
	\begin{equation}
		\eta \equiv \frac{\int\limits_0^1 \big|\mathcal{F}_{\rm irr-on}(\phi)-\mathcal{K}_{\rm best-fit}(\phi)\big|{\rm d}\phi}{\int\limits_0^1 \big|\mathcal{F}_{\rm irr-on}(\phi)\big|{\rm d}\phi},
	\end{equation}
	where $\mathcal{K}_{\rm best-fit}(\phi)$ denotes the K95 model that best fits the $\mathcal{F}_{\rm irr-on}(\phi)$, which stands for $(F_{\rm irr-on}(\phi)/\langle F_{\rm irr-on}\rangle-1)$. The quantity $\eta$ is a ratio between the integral of the residuals from the fit and the integral of the normalized synthetic light curve; therefore, it is independent of the amplitude of the heartbeat. The leftmost column in Figure~\ref{fig:kumar-accuracy} (titled ``K95 fixed") presents the distribution of the mean value of $\eta$, $\langle \eta \rangle$ in the hex-binned $i$\,--\,$e$ plane, obtained for the S-RG and S-MS sets of simulations. For both types of simulations, regardless of the primary's mass, it can be seen that the K95 model with fixed orbital parameters is able to successfully fit the simulated variability when $i\lesssim 15^\circ$. For greater values of $i$, the results are always worse because the contribution of irradiation/reflection to the total light curve can no longer mimic the ellipsoidal variability. This is especially pronounced in the S-MS simulations (lower part of Figure~\ref{fig:kumar-accuracy}) for high-$e$ binaries and $i\gtrsim 40^\circ$. Another important feature is the significant difference in $\langle \eta \rangle$ between the 4 and 10/25\,$M_\odot$ S-MS binaries. The former are generally characterized by a much better quality of the fit. The situation is notably different for the S-RG simulations (upper part of Figure~\ref{fig:kumar-accuracy}). The worst matches lie in a vertical strip with inclinations in the $\sim$20$^\circ$\,--\,50$^\circ$ range, almost independent of the eccentricity of the system. This is in contrast to the effects of the S-MS simulations discussed above. In an obvious way, the models with the primary's mass of 6\,$M_\odot$ exhibit far larger departures from the K95 model when compared to the less massive S-RG systems. This is due to much greater amplitudes of the irradiation/reflection effect for this kind of HBS (see Figure~\ref{fig:irr-effect}, upper part). Nevertheless, the values of $\eta$ are statistically significantly smaller for binaries containing an RG as a primary component than those that harbor two MS stars.
	
	Our next test was analogous to the previous one, except for one major difference. This time, the orbital parameters were treated as free parameters in a least-squares fit together with $S$ and $C$. Thanks to this approach, we were able to track down any systematic biases in the determination of the orbital parameters. The results of our second test are summarized in Figure~\ref{fig:kumar-accuracy}. Similarly to the leftmost column in Figure~\ref{fig:kumar-accuracy}, the column denoted as ``K95 free" shows the distributions of $\langle \eta \rangle$ across the $i$\,--\,$e$ plane. It can be easily seen that the shape of these distributions remained nearly the same as in the previous experiment, but now the values of $\langle \eta \rangle$ are significantly smaller. This is because the K95 model is trying to fit both the ellipsoidal and irradiation/reflection variability. The price to pay for ``forcing" this fit is some systematic errors in the optimized orbital parameters. Let us denote this error for a given orbital parameter, $\mathcal{X}$, as $\Delta \mathcal{X}=\mathcal{X}_{\rm PHOEBE}-\mathcal{X}_{\rm K95}$, where $\mathcal{X}_{\rm PHOEBE}$ is an orbital parameter used to generate a PHOEBE\,2 synthetic light curve, and $\mathcal{X}_{\rm K95}$ corresponds to the parameter obtained from fitting the K95 model to the synthetic data. We note that $\Delta \mathcal{X}>0$ means that $\mathcal{X}$ is being underestimated by the K95 model, while $\Delta \mathcal{X}<0$ suggests the opposite. The right-hand part of Figure~\ref{fig:kumar-accuracy} presents quadriads of panels showing $\Delta e$, $\Delta i$, $\Delta \omega$, and $\Delta T_0/P$ for each primary's mass in the S-RG and S-MS simulations. They are plotted against $i$, which is the parameter most strongly correlated with these $\Delta \mathcal{X}$. The values presented in the corners correspond to the medians of $|\Delta \mathcal{X}|$. There are several conclusions about the behavior of the K95 model that can be drawn from distributions of $\Delta \mathcal{X}$. We discuss them briefly below.
	\begin{itemize}
		\item The systematic errors in determining all four orbital parameters for the S-RG simulations are much smaller than those for the S-MS simulations. Since the majority of our HBSs are systems with the primary being an RG, we infer that the K95 model returns reliable parameters for them. 
		\item The most accurately estimated parameters, regardless of the strength of the irradiation/reflection effect, are $e$ and $T_0$. In the set of S-RG simulations, $|\Delta e|<0.03$ and $|\Delta T_0/P|<0.03$ in almost all cases. For the S-MS simulations, the corresponding ranges are $-0.04< \Delta e<0.1$ and $|\Delta T_0/P|<0.03$.
		\item In general, $i$ is estimated with the lowest accuracy among all four orbital parameters, and its systematic error, $\Delta i$, depends in a nontrivial way on the actual $i$ of the system. For $i\lesssim 15^\circ$, the K95 model has a tendency to overestimate $i$. In turn, $\Delta i$ for the orbits with relatively high values of $i$ can behave in two ways. If the orbit is only slightly eccentric ($e\lesssim 0.2$), it is very likely that the K95 model will return $i\approx 90^\circ$. This can be seen from the diagonal line of points at the bottom of each $\Delta i$ distribution. On the contrary, if the orbit is more eccentric, the K95 model will certainly underestimate the actual value of $i$. This is because the irradiation/reflection effect fills the ``dips" in the heartbeat signal, so the light curve seems to originate from a binary with a fictitious smaller $i$.
		\item As expected, $\Delta \omega$ shows a typical dependence on $i$. The argument of periastron is well constrained for orbits with relatively high $i$'s and $e$'s but becomes poorly constrained for small values of $i$. Eventually, it is undefined for $i=0^\circ$. Let us also emphasize at this point that the K95 model allows for determining $\omega$ with a $180\degree$ ambiguity.
	\end{itemize}
    
	Although the above analysis was performed for HBSs located in the LMC, its results should also remain valid for systems from the Small Magellanic Cloud (SMC) and Galactic bulge (GB). Thanks to the S-MS and S-RG simulations that we performed, we are able to estimate the average accuracy of the orbital parameters obtained by us via modeling the $I_C$-band light curves of OGLE HBSs. Combining the results of all simulations, the median values of $|\Delta e|$, $|\Delta i|$, $|\Delta \omega|$, and $|\Delta T_0/P|$ are equal to about 0.01, 3.0, 2.5, and 0.002, respectively. However, one should be aware that some individual situations can still be characterized by relatively large $\Delta \mathcal{X}$, especially when it comes to the determined $i$. In particular, the massive MS HBSs are vulnerable to such effects in our modeling.

	\section{Detection of TEOs in the OGLE HBSs} \label{sec:teo}
	\subsection{General Properties of TEOs}
	\label{sect:teos}
	The majority of TEOs known so far are forced damped gravity ($g$) modes that may dissipate the total orbital energy and make the system tighter with time. Therefore, TEOs may play a significant role in the dynamical evolution of the binary system. In general, TEOs come in two ``varieties." The first one occurs when the frequency of the eigenmode temporarily coincides with the harmonic, $n$, of the orbital frequency, $f_{\rm orb}$. In such circumstances, which are called a ``chance resonance," the amplitude of a TEO does not exceed the parts per thousand level and is most often significantly smaller. The other variety is associated with the so-called ``resonantly locked modes" (see \citealt{2017MNRAS.472.1538F}; \citealt{2018MNRAS.473.5165H}). These are forced normal modes with frequencies evolving due to the stellar evolution at the same rate and direction as the nearest harmonic of the orbital frequency. Therefore, resonantly locked TEOs have enough time to gain relatively high amplitudes and hence effectively dissipate the total orbital energy. High-amplitude TEOs are expected to be quadrupole ($l=2$) modes with a mainly azimuthal order $m=0$ or~2. Although the history of theoretical studies of the dynamical tides dates back to the 1970s \citep{1970A&A.....4..452Z}, the actual effectiveness of energy dissipation due to TEOs and their impact on the binary's evolution are still a matter of large uncertainties. Studies of large-amplitude TEOs offer a unique opportunity to make progress in this subject.
	
\begin{figure*}[]
		\centering
		\includegraphics[width=0.97\textwidth]{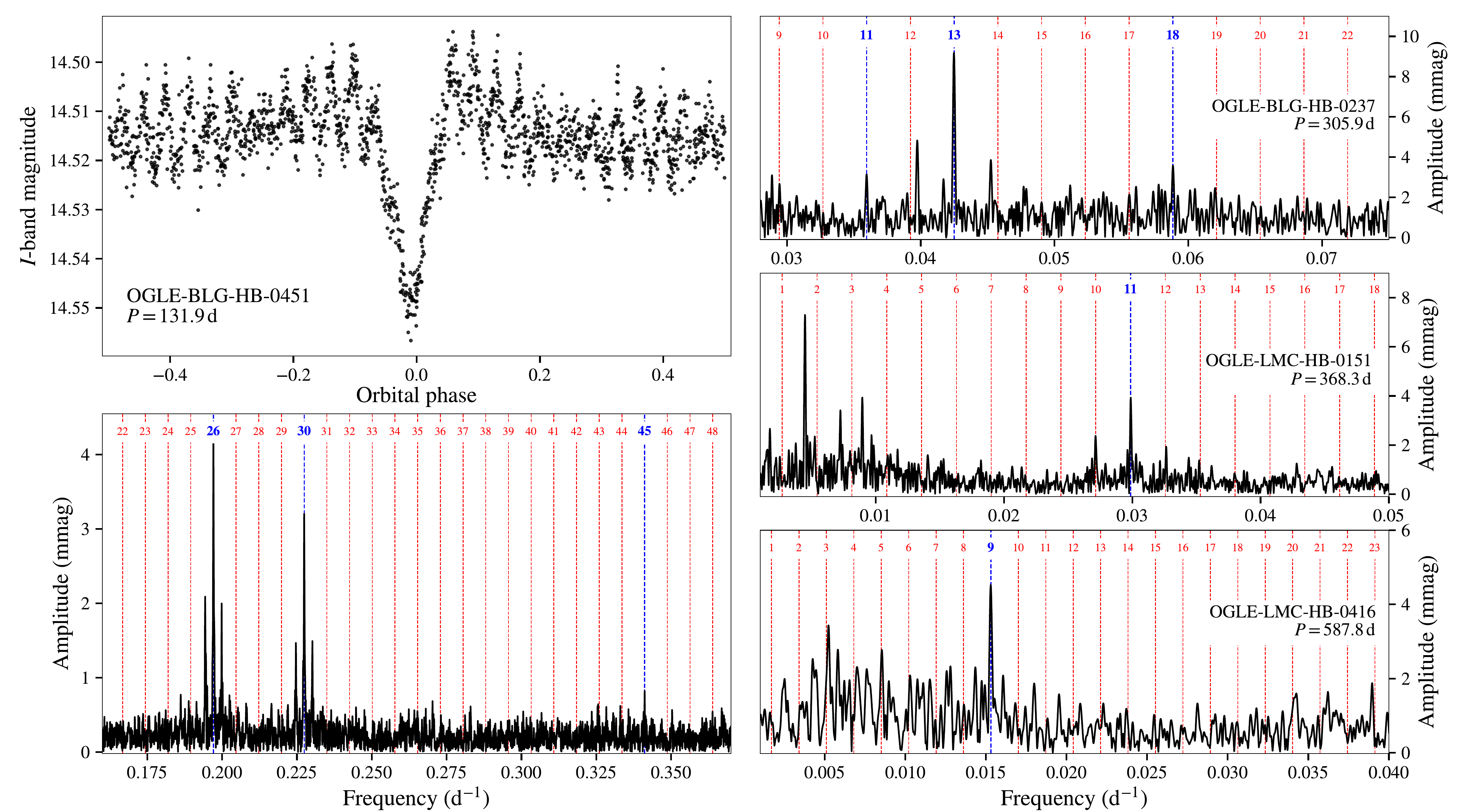}
		\caption{Sample of TEOs detected in the OGLE HBSs. Top left: phase-folded light curve of OGLE-BLG-HB-0451. Zero phase corresponds to the periastron passage. Both the heartbeat (around the periastron) and the high-amplitude TEOs (across the entire range of orbital phases) are clearly visible. Bottom left: Fourier frequency spectrum of the light curve of OGLE-BLG-HB-0451 (shown in the top left panel) after subtraction of the best-fitting K95 model. The vertical dashed lines mark the position of the consecutive harmonics of the orbital frequency. Multiple TEOs at $n=26,\,30$, and 45 are highlighted with vertical blue lines. Side peaks around the TEOs at $n=26$ and 30 are due to aliasing. Right: Fourier frequency spectra calculated analogously to the frequency spectrum presented in the bottom left panel but for the OGLE-BLG-HB-0237, OGLE-LMC-HB-0151, and OGLE-LMC-HB-0416 systems.}
		\label{fig:teo-peri}
	\end{figure*}

	The main observational difference between self-excited pulsations and TEOs is that the latter have frequencies exactly equal to the integer multiples\footnote{Assuming the linear theory of tides. In turn, nonlinear tidal effects can lead to a minor offset between the observed frequency of a TEO and its corresponding harmonic of the orbital frequency.} of $f_{\rm orb}$; therefore, they phase well with the orbital period, $P$ (see the top left panel in Figure~\ref{fig:teo-peri}). Nevertheless, because of the resonant nonlinear mode coupling \citep[NLMC, described extensively by, for instance, ][]{1982AcA....32..147D,1985AcA....35....5D,1988AcA....38...61D}, it is also possible to observe nonharmonic TEOs that are ``daughter" modes of the resonant (harmonic) TEO, which is the ``mother" mode (e.g., \citealt{2020ApJ...896..161G}). The simplest manifestation of this mechanism is the decay of the resonant TEO into two modes, the sum of the frequencies of which equal $n\cdot f_{\rm orb}$. However, one can also expect the quintuplets, septuplets, etc. of frequencies formed via higher-order NLMC and/or non-harmonic TEOs being e.g. a ``granddaughter" modes (i.e. the ``daughter'' modes of some prior ``daughter'' mode).
	
	There are several other features that make TEOs unique when compared to the self-excited oscillations. Therefore, the whole branch of research on this subject is called ``tidal asteroseismology." One of the most extensively studied HBSs in terms of tidal asteroseismology is the aforementioned KOI-54, for which this type of analysis was performed by \cite{2012MNRAS.421..983B}. There are also other papers dedicated solely to this topic, e.g.~\cite{2020ApJ...888...95G}.
	
	While the TEOs observed in the MS stars are mostly $g$ modes, the situation is diametrically different for the RGs, which make up the vast majority of our sample of HBSs. Both the Brunt--V\"ais\"al\"a and Lamb frequencies are very large inside the dense core of an RG; therefore, a nonradial oscillation mode has a dual nature in its interior. It propagates as an acoustic mode in the convective envelope of an RG but as a $g$ mode in its radiative core. This is the reason why these kinds of modes are called mixed modes \citep{2010aste.book.....A}. They were detected in many RGs, mainly thanks to the ultraprecise photometry delivered by the Kepler mission \citep[see, e.g.,][and references therein]{2011Sci...332..205B,2017A&ARv..25....1H}. The mixed modes are generally problematic when it comes to their analysis and modeling, for several reasons. Their wavelengths are extremely short in the core of an RG; hence, the corresponding eigenfunctions are characterized by a large number of nodes in the radial direction (typically of the order of $10^3$). Next, they are expected to undergo severe radiative damping when traveling through the dense core \citep{1971AcA....21..289D} and be subject to significant nonlinear effects in the upper part of the red giant branch (RGB; \citealt{2019ApJ...873...67W}). \cite{2009A&A...506...57D} performed the theoretical investigation of amplitudes and lifetimes of the self-excited radial and nonradial ($l=1,2$) oscillations in the RGs (their ``model C" is the closest to the RG HBSs analyzed in our work). The authors found that the aforementioned properties of mixed modes are a sensitive function of the density contrast between the core and the envelope (i.e., the position in the RGB). The amplitudes and lifetimes of nonradial modes also depend on which part of an RG the mode is trapped in. As expected, the modes trapped in the core should have small amplitudes on the surface, in contrast to the modes trapped in the extended envelope. Considering these facts, one does not expect to observe many nonradial modes in the RG stars, especially located high in the RGB, provided that they are stochastically driven by the turbulent convection. An interesting question arises, however: how will this picture change for the mixed modes that are excited by periodic and coherent variation in tidal forces \citep[see, e.g.,][]{2013MNRAS.429.2425F}? More importantly, what is the impact of tidally excited mixed modes on the dynamical evolution of the HBSs with an evolved component(s)? In order to help answer the questions raised above, we provide the community with the compilation of high-amplitude TEOs detected in the OGLE HBSs (see Appendix \ref{sec:appendix}, Tables~\ref{tab:teos-blg} and~\ref{tab:teos-mc}).
	
	\subsection{Methodology of Search}
	We search for TEOs in our sample of HBSs by means of the Fourier analysis. We performed a standard iterative prewhitening procedure on the residual light curves obtained after the subtraction of the best-fitting K95 model. The vast majority of the residual light curves reveal the presence of long-term variability that, regardless of whether it is physical or not, significantly enhances the signal in the frequency spectra at low frequencies. In order to get rid of these long-term brightness changes, prior to the prewhitening, it was modeled with Akima cubic splines \citep{10.1145/321607.321609} and subtracted from each residual light curve. After calculating the error-weighted Fourier frequency spectra, only peaks with a signal-to-noise ratio (S/N)$\geq 4$ were considered statistically significant. The mean noise level, $N$, was derived as the mean signal in the frequency spectrum in the frequency range 0\,--\,10 day$^{-1}$. The final parameters characterizing the coherent variability were obtained by means of the error-weighted nonlinear least-squares fitting of the truncated Fourier series to the corrected residual light curves. The formal errors of the extracted frequencies and their amplitudes were estimated using the covariance matrix.
	
	The frequency $f$ was considered a TEO if it was sufficiently close to the nearest harmonic of the orbital frequency, i.e., if the following condition was satisfied: $|f/f_{\rm orb}-n|\leq 3\sigma_{f/f_{\rm orb}}$. We estimated the error of $f/f_{\rm orb}$, $\sigma_{f/f_{\rm orb}}$ using a standard error-propagation formula, $\sigma_{f/f_{\rm orb}}=(P^2\sigma_f^2+f^2\sigma_{P}^2)^{1/2}$, where $\sigma_f$ and $\sigma_{P}$ stand for the error of $f$ and $P$, respectively. Recalling that TEOs can also have a nonharmonic nature (provided that they formed via NLMC), we examined if any sum of two nonresonant frequencies, $f_1$ and $f_2$, fulfills the inequality $|(f_1+f_2)/f_{\rm orb}-n|\leq 3\sigma_{(f_1+f_2)/f_{\rm orb}}$, where $\sigma_{(f_1+f_2)/f_{\rm orb}}$ was calculated analogously to $\sigma_{f/f_{\rm orb}}$. We did not look for the presence of nonharmonic TEOs resulting from higher-order NLMC due to the insufficient precision of the fitted frequencies.
	
	\subsection{Results of Search}\label{sect:teos-results}
	The results of our search are presented in Tables~\ref{tab:teos-blg} and~\ref{tab:teos-mc}. In total, we were able to find 52 systems (five on the MS and 47 containing a post-MS star) out of 991 ($\sim$\,5\,\%) that exhibit at least a single TEO, while the total number of detected TEOs amounts to~78. 
	
	Figure~\ref{fig:teo-peri} shows the compilation of frequency spectra of four sample OGLE HBSs with detected TEOs. The left-hand side of Figure~\ref{fig:teo-peri} refers to one of the most prominent TEOs detected by us in OGLE-BLG-HB-0451. The out-of-periastron variability of this object reveals the clear beating pattern between the dominant $n=26$ and 30 TEOs. We also found evidence that 12 nonresonant frequencies present in four systems, namely, OGLE-BLG-HB-0066, OGLE-BLG-HB-0157, OGLE-BLG-HB-0208, and OGLE-BLG-HB-0362, are possible nonharmonic TEOs formed via the NLMC. We did not detect their corresponding ``mother" modes; however, this can be explained with their amplitudes below the detection limit (typically between 0.2 and 3.0 mmag in the analyzed OGLE light curves).
	
	Additionally, we have marked the systems in Tables~\ref{tab:teos-blg} and~\ref{tab:teos-mc} that, in parallel to the TEOs, also show a pronounced intrinsic periodic variability\footnote{Let us emphasize that we did not examine the coherence of these periodic signals and their stability over time.}. This may help in future research about the interaction between tides and intrinsic pulsations. Do tidal interactions suppress self-excited pulsations, or do they not have a major impact on them (e.g., \citealt{10.1093/mnras/stt1041}, \citealt{2020MNRAS.498.5730F})? The cases in which we did not detect any significant periodic variability are also interesting because of the question raised above.
	
\begin{figure}[h]
		\includegraphics[width=0.48\textwidth]{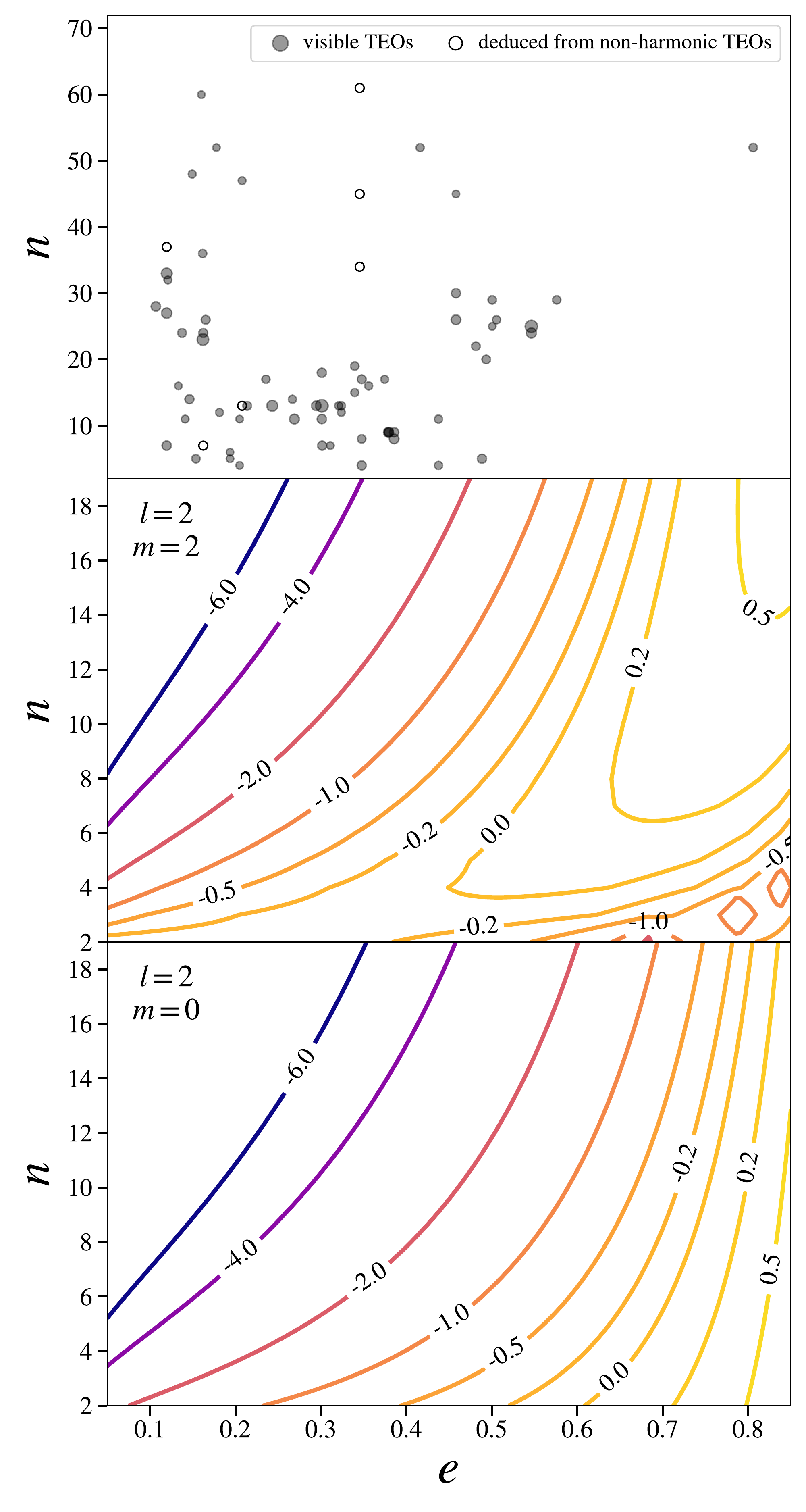}
		\caption{Top: empirical relation between the harmonic number of TEOs detected in the OGLE HBSs (Table~\ref{tab:teos-blg} and Table~\ref{tab:teos-mc}) and their estimated eccentricities. Directly detected TEOs are denoted with filled circles. The higher the amplitude of a TEO, the bigger the circle. Open circles correspond to the TEOs deduced from the presence of `daughter' nonharmonic TEOs. For more details, see Section~\ref{sect:teos-results}. Middle: contour plot of $\log(|X_{nm}|)$ versus $n$ and $e$ for $l=2$, $m=2$ modes. Bottom: same as middle panel but for $m=0$ modes.}
		\label{fig:teo-N-e}
	\end{figure}

	The presented sample of TEOs is the largest homogeneous sample of this kind known so far, which allows us to statistically investigate the dependence between parameters like $n$, $e$, and the amplitude of a TEO. From a theoretical point of view, the amplitude of a TEO depends on several factors that were described in detail by \cite{2017MNRAS.472.1538F} (his Equation (2) and related equations). Nevertheless, one can estimate the range of $n$, which favors excitation of a TEO, considering only the product of the so-called overlap integral, $Q_{kl}$, and the Hansen coefficient, $X_{nm}$. The former describes the spatial coupling between the tidal potential and a given oscillation mode of radial order $k$ and degree $l$. Thus, it generally has the greatest values for small $|k|$ and $l$. Under the assumption of spin--orbit alignment in the system, $X_{nm}=W_{lm}F_{nm}$. Supplementary to $Q_{kl}$, $F_{nm}$ describes the temporal coupling between the mode and characteristic time of periastron passage. It is given by the following expression \citep[][his Equation (5)]{2017MNRAS.472.1538F}:
	\begin{equation}\label{eq:hansen}
		F_{nm}=\frac{1}{\pi}\int\limits_0^\pi\frac{\cos[n(E-e\sin E)-m\varphi(t)]}{(1-e\cos E)^l}\,{\rm d}E.
	\end{equation}
	Here $W_{lm}$ is a constant depending only on the geometry of a given mode. Keeping the assumption about spin-orbit alignment, $W_{20}=-(\pi/5)^{1/2}$, while $W_{22}=(3\pi/10)^{1/2}$. Figure~\ref{fig:teo-N-e} (middle and bottom panels) shows the evolution of $\log(|X_{nm}|)$ with increasing eccentricity for quadrupole modes with $m=0$ and 2. The maximum of $|X_{n0}|$ is always located at $n=1$, regardless of the eccentricity, but the values of $|X_{n0}|$ become greater in the entire range of $n$ for greater eccentricities. The $m=2$ modes are characterized by $|X_{n2}|$, which peaks at $n>1$. Moreover, the position of the maximum moves to higher $n$ with the increasing eccentricity. In principle, $|Q_{kl}|$ and $|X_{nm}|$ peak at different values of $n$; therefore, only their product and its maximum inform about the most favorable $n$ for the occurrence of TEOs. Since $Q_{kl}$ does not depend on $e$, while $X_{nm}$ does, the maximum of $|Q_{kl}X_{nm}|$ shifts toward greater $n$ for increasing eccentricity \citep[see][their Figures 2 and 3]{2012MNRAS.421..983B}. Hence, we can expect that in binaries with greater eccentricity, we will observe, on average, TEOs with higher values of $n$.
	
	The empirical relation between the $n$ of a TEO and the eccentricity can be seen in the top panel of Figure~\ref{fig:teo-N-e}. First of all, theoretical predictions about the positive correlation between $n$ and $e$, which we described above, are reflected in the obtained empirical distribution of TEOs. It seems plausible to claim that the observed distribution of $n$ and $e$ resembles the shape of $|X_{nm}|$ distributions presented in the middle and bottom panels of Figure~\ref{fig:teo-N-e}. The direct comparison between the theoretical distributions of $|Q_{kl}X_{nm}|$ and the observed properties of TEOs would require the calculation of the overlap integrals, which is beyond the scope of this paper. Next, we did not find any clear relation between the amplitude of a TEO, $e$, and $n$, but yet we note that TEOs are not restricted to strongly eccentric binaries. In our sample, we can still observe high-amplitude TEOs even in the systems with $e\approx0.1$. Some of these TEOs are related to the surprisingly large harmonics of the orbital frequency, despite the low value of $e$. Finally, we did not find any obvious dependence between the location of the HBSs on the H-R diagram and the presence of TEOs (see Figures~\ref{fig:CMD} and \ref{fig:HR_diagram}).
	
	The presented collection of evolved OGLE HBSs that exhibit TEOs is a step forward in the application of tidal asteroseismology to the RG stars and studies of the mixed-mode TEOs.
	
	\section{Discussion} \label{sec:results}
	\subsection{Comparison of Obtained $e$, $i$, and $\omega$ Parameters with Previous Works}
	To verify if the resulting parameters are consistent with true physical values, we have compared the solutions for the sample of our HBSs to the ones obtained using different methods, unrelated to the K95 model. For the comparison, we used the results of the work by \cite{2017ApJ...835..209N}. The authors studied 81 ellipsoidal RG binaries cataloged in the OCVS. In the modeling process, based on the 2010 version of the WD code, they used both light (mainly from OGLE-II and OGLE-III) and radial velocity curves using data collected during their previous works (\citealt{2010MNRAS.405.1770N}; \citealt{2012MNRAS.421.2616N}; \citealt{2014AJ....148..118N}). For the details of the modeling process using the WD code, we refer the reader to Section~3 of \cite{2017ApJ...835..209N}. The total sample of 81 ellipsoidal variables consists of 59 systems with circular orbits and 22 with eccentric ones. Among the latter group, 19 stars were present in our catalog. The three remaining systems have a low eccentricity (0.069, 0.061, and 0.054), and we have considered them as classical ellipsoidal variables.
	
	\begin{figure}
		\centering
		\includegraphics[width=0.48\textwidth]{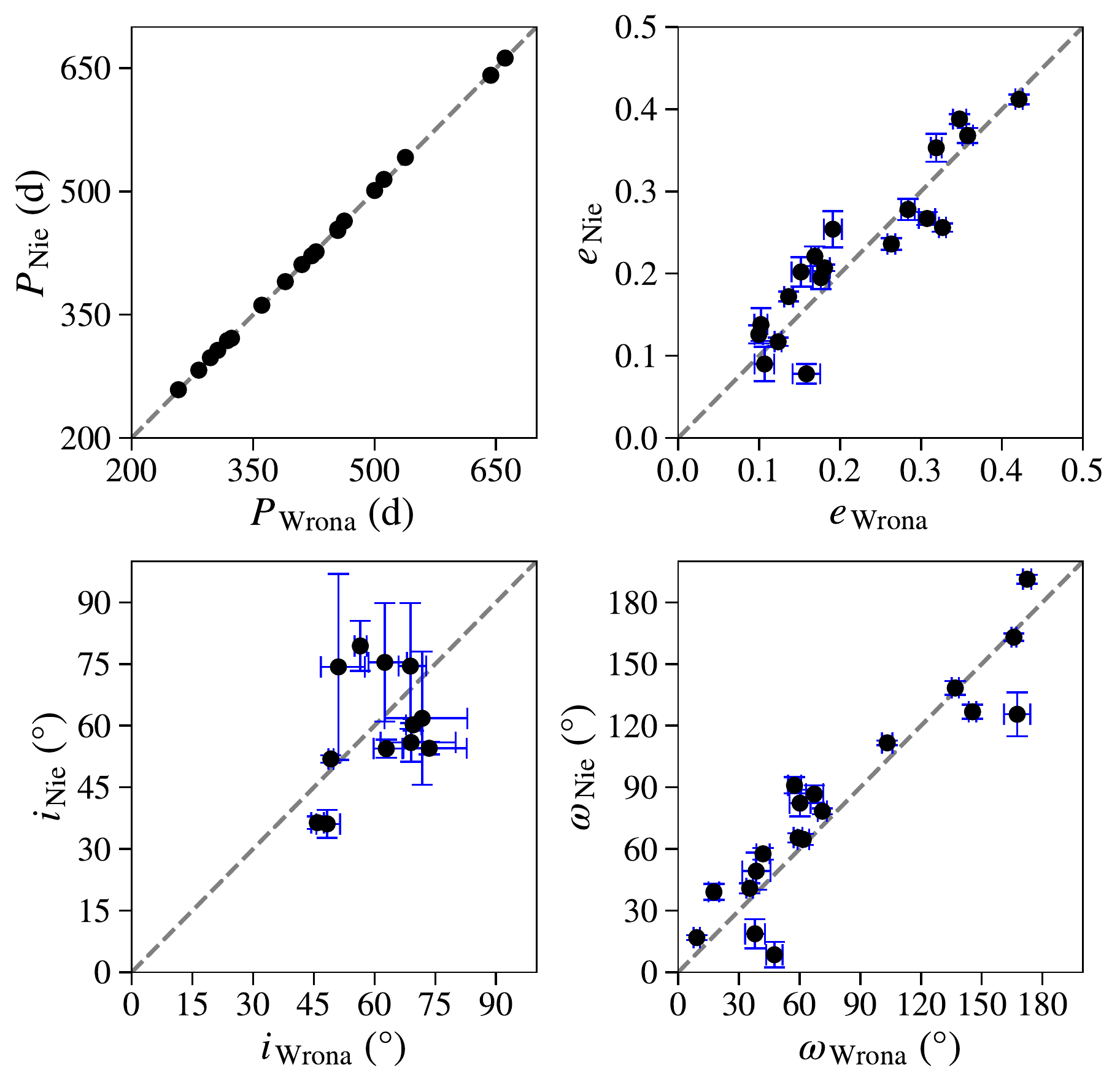}
		\caption{Comparison of the orbital parameters ($P$, $e$, $i$, $\omega$) obtained based on the WD code (\citealt{2017ApJ...835..209N}, vertical axis) and using the K95 model (this work; horizontal axis) for the overlapping sample of HBSs. The dashed gray line represents the $y=x$ line. We do not show error bars for $P$ because their sizes are comparable to the plotted dots.}
		\label{fig:Nie_comparison}
	\end{figure}
	
	The results of the comparison of the orbital parameters $P$, $e$, $i$, and $\omega$ are presented in Figure~\ref{fig:Nie_comparison}. The horizontal and vertical axes represent values obtained during our analysis and the one presented by \cite{2017ApJ...835..209N}, respectively. In the diagram containing inclination, we did not include stars with a $90\degree$ flag from \cite{2017ApJ...835..209N} because it was assigned to the model when the WD code did not converge. Values of $\omega$ from \cite{2017ApJ...835..209N} were reduced to the range $0\degree-180\degree$ that is consistent with the range of $\omega$ used in our analysis. Periods in both works are nearly identical. One can also see a strong correlation for the $e$ and $\omega$ values. Almost all pairs of points are consistent within the $3\sigma$ region. This is also true for the $i$ parameter, but contrary to $e$ and $\omega$ it does not show a clearly visible linear trend. This could be caused by the clumping of the $i$ parameter in the $50\degree-80\degree$ range. Moreover, both sets of data usually have unreliable results for $i\gtrsim70\degree$; values obtained by \cite{2017ApJ...835..209N} often have large uncertainties for such high inclination angles, and the K95 model used in our analysis returns over- or underestimated values of $i$, depending on the orbit's eccentricity (see Section~\ref{sect:irr-effect-kumar}). The results shown in Figure~\ref{fig:Nie_comparison} confirmed that the K95 model is capable of finding reliable orbital parameters for HBSs containing an RG. In Section~\ref{sec:irr_ref}, we also discussed the reliability of the K95 model in the dependence on the irradiation/reflection effect.
	
	\subsection{Distribution of the K95 Model Parameters}
	
	\begin{figure}
		\centering
		\includegraphics[width=0.48\textwidth]{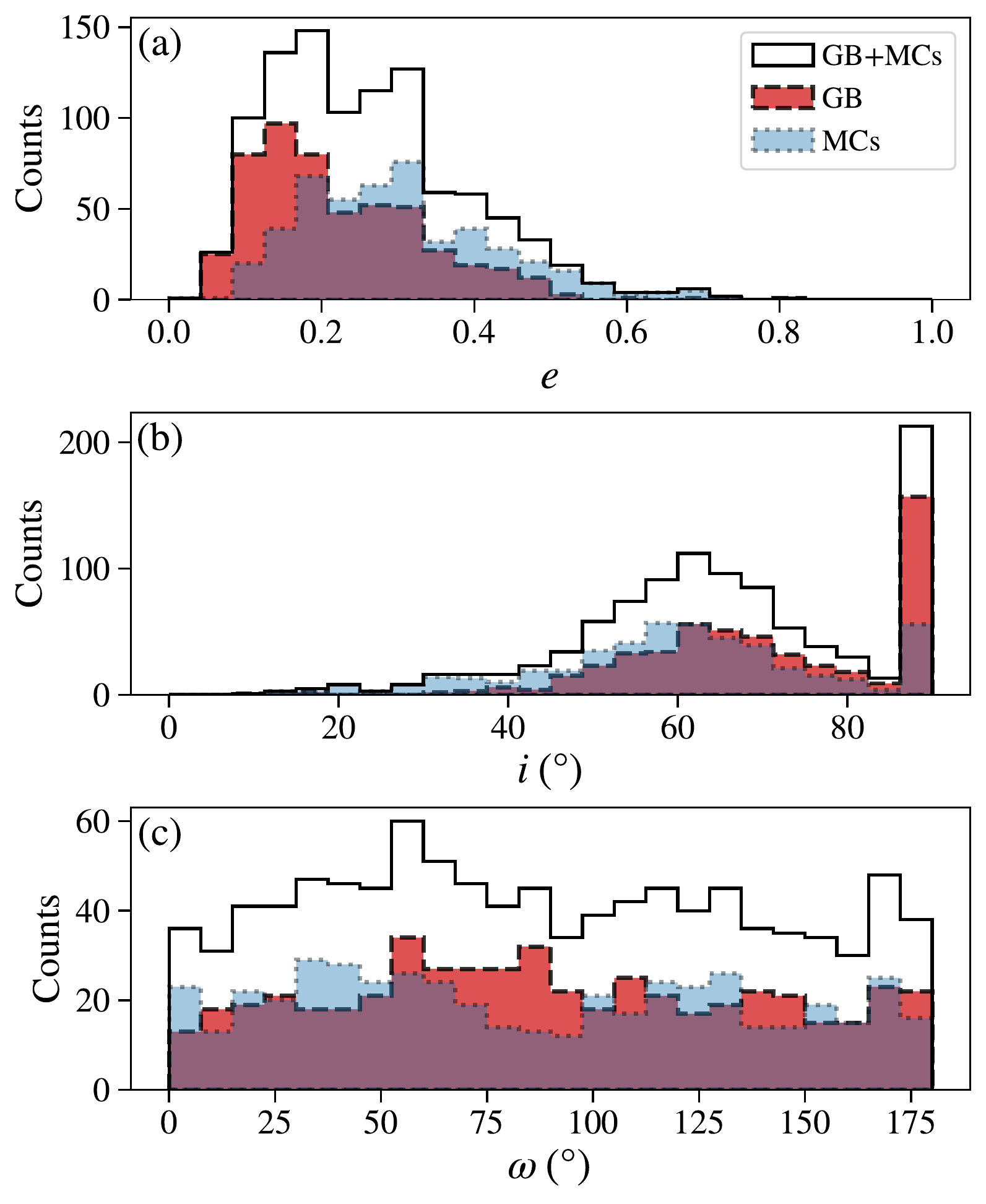}
		\caption{Histograms of orbital parameters for the sample of our HBSs located toward the GB (red boxes) and in the MCs (blue boxes). With the solid black line, we denote the histogram of the combined samples of HBSs from all locations. In panel (a), we present the histogram of the eccentricity, $e$; in panel (b), we present the histogram of inclination, $i$; and in panel (c), we present the histogram of the argument of the periastron, $\omega$.}
		\label{fig:params_hist}
	\end{figure}
	
	In Figure~\ref{fig:params_hist}, we present histograms of $e$, $i$, and $\omega$. Each chart shows a distribution of a given orbital parameter, separately for stars located toward the GB and MCs (colored lines) and combined (black lines). An HBS must have an eccentric orbit by definition, and the higher the value of $e$, the easier it is to distinguish an HBS from a classical ellipsoidal variable; thus, the low number of HBSs with $e\lesssim0.1$ is well understood. The lack of stars with high eccentricity is the result of a natural tendency of binaries to circularize their orbit (e.g., \citealt{1975A&A....41..329Z}; \citealt{1980A&A....92..167H}). In the histogram of $\omega$, we have not noticed any specific features. The distribution of this parameter is nearly uniform. 
	
	Distribution of the observed inclination angles of the HBSs may be influenced by many factors. We discuss some of them below.
	\begin{itemize}
		\item The orbit of a binary system can be oriented in any direction; however, the distribution of the observed inclination angles would not be uniform and would favor higher values of inclination. This is because the probability of observing a system with an inclination angle between $i$ and $i+\dd i$ is proportional to the area of the spherical sector between these angles. The area of the surface element on the unit sphere is proportional to the sine of the inclination angle; therefore, the area of the annular spherical sector bounded by angles $i$ and $i+\dd i$ will be greater near the orbital plane than near the pole. The resulting inclination angle distribution turns out to be uniform in $\cos i$.
		\item \cite{1995ApJ...449..294K} mentioned that the amplitude of the light curve depends mainly on the mass and structure of the star, although if we view a system from different angles, the contribution to the observed light curve from various modes changes. For small $i$, the main contribution to the light curve comes from the $m=0$ mode, while for higher $i$ modes, $m=\pm2$ starts to dominate. The change of the contribution between modes mainly impacts the shape of the light curve, but it also affects the brightness. Thus, if the observer sees only low-amplitude $m=0$ modes, the brightness changes could be too small for proper classification of the star.
		\item Another factor that may skew the $i$ and $\omega$ distribution is the misclassification of stars based on the shape of the light curve. The HBSs, especially with low inclinations, may mimic other types of variability, such as spotted stars (mainly Ap-type variables; e.g., \citealt{2019ApJ...879..114I}) or some kind of Be stars. On the other hand, for high $i$ and for $\omega$ near $90\degree$, the light curve has a striking resemblance to the eclipsing binary (in the case of high eccentricity) or the classical ellipsoidal variable (for lower $e$).
		\item In the histogram of $i$, most prominent is the peak for $i\approx90\degree$. First of all, it is observational bias, caused by the fact that the majority of OGLE HBSs were found in the catalogs of eclipsing and ellipsoidal variables, which exhibit clear minima. Secondly, as we discussed in Section~\ref{sect:irr-effect-kumar}, for low eccentric orbits, due to the irradiation/reflection effect, the K95 model tends to overestimate the value of $i$ and usually returns $i\approx90\degree$, while for more eccentric orbits values of $i$ are underestimated. This inaccuracy of the K95 model also explains a dip in the $i$ distribution between $70\degree$ and $90\degree$.
		\end{itemize}	
	
	\subsection{Color--Magnitude Diagrams}
	
	\begin{figure*}[t]
		\centering
		\includegraphics[scale=0.34]{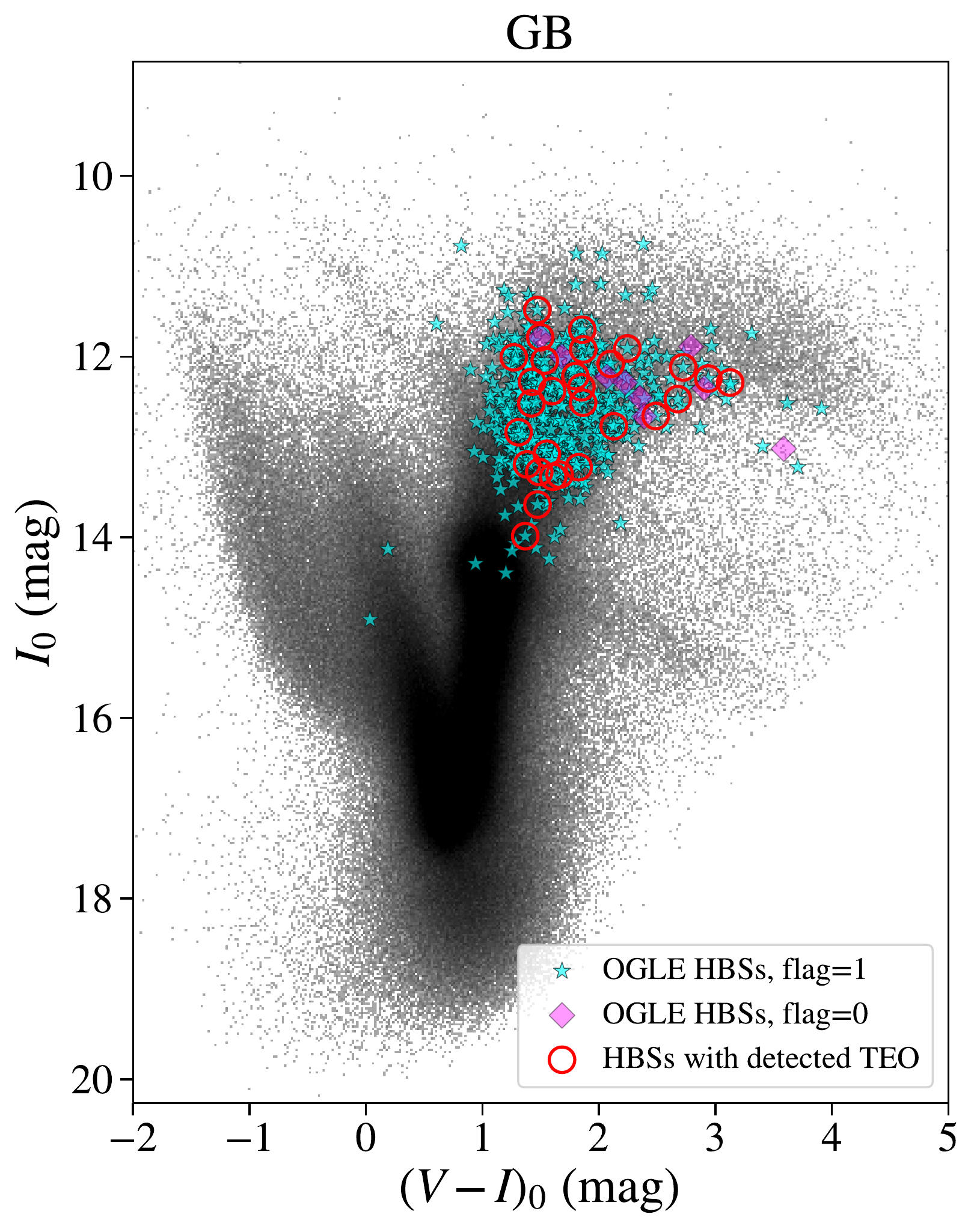}
		\includegraphics[scale=0.34]{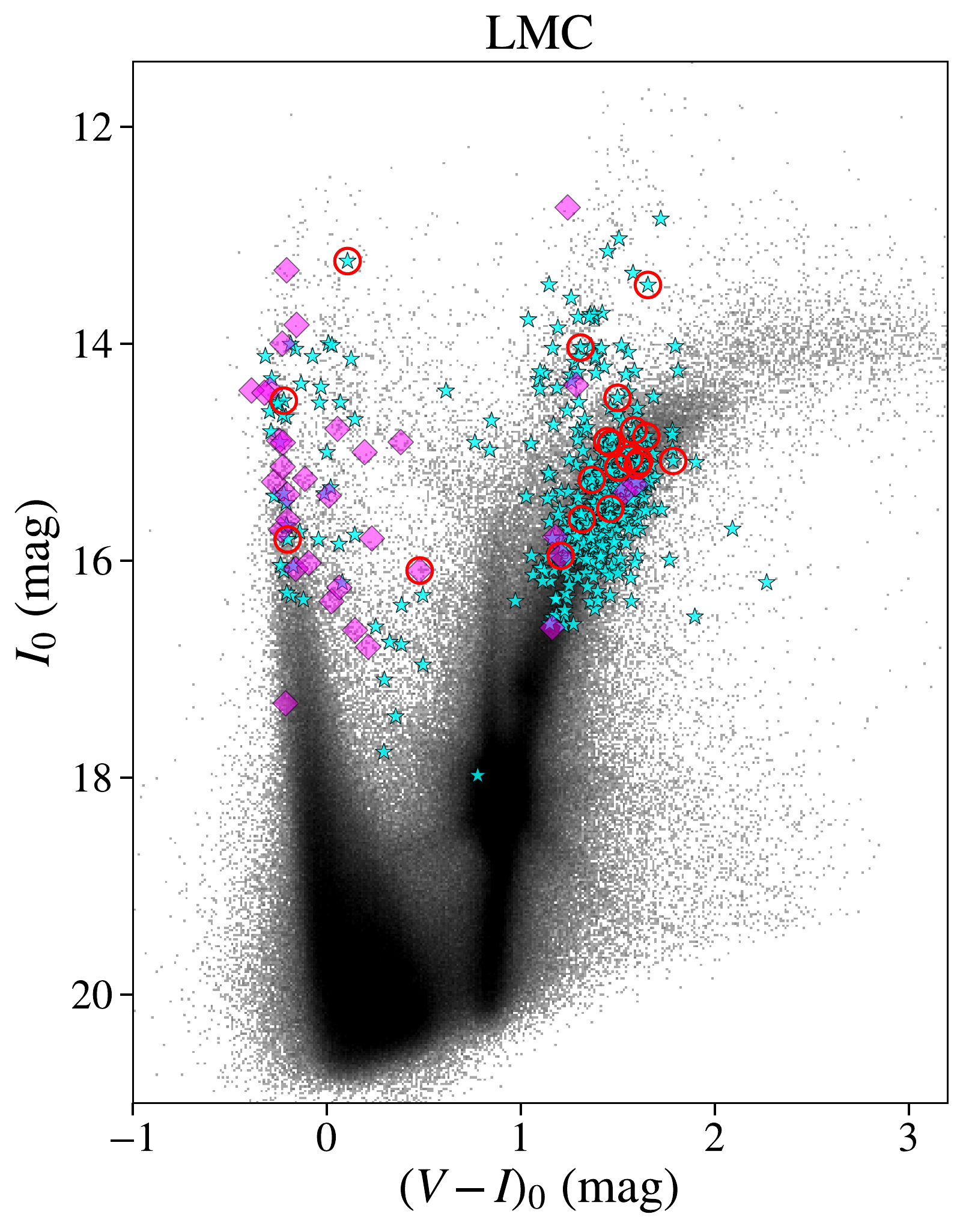}
		\includegraphics[scale=0.34]{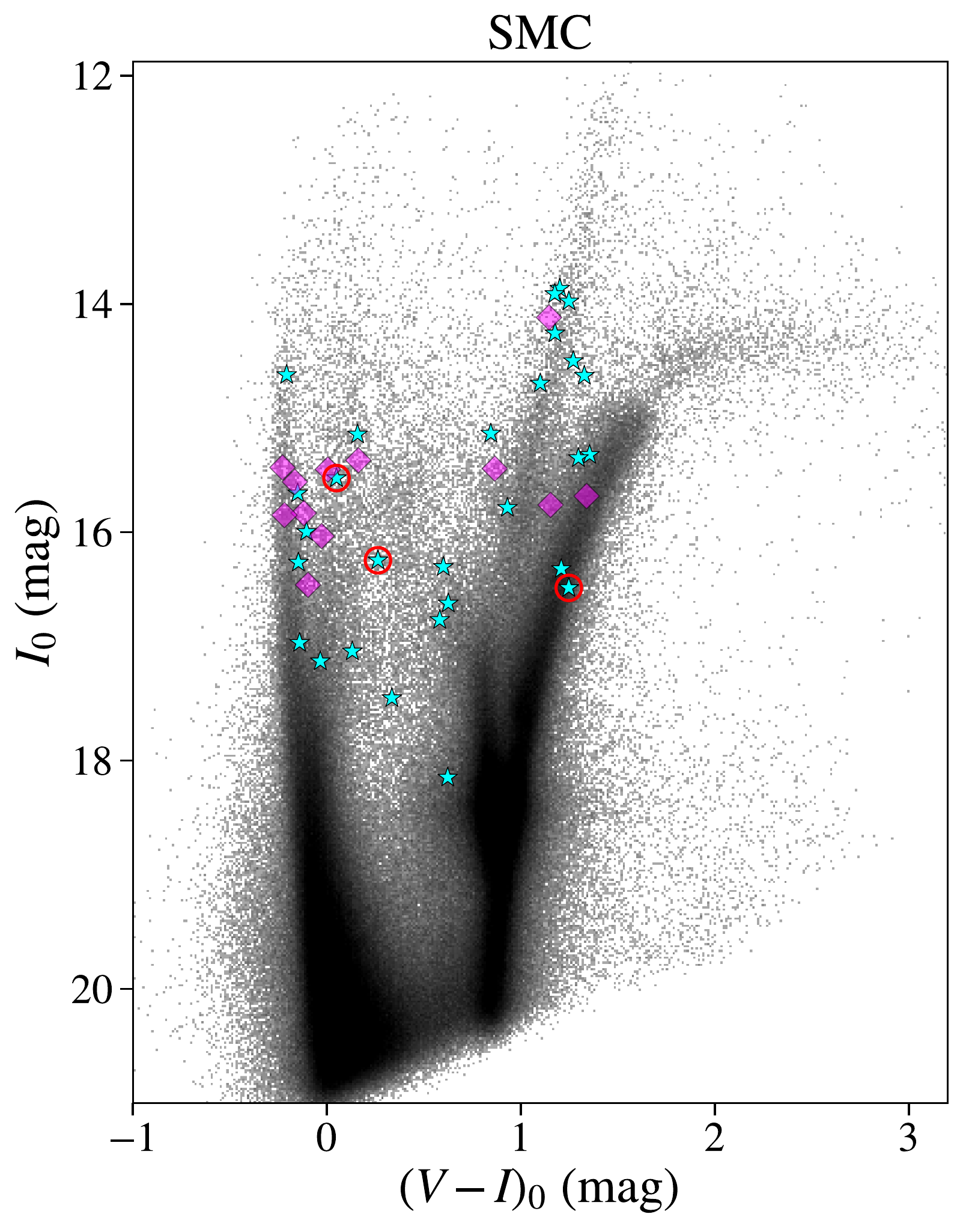}
		
		\caption{The CMDs for the stars (gray points) located toward the GB (left panel), LMC (middle panel), and SMC (right panel) and for the HBSs from the catalog (colored stars and diamonds). The horizontal axis shows the dereddened color index, and the vertical axis is the mean magnitude with subtracted extinction. Stars (flag=1) represent HBSs with the well-fitted K95 model, while diamonds (flag=0) denote stars without a proper model (mainly systems showing deep eclipses). With red circles, we mark HBSs with detected TEOs.}
			
		\label{fig:CMD}
	\end{figure*}
	
	The majority of the previously known HBSs have been found in the Kepler mission database; therefore, those systems are mainly F-, G-, and K-class objects located on the MS. There is also a group of systems containing hotter and usually more evolved stars of classes O, B, and A (e.g., \citealt{2021A&A...647A..12K}, and references therein). About 10\% of the OGLE HBS sample represents that group of stars.
	
	On the other hand, about 90\% of our HBSs belong to the RGB or asymptotic giant branch (AGB) and, less frequently, to the horizontal branch. A large group of RG ellipsoidal variables was described by \cite{2004AcA....54..347S}. The majority of those systems have a circular orbit, but the authors also accentuated a group of systems with high eccentric orbits. Until our work, presented in Paper I, they were the largest collection of RG HBSs (officially cataloged as ellipsoidal variables). 
	
	\subsubsection{Color--Magnitude Diagram for the GB}
	The sample of our HBSs located toward the GB is dominated by stars lying on the RGB and AGB, which are noticeable in the color--magnitude diagram (CMD) in the left panel of Figure~\ref{fig:CMD}. The background for the CMD has been created based on the calibrated photometric maps for several OGLE-IV fields located around the Galactic center. The \textit{I}-band extinction values $A_I$ and color excess of stars $E(V-I)$ were determined using extinction maps presented by \cite{2013ApJ...769...88N}. Extinction toward the GB is very heterogeneous what causes a widening of the color and magnitude distributions in the CMD, which is clearly visible on the red clump (RC; a wide group of points around $[(V-I)_0,I_0]=[1.05, 14.3]$). If there was no extinction, most of the HBSs from our collection would be saturated because the brightness limit for the Warsaw telescope is about 13 mag in the $I$ band. There are only two systems on the MS. Such a small number of HBSs in this region was very unexpected because most of the HBSs known before were in this place in the CMD. The absence of such HBSs in our catalog is probably the result of the data selection.  
	
	\subsubsection{Color--Magnitude Diagram for the MCs}
	In the middle and right panels of Figure~\ref{fig:CMD}, we present the positions of the HBS samples in the CMDs from the LMC and SMC, respectively. Both CMDs have been created similarly to the CMD for the GB. We took calibrated photometric maps for several OGLE-IV fields and combined them, creating the background for the HBS sample. The $E(V-I)$ color excess was calculated using the reddening map of the MCs (\citealt{2021ApJS..252...23S}).
	
	\begin{figure*}[]
		\centering
		\includegraphics[width=0.95\textwidth]{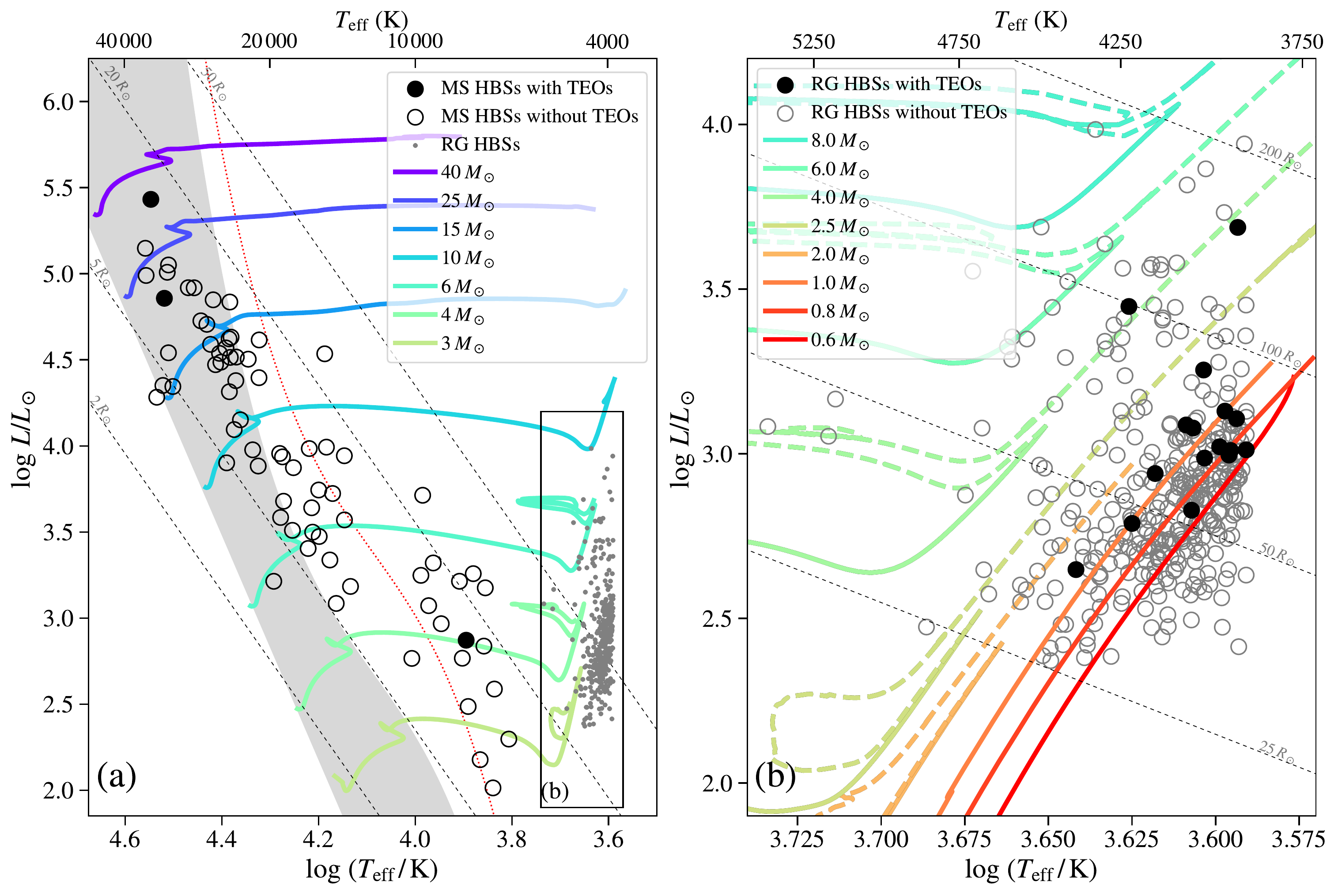}
		\caption{The H-R diagram for the LMC subsample of the OGLE HBSs (panel (a)) and close-up view of the part of the H-R diagram containing the RG HBSs (panel (b)). Large filled and open black circles represent HBSs with or without detected TEOs, respectively. The small gray dots in the left panel are RG HBSs. The shaded gray area indicates the position of the MS. The dotted red line represents the shifted TAMS line. Colored solid lines denote evolutionary tracks generated with MIST v1.2. In the left panel, we show evolution tracks only to the He core-burning phase, while in the right panel, for the masses below or equal to 1~$M_\odot$, we show an evolution track only to the RGB tip (solid lines). For higher masses, we also plot the evolution after this phase (dashed lines). In both panels, with dashed black lines, we show lines of constant radius. See more details in the text.}
		\label{fig:HR_diagram}
	\end{figure*}
	
	Contrary to the distribution of HBSs in the CMD for the GB, here we can easily distinguish at least two large groups of HBSs. The first one consists of hot MS and Hertzsprung-gap stars of spectral type from late O to F, which is consistent with HBSs from the Kepler sample. The second noticeable group (especially for the LMC) is located on the RGB and AGB, which is in agreement with the results for the GB. There are also less numerous groups, e.g., one HBS near RC for each location and a few stars on the horizontal branch.

	\subsection{Evolutionary Status of the OGLE HBSs}
	For a better picture of the evolutionary status of the OGLE HBSs, we describe them in the following two subsections. The first one refers to the stars located on or near the MS, and in the second one, we discuss systems containing an RG star. Here we analyze stars located in the LMC only. The SMC sample is too small for such an analysis. In the GB, extinction is highly nonuniformly distributed, which hampers a precise determination of the color and, in turn, results in very high uncertainties in the effective temperature of stars. Moreover, for the LMC, we can assume that all stars are located at the same distance (the distances between stars are insignificant compared to the distance to the LMC), which allows us to determine an absolute magnitude for each star. In the GB, we do not know the distance for the majority of stars; thus, we are not able to calculate precise absolute magnitudes.
	
	\subsubsection{HBSs with an MS or Hertzsprung-gap Primary}
	The HBSs containing a hot star ($T_{\rm eff} \gtrsim 8000$ K) are located mainly on or near the MS (perhaps the Hertzsprung gap). The HBSs with cooler MS stars usually show very small brightness changes (less than a few thousandths of magnitude; e.g., \citealt{2016AJ....151...68K}), which makes them very difficult to identify in the OGLE data. In panel (a) of Figure~\ref{fig:HR_diagram}, we present the H-R diagram for the bluest part of our HBSs located in the LMC. To calculate the effective temperatures of stars, we used \textit{UBV} photometry obtained by \cite{2002ApJS..141...81M} and \cite{2004AJ....128.1606Z}. We followed the prescription described in \cite{1989AJ.....97..107M},
	\begin{equation}
		\log T_{\rm eff}=\begin{cases}
			3.994-0.267Q+0.364Q^2, \\
			\;\;\;\;\text{if $(B-V)_0 < 0.0$};\\
			3.990-0.510(B-V)_0, \\
			\;\;\;\;\text{if $0.0<(B-V)_0 < 0.2$};\\
			3.960-0.344(B-V)_0, \\
			\;\;\;\;\text{if $0.2<(B-V)_0 < 0.5$},
		\end{cases}
	\end{equation}
	where $Q=(U-B) - 0.72(B-V)$ is the reddening-free index, and $(B-V)_0 = (B-V) - E(B-V)$ is the dereddened color. The color excess $E(B-V)$ is calculated from $E(V-I)$ using $E(B-V)=E(V-I)/1.318$.
	
	For two stars, the $UBV$ photometry was unavailable. In order to estimate photometric temperatures for these stars, we used the $T_{\rm eff}$ - $(V-I)_0$ grid collected in the work of \cite{1998A&A...333..231B} for early-type stars (their Table~1) and the surface gravity $\log g=4.5$. 
	
	To calculate luminosity, we used a simple Pogson's equation,
	\begin{equation}
		\log L/L_\sun = -0.4(M_{\rm bol} - M_{\rm bol,\sun}),
		\label{eq:pogson}
	\end{equation}
	where $M_{\rm bol} = M_I + BC_I$ and $M_{\rm bol, \sun}=4.74$ mag (\citealt{2016AJ....152...41P}) are the bolometric absolute magnitudes of a star and the Sun, respectively. Assuming the distance to the LMC as $d_{\rm LMC}=49.59$ kpc (\citealt{2019Natur.567..200P}) and calculating the bolometric correction, $BC_I$, for the $I$ filter, we find $M_{\rm bol}$ as
	\begin{equation}
		M_{\rm bol} = I_0 - 5\log d_{\rm LMC} +5 + BC_I.
		\label{eq:bol_mag}
	\end{equation}
	Here $I_0$ is the mean $I$-band magnitude with subtracted extinction. We computed $BC_I$ values based on bolometric correction tables available on the MIST project website\footnote{\url{http://waps.cfa.harvard.edu/MIST/index.html}}. We assumed a metallicity $[\rm Fe/ \rm H] = -0.5$ and $\log g = 4.5$.
	
	In Figure~\ref{fig:HR_diagram}, with colored solid lines, we mark the MIST v1.2 evolutionary tracks by \cite{2016ApJ...823..102C} and \cite{2016ApJS..222....8D}, computed with the Modules for Experiments in Stellar Astrophysics (MESA) code (e.g., \citealt{2011ApJS..192....3P}, \citeyear{2013ApJS..208....4P}). For all tracks, we set an initial rotation of $v/v_{\rm crit}=0.4$ and metallicity $[\rm Fe/ \rm H] = -0.4$ (e.g., \citealt{2009AJ....138.1243H}). By comparing the position of HBSs in the H-R diagram with the evolutionary tracks of stars, we can see that HBSs have a wide range of possible ZAMS masses, from about $3M_\odot$ to as high as $40 M_\odot$.
	
	There is also a noticeable trend: the cooler and less luminous the star, the farther it lies from the MS. The MS is shown as the shaded gray area in panel (a) of Figure \ref{fig:HR_diagram}. The lower and upper edges of this region represent the lines of the ZAMS and TAMS, respectively. The second notable observation is that the HBSs form a group along the line that is parallel to the line of constant radius. Moreover, we can see that the majority of HBSs lie in the channel between 5 and 20$\,R_\odot$.
	
	There are at least two possible explanations for the observed departure of less luminous OGLE HBSs from the MS. The first one is an observational bias caused by the dependence between the stellar radius and the amplitude of the heartbeat. In theory, the larger the stellar radius, the larger the tidal deformation of the stellar surface, and thus the larger the amplitude of brightness changes. Stars with $M_{\rm ZAMS}<10M_\odot$ begin their evolution on the MS with a radius not exceeding 4-5$\,R_\odot$; thus, the heartbeat amplitudes may be too low for detection in ground-based surveys like OGLE. 
	
	The second explanation is the increasing light contamination from the companion for less luminous primaries, which may cause a shift in the location on the H-R diagram. If we assume a system where the primary and companion stars have similar effective temperatures, the location of this system on the H-R diagram will be shifted upward relative to the position occupied by the primary itself. The dotted red line in panel (a) of Figure~\ref{fig:HR_diagram} represents the position of systems consisting of two TAMS stars with equal effective temperatures. On the other hand, if the companion has a lower effective temperature, the position of the entire system on the H-R diagram will be shifted to the right relative to the position of the primary. In Figure \ref{fig:HR_diagram}, the markers show the location of the primaries with the assumption that the light contamination from the companion is negligible, which may not be correct. This is probably the case for about half of the hot HBSs because they are also eclipsing binaries with clearly visible primary and secondary eclipses. To determine the amount of contamination from the companion, further analysis is necessary, especially with the use of spectroscopic observations, which would allow for a more accurate estimation of the effective temperature of the main component. However, such an analysis is beyond the scope of this work. 
	
	\subsubsection{HBSs with an RG Primary}
	The most numerous group of OGLE HBSs is located in the RG part of the H-R diagram. Those systems generally consist of a low-mass ($M_{\rm ZAMS}\lesssim2M_\odot$) or intermediate-mass ($2M_\odot\lesssim M_{\rm ZAMS}\lesssim8M_\odot$) primary, which is evolving through the RG region, and a less massive companion, which most likely belongs to the MS. 
	
	Unlike the H-R diagram for the hot stars described in the previous subsection, here we decided to obtain photometric temperatures based on the $(V-I)_0$ color derived directly from the OGLE data. We used an empirical relation from \cite{2000AJ....119.1448H} for giant stars,
	\begin{equation}
		T_{\rm eff} = 8556.22-5235.57\cdot(V-I)_0+1471.09\cdot(V-I)_0^2,
	\end{equation}
	which is reliable for a range $0.70<(V-I)_0<1.68$. Luminosities were calculated using Equations~(\ref{eq:pogson}) and (\ref{eq:bol_mag}). We utilized the bolometric correction from the YBC database\footnote{\url{http://stev.oapd.inaf.it/YBC/}}, described by \cite{2019A&A...632A.105C}. 
	
	In panel (b) of Figure~\ref{fig:HR_diagram}, we present the H-R diagram for the RG part of the HBSs located in the LMC. Similarly to the blue part of the HBSs, here we also generated MIST evolutionary tracks. Solid lines represent stellar evolution until the RGB tip phase, and dashed lines show the subsequent stages (shown only for masses $M_{\rm ZAMS}>1M_\odot$) until the AGB tip. We used a metallicity gradient depending on the initial mass (e.g., \citealt{2009AJ....138.1243H}):
	\begin{equation}
		[\rm Fe / \rm H]=\begin{cases}
			-1, & \text{if $M_{\rm ZAMS} \leqslant 1.0M_\odot$}; \\
			-0.6, & \text{if $1.0M_\odot<M_{\rm ZAMS} \leqslant 2.5M_\odot$}; \\
			-0.5, & \text{if $2.5M_\odot<M_{\rm ZAMS}$}.
		\end{cases}
	\end{equation}
	For each track, we set the initial rotation $v/v_{\rm crit}=0.0$. 
	
	The obtained H-R diagram is fully consistent with the one presented in Figure~4 of \cite{2017ApJ...835..209N}, even though we used a slightly different approach to construct it. \cite{2017ApJ...835..209N} used an evolutionary track from \cite{2008A&A...484..815B}, the $\log L/L_\odot$ and $T_{\rm eff}$ parameters were obtained based on $I$- and $K$-band photometry using the prescription from \cite{2012MNRAS.423.2764N} and, last but not least, their H-R diagram refers mainly to the classical ellipsoidal variables (only 22 systems have an eccentric orbit), while our sample includes only HBSs. In both works, the largest number of stars occupy the region for initial masses of less than $1.85$--$2\,M_\odot$, which indicates that these systems contain the RG star that is evolving through the RGB (with a degenerate He core) or, in the case of stars more massive than the Sun, on the AGB (with a degenerate C/O core). 
	
	There is also a second group of HBSs, which is represented by stars with a larger initial mass. These stars can be in any medium stage of their life, from evolving on the RGB with a nondegenerate He core, through the He core-burning phase, to the evolution on the AGB. For stars with $T_{\rm eff}\lesssim4250$ K ($\log T_{\rm eff}\lesssim3.63$), more favored is an option with AGB evolution because the tip of the RGB and the He core-burning phase for stars with $M_{\rm ZAMS}\lesssim6M_\odot$ do not exceed such temperatures. 
	
	\subsection{Period--Amplitude Diagrams}
\begin{figure}
		\centering
		\includegraphics[width=0.48\textwidth]{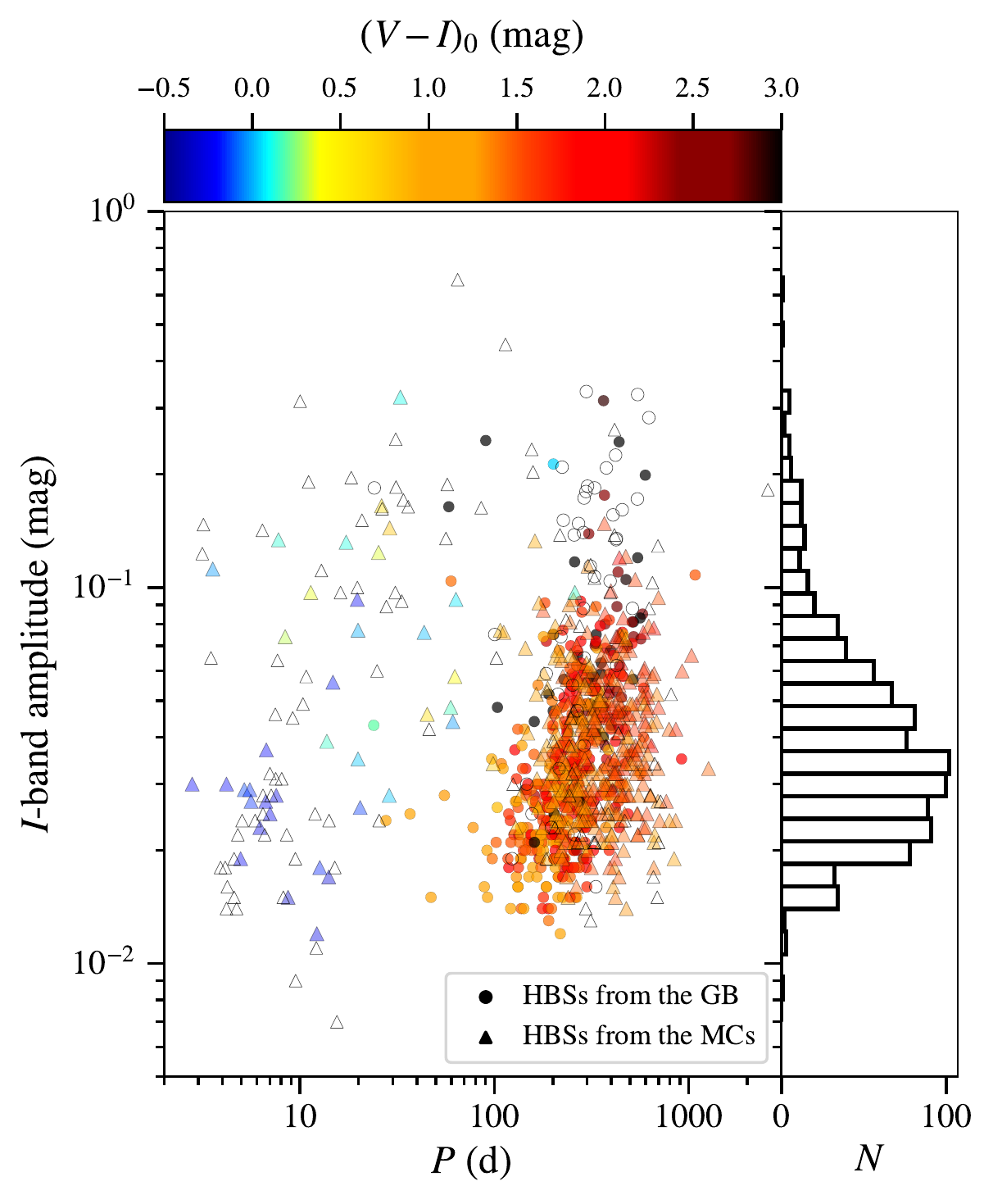}
		\caption{Period--amplitude diagram and histogram of the amplitudes for the OGLE HBSs located in the GB (circles) and in the MCs (triangles). Filled and open markers represent objects with flag=1 and 0, respectively. Flags have the same meaning as in Figure~\ref{fig:CMD}. The color of the markers represents the dereddened color value that corresponds to the effective temperature of the stellar surface.}

	\label{fig:P_ampl}
	\end{figure}

	The orbital period and $I$-band amplitude distribution for the OGLE HBSs are presented in Figure~\ref{fig:P_ampl}. Our sample, contrary to HBSs from the Kepler data, mainly consists of systems with a long orbital period (mostly a few hundred days) and high-amplitude brightness variations as for the HBS (a few hundredths of a magnitude and larger). The color scale reflects the effective temperatures of the stars based on the dereddened $(V-I)_0$ color index. Black ($(V-I)_0 > 3.0$) is an indication of stars located in the regions not included in the extinction maps.
	
	The period and amplitude ranges for the cooler group of HBSs (yellow and red, types from G to M) are similar for all locations, which indicates the common nature of those systems. Hot stars (blue and cyan, types from late O to F), which correspond to HBSs located on the MS or in its vicinity, have a similar mean value of the amplitude but slightly bigger scatter than the cooler group of HBSs. However, the orbital period distribution for hotter stars is extremely different. Those systems have much shorter periods, from a few days to several tens of days. It is the result of the different sizes of the primaries. In Figure \ref{fig:HR_diagram}, we can see that the HBSs located on the blue part of the H-R diagram reach sizes from 5 to 25$\,R_\odot$, while the HBSs from the red part of the H-R diagram have radii from 25 to even 200$\,R_\odot$. Since stars in the system cannot come too close to each other, the smaller the stars, the smaller the minimum distance needed, and thus the shorter the orbital periods. In Figure~\ref{fig:P_ampl}, we can also see a trend in the period--color relations: the higher the value of the color, the longer the orbital period in the system. 
	
	\subsection{PL Relations for RG HBSs}
	 The RG variable stars, such as Miras, OGLE small-amplitude red giants (OSARGs), semiregular variables (SRVs), long secondary periods (LSPs), or long-period eclipsing and ellipsoidal stars exhibit classical PL relations. Within each of those classes, in the PL diagram, the stars assemble into linear sequences, marked with letters from A to E (e.g., \citealt{1999IAUS..191..151W}, \citealt{2004AcA....54..347S}). The sequences differ in slope and spread, and they are also shifted relative to each other. These parameters also depend on the set of filters that were used (e.g., scatter is much smaller for near-IR filters than for optical ones). In the case of HBSs, the most crucial is sequence E, which is formed by eclipsing and ellipsoidal variables.

	\begin{figure*}
	\centering
	\includegraphics[width=0.85\textwidth]{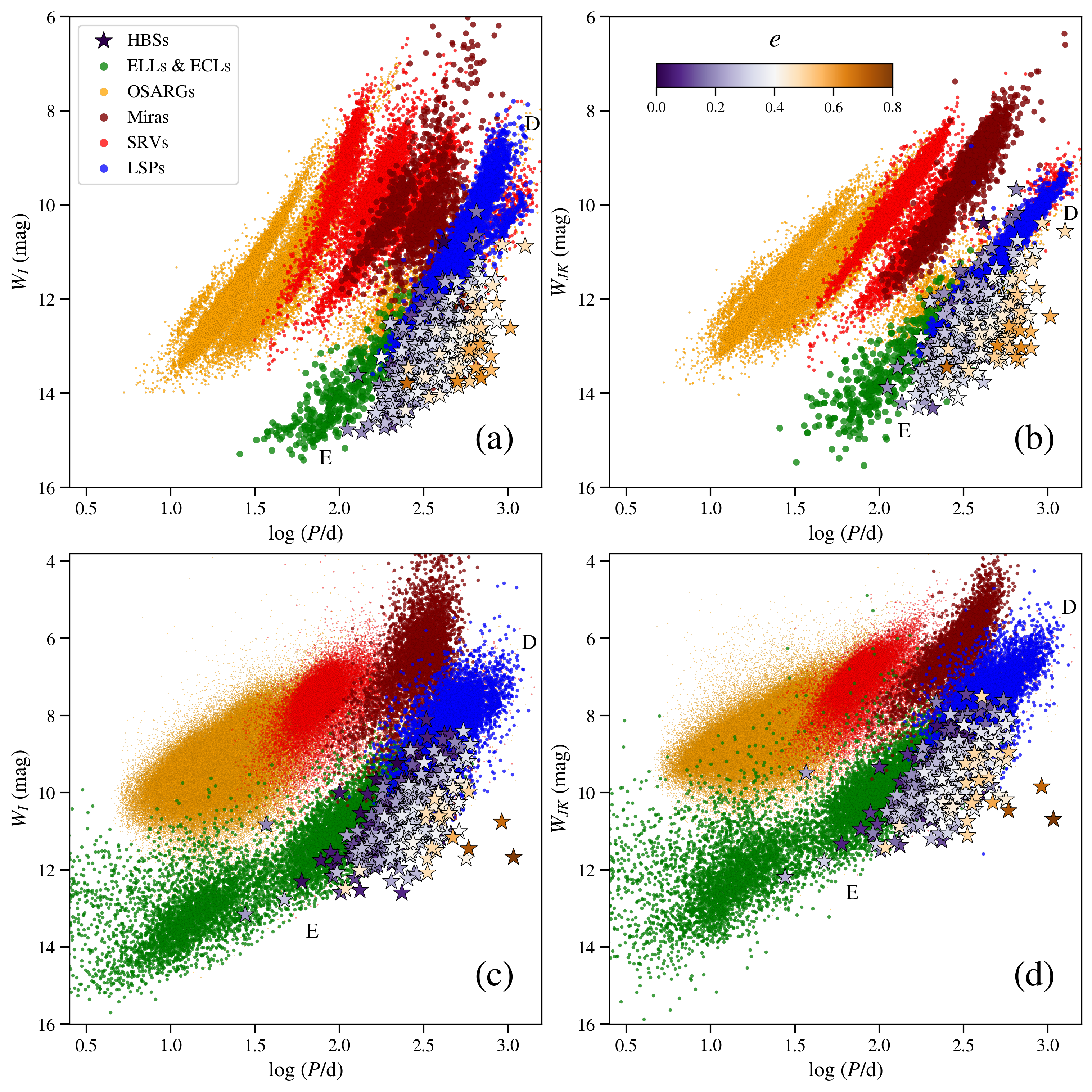}
	\caption{The PL diagrams for long-period variables (yellow, brown, pink, and red points), ellipsoidal and eclipsing binaries (green points), and HBSs (stars), located toward the LMC (panels (a) and (b)) and GB (panels (c) and (d)). On the left-hand side, the brightnesses are shown in the $W_I$ Wesenheit index, and on the other side, they are shown in $W_{JK}$. The color scale of the HBSs indicates orbital eccentricity.}
	\label{fig:PL_diagrams}
	\end{figure*}
	
	In Figure~\ref{fig:PL_diagrams}, we present PL diagrams for RG variables from the OCVS (references are listed in Table~\ref{tab:LPV}), including HBSs from our collection. We plotted PL diagrams separately for the LMC and GB using two types of the extinction-free Wesenheit indices $W_I$ and $W_{JK}$, defined as
	\begin{align}
		W_I &= I - R_{I,V}\cdot(V-I), \\
		W_{JK} &= K_{\rm s} - R_{K_{\rm s}, J}\cdot(J-K_{\rm s}),
	\end{align}
	where $R_{I,V}=A_I/E(V-I)$ and $R_{K_{\rm s}, J}=A_{K_{\rm s}}/E(J-K_{\rm s})$ are the total-to-selective extinction. We adopted $R_{I,V}=1.14$, $R_{K_{\rm s}, J}=0.67$ for the GB sample (e.g., \citealt{2002A&A...381..219D}, \citealt{2015ApJ...811..113P}) and $R_{I,V}=1.55$, $R_{K_{\rm s}, J}=0.686$ for the LMC sample (e.g., \citealt{2009AcA....59..239S}, \citealt{2011A&A...534A..94S}). We do not show PL diagrams for objects located in the SMC because of the low number of RG HBSs. All $I$- and $V$-band data come from the OGLE survey. In the case of MCs, values for the $J$ and $K_{\rm s}$ bands have been taken mainly from IRSF and additionally from 2MASS, and in the case of the GB, we have taken into account only data from the latter survey.
	
	\begin{deluxetable}{rrr}
		
		\label{tab:LPV}	
		\tablecaption{References to the Catalog Papers of Used Long-period Variables}
		
		\tablehead{\colhead{Location} & \colhead{Variability Class} & \colhead{Reference} \\ 
			\colhead{} & \colhead{} & \colhead{} } 
		
		\startdata
		LMC & OSARG & \cite{2009AcA....59..239S} \\
		LMC & Mira & \cite{2009AcA....59..239S} \\
		LMC & SRV & \cite{2009AcA....59..239S} \\
		LMC & LSP & \citeauthor{2009AcA....59..239S} (\citeyear{2009AcA....59..239S}, \citeyear{2021ApJ...911L..22S}) \\
		LMC & ELL & \cite{2016AcA....66..421P} \\
		LMC & ECL & \cite{2016AcA....66..421P} \\
		SMC & OSARG & \cite{2011AcA....61..217S} \\
		SMC & Mira & \cite{2011AcA....61..217S} \\
		SMC & SRV & \cite{2011AcA....61..217S} \\
		SMC & LSP & \cite{2011AcA....61..217S} \\
		SMC & ELL & \cite{2016AcA....66..421P} \\
		SMC & ECL & \cite{2016AcA....66..421P} \\
		BLG & OSARG & \cite{2013AcA....63...21S} \\
		BLG & Mira & \cite{2013AcA....63...21S} \\
		BLG & SRV & \cite{2013AcA....63...21S} \\
		BLG & LSP & \cite{2013AcA....63...21S} \\
		BLG & ELL & \cite{2016AcA....66..405S} \\
		BLG & ECL & \cite{2016AcA....66..405S} \\
		\enddata

	\end{deluxetable}

	In all PL diagrams, we notice a linear trend formed by HBSs: the longer the orbital period, the more luminous the system. The second notable aspect is that HBSs seem to stick to the long-period ($\log P \gtrsim 2$) group of eclipsing and ellipsoidal variables but also to the LSP stars (sequences E and D). This feature was also noticed by \cite{2004AcA....54..347S} and highlighted by \cite{2017ApJ...835..209N}. However, those analyses involved only a small group of HBSs located in the LMC. Here we show that those remarks are genuine for a larger sample of HBSs in the LMC, as well as for HBSs located toward the GB.   
	
	Recently, \cite{2021ApJ...911L..22S} showed that LSP variables (sequence D) are systems that contain an RG and a stellar or substellar companion. In combination with eclipsing and ellipsoidal stars (sequence E), they form a group with a wide range of periods (from a few days to about 3 yr), for which the brightness changes are driven mainly by the interactions between the components of the binary system. Since the RG part of the HBSs forms a group in a similar region of the PL diagram (sequences D and E), it is likely that they are binary systems as well. Moreover, the PL diagrams show that those HBSs are a natural extension of sequence E for longer periods.
	
	In Figure~\ref{fig:PL_diagrams}, the color scale of HBSs reflects the eccentricity of the system, and it is clearly seen that at a given brightness, the longer the period, the larger the eccentricity. In the classical ellipsoidal variables, the separation between stars in the system cannot be too small, because it may entail mass transfer via Roche-lobe overflow, and it cannot be too large, because the amplitude of the brightness variations would be too slight to detect. Now, if we compare two systems with identical luminosities and periods, which indicates a similar size of the semimajor axis, but one system has a circular orbit and the other has an eccentric orbit, we can expect that the second system will be easier to detect because the minimal distance between components will be $a(1-e) < a$, where $a$ is the semimajor axis; thus, the amplitude of the variations will be higher than for the first one.\\
	
	\section{Summary and Conclusions} \label{sec:conc}
	We have presented an analysis of the HBS sample from the OCVS, described in Paper I, based on their photometric properties. The sample consists of 512 and 479 HBSs located toward the GB and MCs, respectively. An $I$-band light-curve variation has been modeled using a simple analytic model of the tidal deformation in the binary during periastron passage described in \cite{1995ApJ...449..294K}. The K95 model parameters for all HBSs were estimated using the \texttt{emcee} Python package, which provides the fitting with the MCMC method. We have presented the distribution of the orbital parameters eccentricity, $e$; inclination angle, $i$; and argument of the periastron, $\omega$. The $\omega$ can be described by a flat distribution, as expected, while $i$ can be described by the normal distribution around $65\degree$ and a separate peak near $90\degree$. The majority of our HBSs have orbits with low or moderate eccentricity ($0.1 \lesssim e \lesssim 0.35$), but we also observe orbits with very high eccentricity, where $e\gtrsim0.5$.
	
	We have shown CMDs and H-R diagrams that indicate that HBSs, similar to the ellipsoidal and eclipsing binaries, are variables that do not belong solely to a specific evolutionary status or position on those diagrams. The most visible are two groups of HBSs. The first group of fewer than 100 systems consists of an early-type primary star lying on the MS or Hertzsprung gap, and the second group, including about 900 systems, most likely contains an RGB or AGB star. By comparing the positions of HBSs on the H-R diagram to the theoretical evolutionary tracks, we see a wide range of the initial mass of the primary, from less than 1 $M_\odot$ to as high as 40 $M_\odot$. However, bear in mind that we assume a separate evolution here, which is most likely not the case for evolved RG stars.
	
	Cool and hot HBSs differ distinctly in the period distribution and slightly in the flux variation amplitudes. The majority of hot HBSs have orbital periods in a range from a few days to 30--40 days, while the RG part of the sample is dominated by long-period binaries with a median value of about 1 yr. In the case of the $I$-band amplitude, the mean value is similar for both groups, but the scatter is larger for hot stars.
	
	Like the classical ellipsoidal variables, the RG HBSs are also grouped on the PL diagrams, extending sequence E to longer periods at a given brightness, which is strong evidence confirming their binary nature. Moreover, thanks to the large number of HBSs, we proved that for a given brightness, the higher the eccentricity, the longer the period. 
	
	Using $I$-band photometry, we also performed time-series analysis and found TEOs in 52 objects with a total of 76 modes. Those oscillations occur at harmonics of orbital frequencies in the range between 4 and 79. We also provide evidence that some of them may have formed due to NLMC. Thanks to this relatively large and homogeneous sample of TEOs, we were able to construct for the first time a diagram showing the positive correlation between the TEO $n$ and eccentricity, as predicted by theory.
	
	\section*{Acknowledgment}
	
	We are grateful to the anonymous referee for many inspiring suggestions that made this manuscript more comprehensible. We thank Prof. Igor Soszy\'nski and Prof. Andrzej Pigulski for the comments that helped to improve this manuscript. This work has been supported by Polish National Science Centre grant OPUS 16 No. 2018/31/B/ST9/00334 and MAESTRO 8 No. 2016/22/A/ST9/00009. P.K.S. acknowledges support from Polish National Science Centre grant PRELUDIUM 18 No. 2019/35/N/ST9/03805. This publication makes use of data products from the Two Micron All Sky Survey, which is a joint project of the University of Massachusetts and the Infrared Processing and Analysis Center/California Institute of Technology, funded by the National Aeronautics and Space Administration and the National Science Foundation. The Infrared Survey Facility (IRSF) project is a collaboration between Nagoya University and the South African Astronomical Observatory (SAAO) supported by the Grants-in-Aid for Scientific Research on Priority Areas (A) (Nos. 10147207 and 10147214) and Optical \& Near-Infrared Astronomy Inter-University Cooperation Program, from the Ministry of Education, Culture, Sports, Science and Technology (MEXT) of Japan and the National Research Foundation (NRF) of South Africa. This research also made use of \texttt{ASTROPY}, a community-developed core Python package for Astronomy (\citealt{2018AJ....156..123A}).
	
	\software{FNPEAKS (Z. Ko\l{}aczkowski, W. Hebisch, G. Kopacki), PHOEBE2 (v2.3; \citealt{2016ApJS..227...29P}, \citealt{2018ApJS..237...26H}, \citealt{2020ApJS..250...34C}, \citealt{2020ApJS..247...63J}), emcee (v3.0.2; \citealt{2013PASP..125..306F}), MIST (v1.2; \citealt{2016ApJ...823..102C}, \citealt{2016ApJS..222....8D}), MESA (\citealt{2011ApJS..192....3P}, \citeyear{2013ApJS..208....4P}), astropy \citep{2018AJ....156..123A}}
	
	\appendix
	\restartappendixnumbering
	\section{Data Tables}\label{sec:appendix}
	In Tables~\ref{tab:teos-blg} and \ref{tab:teos-mc}, we present the results of searching for TEOs in the OGLE HBSs.
	
	\startlongtable
	\begin{deluxetable*}{l l l l l l l}
		\centering
		\tablecaption{TEOs Detected in the OGLE HBSs Located toward the GB}
		\label{tab:teos-blg}
		\tablehead{\colhead{OGLE ID} & \colhead{Frequency} & \colhead{Amplitude} & \colhead{} & \colhead{$n$} & \colhead{$\Delta n$} & \colhead{S/N} \\ 
			\colhead{} & \colhead{(days$^{-1}$)} & \colhead{(mmag)} & \colhead{} & \colhead{} & \colhead{} & \colhead{}} 
		
		\startdata
		OGLE-BLG-HB-0066$\,^{\rm IPV}$&0.069502(8)&6.0(5)&&33&$-$0.006(9)&9.51\\
		&0.056926(10)&5.1(6)&&27&$+$0.023(9)&8.14\\
		&0.014769(13)&3.4(5)&&7&$+$0.011(7)&5.49\\
		&0.059262(10)$^\ast$&4.6(5)&\multirow{2}{*}{$\left.\begin{array}{l}\\\\\end{array}\right\rbrace$}&\multirow{2}{*}{37}&\multirow{2}{*}{$-$0.012(11)}&7.41\\
		&0.018654(12)$^\ast$&3.9(5)&&&&6.21\\
		\noalign{\smallskip}
		OGLE-BLG-HB-0081$\,^{\rm IPV}$&0.048188(9)&1.94(26)&&52&$+$0.002(9)&5.80\\
		\noalign{\smallskip}
		OGLE-BLG-HB-0091$\,^{\rm IPV}$&0.044299(9)&7.2(8)&&23&$+$0.00005$\pm$0.01&8.43\\
		\noalign{\smallskip}
		OGLE-BLG-HB-0095$\,^{\rm IPV}$&0.123049(14)&1.17(17)&&32&$+$0.02(7)&5.45\\
		\noalign{\smallskip}
		OGLE-BLG-HB-0143&0.202440(9)&2.31(25)&&22&$-$0.0040(20)&7.10\\
		\noalign{\smallskip}
		OGLE-BLG-HB-0145$\,^{\rm IPV}$&0.034637(13)&0.92(6)&&11&$+$0.001(4)&14.21\\
		\noalign{\smallskip}
		OGLE-BLG-HB-0147$\,^{\rm IPV}$&0.037320(13)&1.28(18)&&17&$-$0.002(6)&5.81\\
		\noalign{\smallskip}
		OGLE-BLG-HB-0156&0.0791056(28)&2.01(6)&&29&$-$0.0013(12)&27.02\\
		\noalign{\smallskip}
		OGLE-BLG-HB-0157$\,^{\rm IPV}$&0.051040(9)$^\ast$&3.9(4)&\multirow{2}{*}{$\left.\begin{array}{l}\\\\\end{array}\right\rbrace$}&\multirow{2}{*}{61}&\multirow{2}{*}{$+$0.008$\pm$0.015}&9.52\\
		&0.054405(9)$^\ast$&3.6(3)&&&&8.77\\
		&0.037978(7)$^\ast$&4.9(3)&\multirow{2}{*}{$\left.\begin{array}{l}\\\\\end{array}\right\rbrace$}&\multirow{2}{*}{34}&\multirow{2}{*}{$+$0.007$\pm$0.011}&11.76\\
		&0.020775(12)$^\ast$&2.7(3)&&&&6.52\\
		&0.037362(8)$^\ast$&3.9(3)&\multirow{2}{*}{$\left.\begin{array}{l}\\\\\end{array}\right\rbrace$}&\multirow{2}{*}{45}&\multirow{2}{*}{$-$0.0009$\pm$0.012}&9.75\\
		&0.040414(9)$^\ast$&3.5(3)&&&&8.58\\
		\noalign{\smallskip}
		OGLE-BLG-HB-0160$\,^{\rm IPV}$&0.024130(12)&0.75(6)&&7&$+$0.011(4)&10.23\\
		\noalign{\smallskip}
		OGLE-BLG-HB-0208$\,^{\rm IPV}$&0.146553(19)&1.10(21)&&47&$+$0.008(9)&4.15\\
		&0.018556(9)$^\ast$&2.09(22)&\multirow{2}{*}{$\left.\begin{array}{l}\\\\\end{array}\right\rbrace$}&\multirow{2}{*}{13}&\multirow{2}{*}{$-$0.002(5)}&7.88\\
		&0.021967(13)$^\ast$&1.65(21)&&&&6.20\\
		\noalign{\smallskip}
		OGLE-BLG-HB-0209$\,^{\rm IPV}$&0.055979(14)&0.70(9)&&25&$+$0.020(7)&7.77\\
		\noalign{\smallskip}
		OGLE-BLG-HB-0211$\,^{\rm IPV}$&0.077900(17)&0.71(13)&&16&$-$0.007(4)&4.25\\
		\noalign{\smallskip}
		OGLE-BLG-HB-0225$\,^{\rm IPV}$&0.017026(9)&0.70(5)&&5&$-$0.0072(28)&13.26\\
		&0.020405(11)&0.65(5)&&6&$+$0.001(3)&12.29\\
		\noalign{\smallskip}
		OGLE-BLG-HB-0234$\,^{\rm IPV}$&0.007065(9)&1.55(14)&&4&$-$0.010(5)&8.66\\
		&0.019491(9)&1.60(15)&&11&$+$0.008(6)&8.93\\
		\noalign{\smallskip}
		OGLE-BLG-HB-0237$\,^{\rm IPV}$&0.042487(5)&9.4(5)&&13&$-$0.0035(18)&14.68\\
		&0.035964(13)&3.4(5)&&11&$+$0.001(4)&5.38\\
		&0.058859(13)&3.3(5)&&18&$+$0.005(4)&5.20\\
		\noalign{\smallskip}
		OGLE-BLG-HB-0261$\,^{\rm IPV}$&0.372014(14)&0.55(8)&&60&$+$0.004(4)&5.59\\
		\noalign{\smallskip}
		OGLE-BLG-HB-0273$\,^{\rm IPV}$&0.041733(8)&0.68(7)&&11&$-$0.0096(23)&8.12\\
		&0.015196(9)&0.61(7)&&4&$+$0.0020(25)&7.23\\
		\noalign{\smallskip}
		OGLE-BLG-HB-0286$\,^{\rm IPV}$&0.207209(12)&0.65(9)&&52&$+$0.009(5)&7.31\\
		\noalign{\smallskip}
		OGLE-BLG-HB-0298&0.0582328(27)&6.31(20)&&13&$-$0.0054(7)&33.81\\
		\noalign{\smallskip}
		OGLE-BLG-HB-0310$\,^{\rm IPV}$&0.065275(6)&3.55(23)&&28&$+$0.019(7)&12.54\\
		\noalign{\smallskip}
		OGLE-BLG-HB-0315&0.12390(4)&1.51(22)&&52&$+$0.04(4)&5.57\\
		\noalign{\smallskip}
		OGLE-BLG-HB-0346$\,^{\rm IPV}$&0.070270(7)&1.94(15)&&13&$+$0.0006$\pm$0.0016&10.83\\
		&0.064861(14)&0.98(14)&&12&$-$0.00022$\pm$0.0028&5.48\\
		\noalign{\smallskip}
		OGLE-BLG-HB-0357$\,^{\rm IPV}$&0.196011(18)&1.32(16)&&48&$+$0.018(7)&7.95\\
		\noalign{\smallskip}
		OGLE-BLG-HB-0362$\,^{\rm IPV}$&0.071917(6)&2.76(13)&&24&$-$0.0031(24)&21.71\\
		&0.009237(7)$^\ast$&2.38(13)&\multirow{2}{*}{$\left.\begin{array}{l}\\\\\end{array}\right\rbrace$}&\multirow{2}{*}{7}&\multirow{2}{*}{$-$0.007(4)}&18.76\\
		&0.011720(9)$^\ast$&1.89(13)&&&&14.90\\
		\noalign{\smallskip}
		OGLE-BLG-HB-0435$\,^{\rm IPV}$&0.032134(9)&2.94(27)&&14&$+$0.008(6)&8.86\\
		\noalign{\smallskip}
		OGLE-BLG-HB-0451&0.197094(3)&4.04(13)&&26&$-$0.0016(12)&23.38\\
		&0.227416(4)&3.04(14)&&30&$-$0.0019(15)&17.56\\
		&0.341157(17)&0.75(14)&&45&$+$0.002(3)&4.31\\
		\noalign{\smallskip}
		OGLE-BLG-HB-0463&0.069864(9)&1.46(13)&&16&$-$0.0002$\pm$0.0023&9.06\\
		\noalign{\smallskip}
		OGLE-BLG-HB-0486&0.020009(8)&2.79(26)&&4&$+$0.0010(17)&8.41\\
		&0.084990(9)&2.50(26)&&17&$-$0.0050(23)&7.56\\
		&0.040022(13)&1.90(26)&&8&$+$0.0030(26)&5.72
		\enddata
		\tablecomments{A superscript IPV indicates a pronounced intrinsic periodic variability coexisting with TEOs.
		$^\ast$ Possible nonharmonic TEOs present due to the NLMC. In such a case, the pair of frequencies denoted with asterisks and enclosed in a brace are suspected to be the ``daughter" modes of the harmonic ``mother" TEO, with $n$ provided after the brace.}
	\end{deluxetable*}

	\startlongtable
	\begin{deluxetable*}{l l l r l l l}
		\centering
		\tabletypesize{\small}
		
		\tablecaption{The same as Table~\ref{tab:teos-blg} but for OGLE HBSs located in the MCs.}
		\label{tab:teos-mc}
		
		\tablehead{\colhead{OGLE ID} & \colhead{Frequency} & \colhead{Amplitude} & \colhead{} & \colhead{$n$} & \colhead{$\Delta n$} & \colhead{S/N} \\ 
			\colhead{} & \colhead{(days$^{-1}$)} & \colhead{(mmag)} & \colhead{} & \colhead{} & \colhead{} & \colhead{}} 
		
		\startdata
		OGLE-LMC-HB-0006&0.059295(18)&1.29(25)&&12&$-$0.003(4)&4.29\\
		\noalign{\smallskip}
		OGLE-LMC-HB-0044$^{\,\rm IPV}$&0.037458(16)&1.72(26)&&26&$+$0.007$\pm$0.014&4.87\\
		\noalign{\smallskip}
		OGLE-LMC-HB-0101$^{\,\rm IPV}$&0.051118(20)&1.94(25)&&19&$+$0.003$\pm$0.010&6.31\\
		&0.040375(29)&1.35(24)&&15&$+$0.009$\pm$0.012&4.37\\
		\noalign{\smallskip}
		OGLE-LMC-HB-0109$^{\,\rm IPV}$&0.035508(11)&1.60(17)&&17&$+$0.010(12)&7.52\\
		\noalign{\smallskip}
		OGLE-LMC-HB-0151$^{\,\rm IPV}$&0.029874(4)&4.16(24)&&11&$+$0.0025(18)&13.71\\
		\noalign{\smallskip}
		OGLE-LMC-HB-0152$^{\,\rm IPV}$&0.040469(17)&2.6(4)&&24&$+$0.027(18)&5.87\\
		\noalign{\smallskip}
		OGLE-LMC-HB-0207&1.341280(9)&1.51(23)&&9&$-$0.00015(7)&5.18\\
		\noalign{\smallskip}
		OGLE-LMC-HB-0208$^{\,\rm IPV}$&0.036602(14)&1.9(3)&&29&$-$0.003$\pm$0.015&4.86\\
		\noalign{\smallskip}
		OGLE-LMC-HB-0209$^{\,\rm IPV}$&0.040007(9)&1.48(21)&&14&$-$0.003(3)&5.67\\
		\noalign{\smallskip}
		OGLE-LMC-HB-0221$^{\,\rm IPV}$&0.076413(9)&1.77(24)&&36&$-$0.006(8)&6.12\\
		\noalign{\smallskip}
		OGLE-LMC-HB-0223$^{\,\rm IPV}$&0.059357(15)&2.6(4)&&26&$+$0.003$\pm$0.014&5.76\\
		\noalign{\smallskip}
		OGLE-LMC-HB-0231$^{\,\rm IPV}$&0.076427(9)&1.62(26)&&79&$+$0.008$\pm$0.016&5.20\\
		\noalign{\smallskip}
		OGLE-LMC-HB-0236$^{\,\rm IPV}$&0.028232(5)&3.88(29)&&13&$+$0.0004$\pm$0.0023&11.03\\
		\noalign{\smallskip}
		OGLE-LMC-HB-0254&0.761545(14)&8.8(8)&&25&$+$0.0009(6)&8.73\\
		&0.731068(27)&4.7(8)&&24&$+$0.0004(9)&4.66\\
		\noalign{\smallskip}
		OGLE-LMC-HB-0287&0.032715(15)&3.5(5)&&9&$-$0.005(4)&5.57\\
		\noalign{\smallskip}
		OGLE-LMC-HB-0308&0.149763(9)&3.1(3)&&5&$-$0.0006(3)&7.54\\
		\noalign{\smallskip}
		OGLE-LMC-HB-0350&1.413994(6)&2.99(20)&&7&$-$0.00017(6)&12.77\\
		\noalign{\smallskip}
		OGLE-LMC-HB-0351&0.041056(9)&1.47(19)&&13&$-$0.004(3)&6.10\\
		\noalign{\smallskip}
		OGLE-LMC-HB-0385$^{\,\rm IPV}$&0.028620(19)&2.21(25)&&20&$+$0.00012$\pm$0.014&7.06\\
		\noalign{\smallskip}
		OGLE-LMC-HB-0416$^{\,\rm IPV}$&0.015303(8)&4.4(4)&&9&$-$0.005(6)&8.92\\
		\noalign{\smallskip}
		OGLE-SMC-HB-0015&0.157140(8)&3.3(4)&&9&$+$0.0005(4)&6.75\\
		&0.139663(7)&3.9(4)&&8&$-$0.0006(4)&8.10\\
		\noalign{\smallskip}
		OGLE-SMC-HB-0018&0.089935(13)&2.4(4)&&13&$-$0.0070(20)&5.17\\
		\noalign{\smallskip}
		OGLE-SMC-HB-0019&0.161628(7)&2.00(22)&&5&$+$0.00027(21)&7.35
		\enddata
		\tablecomments{A superscript IPV indicates a pronounced intrinsic periodic variability coexisting with TEOs.}
	\end{deluxetable*}

	\bibliography{main}

\begin{thebibliography}{}
\expandafter\ifx\csname natexlab\endcsname\relax\def\natexlab#1{#1}\fi
\providecommand{\url}[1]{\href{#1}{#1}}
\providecommand{\dodoi}[1]{doi:~\href{http://doi.org/#1}{\nolinkurl{#1}}}
\providecommand{\doeprint}[1]{\href{http://ascl.net/#1}{\nolinkurl{http://ascl.net/#1}}}
\providecommand{\doarXiv}[1]{\href{https://arxiv.org/abs/#1}{\nolinkurl{https://arxiv.org/abs/#1}}}

\bibitem[{{Aerts} {et~al.}(2010){Aerts}, {Christensen-Dalsgaard}, \&
  {Kurtz}}]{2010aste.book.....A}
{Aerts}, C., {Christensen-Dalsgaard}, J., \& {Kurtz}, D.~W. 2010,
  {Asteroseismology}

\bibitem[{Akima(1970)}]{10.1145/321607.321609}
Akima, H. 1970, J. ACM, 17, 589–602, \dodoi{10.1145/321607.321609}

\bibitem[{{Alard} \& {Lupton}(1998)}]{1998ApJ...503..325A}
{Alard}, C., \& {Lupton}, R.~H. 1998, \apj, 503, 325, \dodoi{10.1086/305984}

\bibitem[{{Astropy Collaboration} {et~al.}(2018){Astropy Collaboration},
  {Price-Whelan}, {Sip{\H{o}}cz}, {G{\"u}nther}, {Lim}, {Crawford}, {Conseil},
  {Shupe}, {Craig}, {Dencheva}, {Ginsburg}, {VanderPlas}, {Bradley},
  {P{\'e}rez-Su{\'a}rez}, {de Val-Borro}, {Aldcroft}, {Cruz}, {Robitaille},
  {Tollerud}, {Ardelean}, {Babej}, {Bach}, {Bachetti}, {Bakanov}, {Bamford},
  {Barentsen}, {Barmby}, {Baumbach}, {Berry}, {Biscani}, {Boquien}, {Bostroem},
  {Bouma}, {Brammer}, {Bray}, {Breytenbach}, {Buddelmeijer}, {Burke},
  {Calderone}, {Cano Rodr{\'\i}guez}, {Cara}, {Cardoso}, {Cheedella}, {Copin},
  {Corrales}, {Crichton}, {D'Avella}, {Deil}, {Depagne}, {Dietrich}, {Donath},
  {Droettboom}, {Earl}, {Erben}, {Fabbro}, {Ferreira}, {Finethy}, {Fox},
  {Garrison}, {Gibbons}, {Goldstein}, {Gommers}, {Greco}, {Greenfield},
  {Groener}, {Grollier}, {Hagen}, {Hirst}, {Homeier}, {Horton}, {Hosseinzadeh},
  {Hu}, {Hunkeler}, {Ivezi{\'c}}, {Jain}, {Jenness}, {Kanarek}, {Kendrew},
  {Kern}, {Kerzendorf}, {Khvalko}, {King}, {Kirkby}, {Kulkarni}, {Kumar},
  {Lee}, {Lenz}, {Littlefair}, {Ma}, {Macleod}, {Mastropietro}, {McCully},
  {Montagnac}, {Morris}, {Mueller}, {Mumford}, {Muna}, {Murphy}, {Nelson},
  {Nguyen}, {Ninan}, {N{\"o}the}, {Ogaz}, {Oh}, {Parejko}, {Parley}, {Pascual},
  {Patil}, {Patil}, {Plunkett}, {Prochaska}, {Rastogi}, {Reddy Janga},
  {Sabater}, {Sakurikar}, {Seifert}, {Sherbert}, {Sherwood-Taylor}, {Shih},
  {Sick}, {Silbiger}, {Singanamalla}, {Singer}, {Sladen}, {Sooley},
  {Sornarajah}, {Streicher}, {Teuben}, {Thomas}, {Tremblay}, {Turner},
  {Terr{\'o}n}, {van Kerkwijk}, {de la Vega}, {Watkins}, {Weaver}, {Whitmore},
  {Woillez}, {Zabalza}, \& {Astropy Contributors}}]{2018AJ....156..123A}
{Astropy Collaboration}, {Price-Whelan}, A.~M., {Sip{\H{o}}cz}, B.~M., {et~al.}
  2018, \aj, 156, 123, \dodoi{10.3847/1538-3881/aabc4f}

\bibitem[{{Beck} {et~al.}(2011){Beck}, {Bedding}, {Mosser}, {Stello}, {Garcia},
  {Kallinger}, {Hekker}, {Elsworth}, {Frandsen}, {Carrier}, {De Ridder},
  {Aerts}, {White}, {Huber}, {Dupret}, {Montalb{\'a}n}, {Miglio}, {Noels},
  {Chaplin}, {Kjeldsen}, {Christensen-Dalsgaard}, {Gilliland}, {Brown},
  {Kawaler}, {Mathur}, \& {Jenkins}}]{2011Sci...332..205B}
{Beck}, P.~G., {Bedding}, T.~R., {Mosser}, B., {et~al.} 2011, Science, 332,
  205, \dodoi{10.1126/science.1201939}

\bibitem[{{Beck} {et~al.}(2014){Beck}, {Hambleton}, {Vos}, {Kallinger},
  {Bloemen}, {Tkachenko}, {Garc{\'\i}a}, {{\O}stensen}, {Aerts}, {Kurtz}, {De
  Ridder}, {Hekker}, {Pavlovski}, {Mathur}, {De Smedt}, {Derekas}, {Corsaro},
  {Mosser}, {Van Winckel}, {Huber}, {Degroote}, {Davies}, {Pr{\v{s}}a},
  {Debosscher}, {Elsworth}, {Nemeth}, {Siess}, {Schmid}, {P{\'a}pics}, {de
  Vries}, {van Marle}, {Marcos-Arenal}, \& {Lobel}}]{2014A&A...564A..36B}
{Beck}, P.~G., {Hambleton}, K., {Vos}, J., {et~al.} 2014, \aap, 564, A36,
  \dodoi{10.1051/0004-6361/201322477}

\bibitem[{{Bertelli} {et~al.}(2008){Bertelli}, {Girardi}, {Marigo}, \&
  {Nasi}}]{2008A&A...484..815B}
{Bertelli}, G., {Girardi}, L., {Marigo}, P., \& {Nasi}, E. 2008, \aap, 484,
  815, \dodoi{10.1051/0004-6361:20079165}

\bibitem[{{Bessell} {et~al.}(1998){Bessell}, {Castelli}, \&
  {Plez}}]{1998A&A...333..231B}
{Bessell}, M.~S., {Castelli}, F., \& {Plez}, B. 1998, \aap, 333, 231

\bibitem[{{Borucki} {et~al.}(2010){Borucki}, {Koch}, {Basri}, {Batalha},
  {Brown}, {Caldwell}, {Caldwell}, {Christensen-Dalsgaard}, {Cochran},
  {DeVore}, {Dunham}, {Dupree}, {Gautier}, {Geary}, {Gilliland}, {Gould},
  {Howell}, {Jenkins}, {Kondo}, {Latham}, {Marcy}, {Meibom}, {Kjeldsen},
  {Lissauer}, {Monet}, {Morrison}, {Sasselov}, {Tarter}, {Boss}, {Brownlee},
  {Owen}, {Buzasi}, {Charbonneau}, {Doyle}, {Fortney}, {Ford}, {Holman},
  {Seager}, {Steffen}, {Welsh}, {Rowe}, {Anderson}, {Buchhave}, {Ciardi},
  {Walkowicz}, {Sherry}, {Horch}, {Isaacson}, {Everett}, {Fischer}, {Torres},
  {Johnson}, {Endl}, {MacQueen}, {Bryson}, {Dotson}, {Haas}, {Kolodziejczak},
  {Van Cleve}, {Chandrasekaran}, {Twicken}, {Quintana}, {Clarke}, {Allen},
  {Li}, {Wu}, {Tenenbaum}, {Verner}, {Bruhweiler}, {Barnes}, \&
  {Prsa}}]{2010Sci...327..977B}
{Borucki}, W.~J., {Koch}, D., {Basri}, G., {et~al.} 2010, Science, 327, 977,
  \dodoi{10.1126/science.1185402}

\bibitem[{{Burkart} {et~al.}(2012){Burkart}, {Quataert}, {Arras}, \&
  {Weinberg}}]{2012MNRAS.421..983B}
{Burkart}, J., {Quataert}, E., {Arras}, P., \& {Weinberg}, N.~N. 2012, \mnras,
  421, 983, \dodoi{10.1111/j.1365-2966.2011.20344.x}

\bibitem[{{Castelli} \& {Kurucz}(2003)}]{2003IAUS..210P.A20C}
{Castelli}, F., \& {Kurucz}, R.~L. 2003, in Modelling of Stellar Atmospheres,
  ed. N.~{Piskunov}, W.~W. {Weiss}, \& D.~F. {Gray}, Vol. 210, A20.
\newblock \doarXiv{astro-ph/0405087}

\bibitem[{{Chen} {et~al.}(2019){Chen}, {Girardi}, {Fu}, {Bressan}, {Aringer},
  {Dal Tio}, {Pastorelli}, {Marigo}, {Costa}, \& {Zhang}}]{2019A&A...632A.105C}
{Chen}, Y., {Girardi}, L., {Fu}, X., {et~al.} 2019, \aap, 632, A105,
  \dodoi{10.1051/0004-6361/201936612}

\bibitem[{{Choi} {et~al.}(2016){Choi}, {Dotter}, {Conroy}, {Cantiello},
  {Paxton}, \& {Johnson}}]{2016ApJ...823..102C}
{Choi}, J., {Dotter}, A., {Conroy}, C., {et~al.} 2016, \apj, 823, 102,
  \dodoi{10.3847/0004-637X/823/2/102}

\bibitem[{{Conroy} {et~al.}(2020){Conroy}, {Kochoska}, {Hey}, {Pablo},
  {Hambleton}, {Jones}, {Giammarco}, {Abdul-Masih}, \&
  {Pr{\v{s}}a}}]{2020ApJS..250...34C}
{Conroy}, K.~E., {Kochoska}, A., {Hey}, D., {et~al.} 2020, \apjs, 250, 34,
  \dodoi{10.3847/1538-4365/abb4e2}

\bibitem[{{Dotter}(2016)}]{2016ApJS..222....8D}
{Dotter}, A. 2016, \apjs, 222, 8, \dodoi{10.3847/0067-0049/222/1/8}

\bibitem[{{Dupret} {et~al.}(2009){Dupret}, {Belkacem}, {Samadi}, {Montalban},
  {Moreira}, {Miglio}, {Godart}, {Ventura}, {Ludwig}, {Grigahc{\`e}ne},
  {Goupil}, {Noels}, \& {Caffau}}]{2009A&A...506...57D}
{Dupret}, M.~A., {Belkacem}, K., {Samadi}, R., {et~al.} 2009, \aap, 506, 57,
  \dodoi{10.1051/0004-6361/200911713}

\bibitem[{{Dutra} {et~al.}(2002){Dutra}, {Santiago}, \&
  {Bica}}]{2002A&A...381..219D}
{Dutra}, C.~M., {Santiago}, B.~X., \& {Bica}, E. 2002, \aap, 381, 219,
  \dodoi{10.1051/0004-6361:20011541}

\bibitem[{{Dziembowski}(1982)}]{1982AcA....32..147D}
{Dziembowski}, W. 1982, \actaa, 32, 147

\bibitem[{{Dziembowski} \& {Kr\'olikowska}(1985)}]{1985AcA....35....5D}
{Dziembowski}, W., \& {Kr\'olikowska}, M. 1985, \actaa, 35, 5

\bibitem[{{Dziembowski} {et~al.}(1988){Dziembowski}, {Kr\'olikowska}, \&
  {Kosovichev}}]{1988AcA....38...61D}
{Dziembowski}, W., {Kr\'olikowska}, M., \& {Kosovichev}, A. 1988, \actaa, 38,
  61

\bibitem[{{Dziembowski}(1971)}]{1971AcA....21..289D}
{Dziembowski}, W.~A. 1971, \actaa, 21, 289

\bibitem[{{Foreman-Mackey} {et~al.}(2013){Foreman-Mackey}, {Hogg}, {Lang}, \&
  {Goodman}}]{2013PASP..125..306F}
{Foreman-Mackey}, D., {Hogg}, D.~W., {Lang}, D., \& {Goodman}, J. 2013, \pasp,
  125, 306, \dodoi{10.1086/670067}

\bibitem[{{Fuller}(2017)}]{2017MNRAS.472.1538F}
{Fuller}, J. 2017, \mnras, 472, 1538, \dodoi{10.1093/mnras/stx2135}

\bibitem[{{Fuller} {et~al.}(2013){Fuller}, {Derekas}, {Borkovits}, {Huber},
  {Bedding}, \& {Kiss}}]{2013MNRAS.429.2425F}
{Fuller}, J., {Derekas}, A., {Borkovits}, T., {et~al.} 2013, \mnras, 429, 2425,
  \dodoi{10.1093/mnras/sts511}

\bibitem[{{Fuller} {et~al.}(2020){Fuller}, {Kurtz}, {Handler}, \&
  {Rappaport}}]{2020MNRAS.498.5730F}
{Fuller}, J., {Kurtz}, D.~W., {Handler}, G., \& {Rappaport}, S. 2020, \mnras,
  498, 5730, \dodoi{10.1093/mnras/staa2376}

\bibitem[{{Fuller} \& {Lai}(2012)}]{2012MNRAS.420.3126F}
{Fuller}, J., \& {Lai}, D. 2012, \mnras, 420, 3126,
  \dodoi{10.1111/j.1365-2966.2011.20237.x}

\bibitem[{{Guo}(2020)}]{2020ApJ...896..161G}
{Guo}, Z. 2020, \apj, 896, 161, \dodoi{10.3847/1538-4357/ab911f}

\bibitem[{Guo(2021)}]{10.3389/fspas.2021.663026}
Guo, Z. 2021, Frontiers in Astronomy and Space Sciences, 8, 67,
  \dodoi{10.3389/fspas.2021.663026}

\bibitem[{{Guo} {et~al.}(2020){Guo}, {Shporer}, {Hambleton}, \&
  {Isaacson}}]{2020ApJ...888...95G}
{Guo}, Z., {Shporer}, A., {Hambleton}, K., \& {Isaacson}, H. 2020, \apj, 888,
  95, \dodoi{10.3847/1538-4357/ab58c2}

\bibitem[{{Hambleton} {et~al.}(2016){Hambleton}, {Kurtz}, {Pr{\v{s}}a},
  {Quinn}, {Fuller}, {Murphy}, {Thompson}, {Latham}, \&
  {Shporer}}]{2016MNRAS.463.1199H}
{Hambleton}, K., {Kurtz}, D.~W., {Pr{\v{s}}a}, A., {et~al.} 2016, \mnras, 463,
  1199, \dodoi{10.1093/mnras/stw1970}

\bibitem[{{Hambleton} {et~al.}(2018){Hambleton}, {Fuller}, {Thompson},
  {Pr{\v{s}}a}, {Kurtz}, {Shporer}, {Isaacson}, {Howard}, {Endl}, {Cochran}, \&
  {Murphy}}]{2018MNRAS.473.5165H}
{Hambleton}, K., {Fuller}, J., {Thompson}, S., {et~al.} 2018, \mnras, 473,
  5165, \dodoi{10.1093/mnras/stx2673}

\bibitem[{{Handler} {et~al.}(2002){Handler}, {Balona}, {Shobbrook}, {Koen},
  {Bruch}, {Romero-Colmenero}, {Pamyatnykh}, {Willems}, {Eyer}, {James}, \&
  {Maas}}]{2002MNRAS.333..262H}
{Handler}, G., {Balona}, L.~A., {Shobbrook}, R.~R., {et~al.} 2002, \mnras, 333,
  262, \dodoi{10.1046/j.1365-8711.2002.05295.x}

\bibitem[{{Harris} \& {Zaritsky}(2009)}]{2009AJ....138.1243H}
{Harris}, J., \& {Zaritsky}, D. 2009, \aj, 138, 1243,
  \dodoi{10.1088/0004-6256/138/5/1243}

\bibitem[{{Hauschildt} {et~al.}(1997){Hauschildt}, {Baron}, \&
  {Allard}}]{1997ApJ...483..390H}
{Hauschildt}, P.~H., {Baron}, E., \& {Allard}, F. 1997, \apj, 483, 390,
  \dodoi{10.1086/304233}

\bibitem[{{Hekker} \& {Christensen-Dalsgaard}(2017)}]{2017A&ARv..25....1H}
{Hekker}, S., \& {Christensen-Dalsgaard}, J. 2017, \aapr, 25, 1,
  \dodoi{10.1007/s00159-017-0101-x}

\bibitem[{{Horvat} {et~al.}(2018){Horvat}, {Conroy}, {Pablo}, {Hambleton},
  {Kochoska}, {Giammarco}, \& {Pr{\v{s}}a}}]{2018ApJS..237...26H}
{Horvat}, M., {Conroy}, K.~E., {Pablo}, H., {et~al.} 2018, \apjs, 237, 26,
  \dodoi{10.3847/1538-4365/aacd0f}

\bibitem[{{Houdashelt} {et~al.}(2000){Houdashelt}, {Bell}, \&
  {Sweigart}}]{2000AJ....119.1448H}
{Houdashelt}, M.~L., {Bell}, R.~A., \& {Sweigart}, A.~V. 2000, \aj, 119, 1448,
  \dodoi{10.1086/301243}

\bibitem[{{Husser} {et~al.}(2013){Husser}, {Wende-von Berg}, {Dreizler},
  {Homeier}, {Reiners}, {Barman}, \& {Hauschildt}}]{2013A&A...553A...6H}
{Husser}, T.~O., {Wende-von Berg}, S., {Dreizler}, S., {et~al.} 2013, \aap,
  553, A6, \dodoi{10.1051/0004-6361/201219058}

\bibitem[{{Hut}(1980)}]{1980A&A....92..167H}
{Hut}, P. 1980, \aap, 92, 167

\bibitem[{{Iwanek} {et~al.}(2019){Iwanek}, {Soszy{\'n}ski}, {Skowron},
  {Udalski}, {St{\k{e}}pie{\'n}}, {Koz{\l}owski}, {Mr{\'o}z}, {Poleski},
  {Skowron}, {Szyma{\'n}ski}, {Pietrukowicz}, {Ulaczyk}, {Wyrzykowski},
  {Kruszy{\'n}ska}, \& {Rybicki}}]{2019ApJ...879..114I}
{Iwanek}, P., {Soszy{\'n}ski}, I., {Skowron}, J., {et~al.} 2019, \apj, 879,
  114, \dodoi{10.3847/1538-4357/ab23f6}

\bibitem[{{Jayasinghe} {et~al.}(2019){Jayasinghe}, {Stanek}, {Kochanek},
  {Thompson}, {Shappee}, \& {Fausnaugh}}]{2019MNRAS.489.4705J}
{Jayasinghe}, T., {Stanek}, K.~Z., {Kochanek}, C.~S., {et~al.} 2019, \mnras,
  489, 4705, \dodoi{10.1093/mnras/stz2460}

\bibitem[{{Jayasinghe} {et~al.}(2021){Jayasinghe}, {Kochanek}, {Strader},
  {Stanek}, {Vallely}, {Thompson}, {Hinkle}, {Shappee}, {Dupree}, {Auchettl},
  {Chomiuk}, {Aydi}, {Dage}, {Hughes}, {Shishkovsky}, {Sokolovsky}, {Swihart},
  {Voggel}, \& {Thompson}}]{2021MNRAS.506.4083J}
{Jayasinghe}, T., {Kochanek}, C.~S., {Strader}, J., {et~al.} 2021, \mnras, 506,
  4083, \dodoi{10.1093/mnras/stab1920}

\bibitem[{{Jones} {et~al.}(2020){Jones}, {Conroy}, {Horvat}, {Giammarco},
  {Kochoska}, {Pablo}, {Brown}, {Sowicka}, \&
  {Pr{\v{s}}a}}]{2020ApJS..247...63J}
{Jones}, D., {Conroy}, K.~E., {Horvat}, M., {et~al.} 2020, \apjs, 247, 63,
  \dodoi{10.3847/1538-4365/ab7927}

\bibitem[{{Kato} {et~al.}(2007){Kato}, {Nagashima}, {Nagayama}, {Kurita},
  {Koerwer}, {Kawai}, {Yamamuro}, {Zenno}, {Nishiyama}, {Baba}, {Kadowaki},
  {Haba}, {Hatano}, {Shimizu}, {Nishimura}, {Nagata}, {Sato}, {Murai},
  {Kawazu}, {Nakajima}, {Nakaya}, {Kandori}, {Kusakabe}, {Ishihara},
  {Kaneyasu}, {Hashimoto}, {Tamura}, {Tanab{\'e}}, {Ita}, {Matsunaga},
  {Nakada}, {Sugitani}, {Wakamatsu}, {Glass}, {Feast}, {Menzies}, {Whitelock},
  {Fourie}, {Stoffels}, {Evans}, \& {Hasegawa}}]{2007PASJ...59..615K}
{Kato}, D., {Nagashima}, C., {Nagayama}, T., {et~al.} 2007, \pasj, 59, 615,
  \dodoi{10.1093/pasj/59.3.615}

\bibitem[{{Kirk} {et~al.}(2016){Kirk}, {Conroy}, {Pr{\v{s}}a}, {Abdul-Masih},
  {Kochoska}, {Matijevi{\v{c}}}, {Hambleton}, {Barclay}, {Bloemen}, {Boyajian},
  {Doyle}, {Fulton}, {Hoekstra}, {Jek}, {Kane}, {Kostov}, {Latham}, {Mazeh},
  {Orosz}, {Pepper}, {Quarles}, {Ragozzine}, {Shporer}, {Southworth},
  {Stassun}, {Thompson}, {Welsh}, {Agol}, {Derekas}, {Devor}, {Fischer},
  {Green}, {Gropp}, {Jacobs}, {Johnston}, {LaCourse}, {Saetre}, {Schwengeler},
  {Toczyski}, {Werner}, {Garrett}, {Gore}, {Martinez}, {Spitzer}, {Stevick},
  {Thomadis}, {Vrijmoet}, {Yenawine}, {Batalha}, \&
  {Borucki}}]{2016AJ....151...68K}
{Kirk}, B., {Conroy}, K., {Pr{\v{s}}a}, A., {et~al.} 2016, \aj, 151, 68,
  \dodoi{10.3847/0004-6256/151/3/68}

\bibitem[{{Ko{\l}aczek-Szyma{\'n}ski}
  {et~al.}(2021){Ko{\l}aczek-Szyma{\'n}ski}, {Pigulski}, {Michalska},
  {Mo{\'z}dzierski}, \& {R{\'o}{\.z}a{\'n}ski}}]{2021A&A...647A..12K}
{Ko{\l}aczek-Szyma{\'n}ski}, P.~A., {Pigulski}, A., {Michalska}, G.,
  {Mo{\'z}dzierski}, D., \& {R{\'o}{\.z}a{\'n}ski}, T. 2021, \aap, 647, A12,
  \dodoi{10.1051/0004-6361/202039553}

\bibitem[{{Ko{\l}aczek-Szyma{\'n}ski}
  {et~al.}(2022){Ko{\l}aczek-Szyma{\'n}ski}, {Pigulski}, {Wrona}, {Ratajczak},
  \& {Udalski}}]{2022A&A...659A..47K}
{Ko{\l}aczek-Szyma{\'n}ski}, P.~A., {Pigulski}, A., {Wrona}, M., {Ratajczak},
  M., \& {Udalski}, A. 2022, \aap, 659, A47,
  \dodoi{10.1051/0004-6361/202142171}

\bibitem[{{Kumar} {et~al.}(1995){Kumar}, {Ao}, \&
  {Quataert}}]{1995ApJ...449..294K}
{Kumar}, P., {Ao}, C.~O., \& {Quataert}, E.~J. 1995, \apj, 449, 294,
  \dodoi{10.1086/176055}

\bibitem[{{Maceroni} {et~al.}(2009){Maceroni}, {Montalb{\'a}n}, {Michel},
  {Harmanec}, {Prsa}, {Briquet}, {Niemczura}, {Morel}, {Ladjal}, {Auvergne},
  {Baglin}, {Baudin}, {Catala}, {Samadi}, \& {Aerts}}]{2009A&A...508.1375M}
{Maceroni}, C., {Montalb{\'a}n}, J., {Michel}, E., {et~al.} 2009, \aap, 508,
  1375, \dodoi{10.1051/0004-6361/200913311}

\bibitem[{{Massey}(2002)}]{2002ApJS..141...81M}
{Massey}, P. 2002, \apjs, 141, 81, \dodoi{10.1086/338286}

\bibitem[{{Massey} {et~al.}(1989){Massey}, {Garmany}, {Silkey}, \&
  {Degioia-Eastwood}}]{1989AJ.....97..107M}
{Massey}, P., {Garmany}, C.~D., {Silkey}, M., \& {Degioia-Eastwood}, K. 1989,
  \aj, 97, 107, \dodoi{10.1086/114961}

\bibitem[{{Morris}(1985)}]{1985ApJ...295..143M}
{Morris}, S.~L. 1985, \apj, 295, 143, \dodoi{10.1086/163359}

\bibitem[{{Nataf} {et~al.}(2013){Nataf}, {Gould}, {Fouqu{\'e}}, {Gonzalez},
  {Johnson}, {Skowron}, {Udalski}, {Szyma{\'n}ski}, {Kubiak},
  {Pietrzy{\'n}ski}, {Soszy{\'n}ski}, {Ulaczyk}, {Wyrzykowski}, \&
  {Poleski}}]{2013ApJ...769...88N}
{Nataf}, D.~M., {Gould}, A., {Fouqu{\'e}}, P., {et~al.} 2013, \apj, 769, 88,
  \dodoi{10.1088/0004-637X/769/2/88}

\bibitem[{{Nicholls} \& {Wood}(2012)}]{2012MNRAS.421.2616N}
{Nicholls}, C.~P., \& {Wood}, P.~R. 2012, \mnras, 421, 2616,
  \dodoi{10.1111/j.1365-2966.2012.20492.x}

\bibitem[{{Nicholls} {et~al.}(2010){Nicholls}, {Wood}, \&
  {Cioni}}]{2010MNRAS.405.1770N}
{Nicholls}, C.~P., {Wood}, P.~R., \& {Cioni}, M. R.~L. 2010, \mnras, 405, 1770,
  \dodoi{10.1111/j.1365-2966.2010.16548.x}

\bibitem[{{Nie} \& {Wood}(2014)}]{2014AJ....148..118N}
{Nie}, J.~D., \& {Wood}, P.~R. 2014, \aj, 148, 118,
  \dodoi{10.1088/0004-6256/148/6/118}

\bibitem[{{Nie} {et~al.}(2012){Nie}, {Wood}, \&
  {Nicholls}}]{2012MNRAS.423.2764N}
{Nie}, J.~D., {Wood}, P.~R., \& {Nicholls}, C.~P. 2012, \mnras, 423, 2764,
  \dodoi{10.1111/j.1365-2966.2012.21087.x}

\bibitem[{{Nie} {et~al.}(2017){Nie}, {Wood}, \&
  {Nicholls}}]{2017ApJ...835..209N}
---. 2017, \apj, 835, 209, \dodoi{10.3847/1538-4357/835/2/209}

\bibitem[{{Pawlak} {et~al.}(2016){Pawlak}, {Soszy{\'n}ski}, {Udalski},
  {Szyma{\'n}ski}, {Wyrzykowski}, {Ulaczyk}, {Poleski}, {Pietrukowicz},
  {Koz{\l}owski}, {Skowron}, {Skowron}, {Mr{\'o}z}, \&
  {Hamanowicz}}]{2016AcA....66..421P}
{Pawlak}, M., {Soszy{\'n}ski}, I., {Udalski}, A., {et~al.} 2016, \actaa, 66,
  421.
\newblock \doarXiv{1612.06394}

\bibitem[{{Paxton} {et~al.}(2011){Paxton}, {Bildsten}, {Dotter}, {Herwig},
  {Lesaffre}, \& {Timmes}}]{2011ApJS..192....3P}
{Paxton}, B., {Bildsten}, L., {Dotter}, A., {et~al.} 2011, \apjs, 192, 3,
  \dodoi{10.1088/0067-0049/192/1/3}

\bibitem[{{Paxton} {et~al.}(2013){Paxton}, {Cantiello}, {Arras}, {Bildsten},
  {Brown}, {Dotter}, {Mankovich}, {Montgomery}, {Stello}, {Timmes}, \&
  {Townsend}}]{2013ApJS..208....4P}
{Paxton}, B., {Cantiello}, M., {Arras}, P., {et~al.} 2013, \apjs, 208, 4,
  \dodoi{10.1088/0067-0049/208/1/4}

\bibitem[{{Pietrukowicz} {et~al.}(2015){Pietrukowicz}, {Koz{\l}owski},
  {Skowron}, {Soszy{\'n}ski}, {Udalski}, {Poleski}, {Wyrzykowski},
  {Szyma{\'n}ski}, {Pietrzy{\'n}ski}, {Ulaczyk}, {Mr{\'o}z}, {Skowron}, \&
  {Kubiak}}]{2015ApJ...811..113P}
{Pietrukowicz}, P., {Koz{\l}owski}, S., {Skowron}, J., {et~al.} 2015, \apj,
  811, 113, \dodoi{10.1088/0004-637X/811/2/113}

\bibitem[{{Pietrzy{\'n}ski} {et~al.}(2019){Pietrzy{\'n}ski}, {Graczyk},
  {Gallenne}, {Gieren}, {Thompson}, {Pilecki}, {Karczmarek}, {G{\'o}rski},
  {Suchomska}, {Taormina}, {Zgirski}, {Wielg{\'o}rski}, {Ko{\l}aczkowski},
  {Konorski}, {Villanova}, {Nardetto}, {Kervella}, {Bresolin}, {Kudritzki},
  {Storm}, {Smolec}, \& {Narloch}}]{2019Natur.567..200P}
{Pietrzy{\'n}ski}, G., {Graczyk}, D., {Gallenne}, A., {et~al.} 2019, \nat, 567,
  200, \dodoi{10.1038/s41586-019-0999-4}

\bibitem[{{Pourbaix} {et~al.}(2004){Pourbaix}, {Tokovinin}, {Batten}, {Fekel},
  {Hartkopf}, {Levato}, {Morrell}, {Torres}, \& {Udry}}]{2004A&A...424..727P}
{Pourbaix}, D., {Tokovinin}, A.~A., {Batten}, A.~H., {et~al.} 2004, \aap, 424,
  727, \dodoi{10.1051/0004-6361:20041213}

\bibitem[{{Pr{\v{s}}a} \& {Zwitter}(2005)}]{2005ApJ...628..426P}
{Pr{\v{s}}a}, A., \& {Zwitter}, T. 2005, \apj, 628, 426, \dodoi{10.1086/430591}

\bibitem[{{Pr{\v{s}}a} {et~al.}(2016{\natexlab{a}}){Pr{\v{s}}a}, {Conroy},
  {Horvat}, {Pablo}, {Kochoska}, {Bloemen}, {Giammarco}, {Hambleton}, \&
  {Degroote}}]{2016ApJS..227...29P}
{Pr{\v{s}}a}, A., {Conroy}, K.~E., {Horvat}, M., {et~al.} 2016{\natexlab{a}},
  \apjs, 227, 29, \dodoi{10.3847/1538-4365/227/2/29}

\bibitem[{{Pr{\v{s}}a} {et~al.}(2016{\natexlab{b}}){Pr{\v{s}}a}, {Harmanec},
  {Torres}, {Mamajek}, {Asplund}, {Capitaine}, {Christensen-Dalsgaard},
  {Depagne}, {Haberreiter}, {Hekker}, {Hilton}, {Kopp}, {Kostov}, {Kurtz},
  {Laskar}, {Mason}, {Milone}, {Montgomery}, {Richards}, {Schmutz}, {Schou}, \&
  {Stewart}}]{2016AJ....152...41P}
{Pr{\v{s}}a}, A., {Harmanec}, P., {Torres}, G., {et~al.} 2016{\natexlab{b}},
  \aj, 152, 41, \dodoi{10.3847/0004-6256/152/2/41}

\bibitem[{{Skowron} {et~al.}(2021){Skowron}, {Skowron}, {Udalski},
  {Szyma{\'n}ski}, {Soszy{\'n}ski}, {Wyrzykowski}, {Ulaczyk}, {Poleski},
  {Koz{\l}owski}, {Pietrukowicz}, {Mr{\'o}z}, {Rybicki}, {Iwanek}, {Wrona}, \&
  {Gromadzki}}]{2021ApJS..252...23S}
{Skowron}, D.~M., {Skowron}, J., {Udalski}, A., {et~al.} 2021, \apjs, 252, 23,
  \dodoi{10.3847/1538-4365/abcb81}

\bibitem[{{Skowron} {et~al.}(2016){Skowron}, {Udalski}, {Koz{\l}owski},
  {Szyma{\'n}ski}, {Mr{\'o}z}, {Wyrzykowski}, {Poleski}, {Pietrukowicz},
  {Ulaczyk}, {Pawlak}, \& {Soszy{\'n}ski}}]{2016AcA....66....1S}
{Skowron}, J., {Udalski}, A., {Koz{\l}owski}, S., {et~al.} 2016, \actaa, 66, 1.
\newblock \doarXiv{1604.01966}

\bibitem[{{Skrutskie} {et~al.}(2006){Skrutskie}, {Cutri}, {Stiening},
  {Weinberg}, {Schneider}, {Carpenter}, {Beichman}, {Capps}, {Chester},
  {Elias}, {Huchra}, {Liebert}, {Lonsdale}, {Monet}, {Price}, {Seitzer},
  {Jarrett}, {Kirkpatrick}, {Gizis}, {Howard}, {Evans}, {Fowler}, {Fullmer},
  {Hurt}, {Light}, {Kopan}, {Marsh}, {McCallon}, {Tam}, {Van Dyk}, \&
  {Wheelock}}]{2006AJ....131.1163S}
{Skrutskie}, M.~F., {Cutri}, R.~M., {Stiening}, R., {et~al.} 2006, \aj, 131,
  1163, \dodoi{10.1086/498708}

\bibitem[{{Skrutskie} {et~al.}(2019){Skrutskie}, {Cutri}, {Stiening},
  {Weinberg}, {Schneider}, {Carpenter}, {Beichman}, {Capps}, {Chester},
  {Elias}, {Huchra}, {Liebert}, {Lonsdale}, {Monet}, {Price}, {Seitzer},
  {Jarrett}, {Kirkpatrick}, {Gizis}, {Howard}, {Evans}, {Fowler}, {Fullmer},
  {Hurt}, {Light}, {Kopan}, {Marsh}, {McCallon}, {Tam}, {Van Dyk}, \&
  {Wheelock}}]{https://doi.org/10.26131/irsa2}
---. 2019, 2MASS All-Sky Point Source Catalog,  IPAC, \dodoi{10.26131/IRSA2}

\bibitem[{{Soszy{\'n}ski} {et~al.}(2004){Soszy{\'n}ski}, {Udalski}, {Kubiak},
  {Szymanski}, {Pietrzynski}, {Zebrun}, {Szewczyk}, {Wyrzykowski}, \&
  {Dziembowski}}]{2004AcA....54..347S}
{Soszy{\'n}ski}, I., {Udalski}, A., {Kubiak}, M., {et~al.} 2004, \actaa, 54,
  347.
\newblock \doarXiv{astro-ph/0412505}

\bibitem[{{Soszy{\'n}ski} {et~al.}(2009){Soszy{\'n}ski}, {Udalski},
  {Szyma{\'n}ski}, {Kubiak}, {Pietrzy{\'n}ski}, {Wyrzykowski}, {Szewczyk},
  {Ulaczyk}, \& {Poleski}}]{2009AcA....59..239S}
{Soszy{\'n}ski}, I., {Udalski}, A., {Szyma{\'n}ski}, M.~K., {et~al.} 2009,
  \actaa, 59, 239.
\newblock \doarXiv{0910.1354}

\bibitem[{{Soszy{\'n}ski} {et~al.}(2011){Soszy{\'n}ski}, {Udalski},
  {Szyma{\'n}ski}, {Kubiak}, {Pietrzy{\'n}ski}, {Wyrzykowski}, {Ulaczyk},
  {Poleski}, {Koz{\l}owski}, \& {Pietrukowicz}}]{2011AcA....61..217S}
---. 2011, \actaa, 61, 217.
\newblock \doarXiv{1109.1143}

\bibitem[{{Soszy{\'n}ski} {et~al.}(2013){Soszy{\'n}ski}, {Udalski},
  {Szyma{\'n}ski}, {Kubiak}, {Pietrzy{\'n}ski}, {Wyrzykowski}, {Ulaczyk},
  {Poleski}, {Koz{\l}owski}, {Pietrukowicz}, \&
  {Skowron}}]{2013AcA....63...21S}
---. 2013, \actaa, 63, 21.
\newblock \doarXiv{1304.2787}

\bibitem[{{Soszy{\'n}ski} {et~al.}(2016){Soszy{\'n}ski}, {Pawlak},
  {Pietrukowicz}, {Udalski}, {Szyma{\'n}ski}, {Wyrzykowski}, {Ulaczyk},
  {Poleski}, {Koz{\l}owski}, {Skowron}, {Skowron}, {Mr{\'o}z}, \&
  {Hamanowicz}}]{2016AcA....66..405S}
{Soszy{\'n}ski}, I., {Pawlak}, M., {Pietrukowicz}, P., {et~al.} 2016, \actaa,
  66, 405.
\newblock \doarXiv{1701.03105}

\bibitem[{{Soszy{\'n}ski} {et~al.}(2021){Soszy{\'n}ski}, {Olechowska},
  {Ratajczak}, {Iwanek}, {Skowron}, {Mr{\'o}z}, {Pietrukowicz}, {Udalski},
  {Szyma{\'n}ski}, {Skowron}, {Gromadzki}, {Poleski}, {Koz{\l}owski}, {Wrona},
  {Ulaczyk}, \& {Rybicki}}]{2021ApJ...911L..22S}
{Soszy{\'n}ski}, I., {Olechowska}, A., {Ratajczak}, M., {et~al.} 2021, \apjl,
  911, L22, \dodoi{10.3847/2041-8213/abf3c9}

\bibitem[{Springer \& Shaviv(2013)}]{10.1093/mnras/stt1041}
Springer, O.~M., \& Shaviv, N.~J. 2013, Monthly Notices of the Royal
  Astronomical Society, 434, 1869, \dodoi{10.1093/mnras/stt1041}

\bibitem[{{Storm} {et~al.}(2011){Storm}, {Gieren}, {Fouqu{\'e}}, {Barnes},
  {Pietrzy{\'n}ski}, {Nardetto}, {Weber}, {Granzer}, \&
  {Strassmeier}}]{2011A&A...534A..94S}
{Storm}, J., {Gieren}, W., {Fouqu{\'e}}, P., {et~al.} 2011, \aap, 534, A94,
  \dodoi{10.1051/0004-6361/201117155}

\bibitem[{{Thompson} {et~al.}(2012){Thompson}, {Everett}, {Mullally},
  {Barclay}, {Howell}, {Still}, {Rowe}, {Christiansen}, {Kurtz}, {Hambleton},
  {Twicken}, {Ibrahim}, \& {Clarke}}]{2012ApJ...753...86T}
{Thompson}, S.~E., {Everett}, M., {Mullally}, F., {et~al.} 2012, \apj, 753, 86,
  \dodoi{10.1088/0004-637X/753/1/86}

\bibitem[{{Udalski} {et~al.}(2015){Udalski}, {Szyma{\'n}ski}, \&
  {Szyma{\'n}ski}}]{2015AcA....65....1U}
{Udalski}, A., {Szyma{\'n}ski}, M.~K., \& {Szyma{\'n}ski}, G. 2015, \actaa, 65,
  1.
\newblock \doarXiv{1504.05966}

\bibitem[{{Weinberg} \& {Arras}(2019)}]{2019ApJ...873...67W}
{Weinberg}, N.~N., \& {Arras}, P. 2019, \apj, 873, 67,
  \dodoi{10.3847/1538-4357/ab0204}

\bibitem[{{Welsh} {et~al.}(2011){Welsh}, {Orosz}, {Aerts}, {Brown},
  {Brugamyer}, {Cochran}, {Gilliland}, {Guzik}, {Kurtz}, {Latham}, {Marcy},
  {Quinn}, {Zima}, {Allen}, {Batalha}, {Bryson}, {Buchhave}, {Caldwell},
  {Gautier}, {Howell}, {Kinemuchi}, {Ibrahim}, {Isaacson}, {Jenkins}, {Prsa},
  {Still}, {Street}, {Wohler}, {Koch}, \& {Borucki}}]{2011ApJS..197....4W}
{Welsh}, W.~F., {Orosz}, J.~A., {Aerts}, C., {et~al.} 2011, \apjs, 197, 4,
  \dodoi{10.1088/0067-0049/197/1/4}

\bibitem[{{Wilson}(1990)}]{1990ApJ...356..613W}
{Wilson}, R.~E. 1990, \apj, 356, 613, \dodoi{10.1086/168867}

\bibitem[{{Wilson} \& {Devinney}(1971)}]{1971ApJ...166..605W}
{Wilson}, R.~E., \& {Devinney}, E.~J. 1971, \apj, 166, 605,
  \dodoi{10.1086/150986}

\bibitem[{{Wood} {et~al.}(1999){Wood}, {Alcock}, {Allsman}, {Alves}, {Axelrod},
  {Becker}, {Bennett}, {Cook}, {Drake}, {Freeman}, {Griest}, {King}, {Lehner},
  {Marshall}, {Minniti}, {Peterson}, {Pratt}, {Quinn}, {Stubbs}, {Sutherland},
  {Tomaney}, {Vandehei}, \& {Welch}}]{1999IAUS..191..151W}
{Wood}, P.~R., {Alcock}, C., {Allsman}, R.~A., {et~al.} 1999, in Asymptotic
  Giant Branch Stars, ed. T.~{Le Bertre}, A.~{Lebre}, \& C.~{Waelkens}, Vol.
  191, 151

\bibitem[{{Wo{\'z}niak}(2000)}]{2000AcA....50..421W}
{Wo{\'z}niak}, P.~R. 2000, \actaa, 50, 421.
\newblock \doarXiv{astro-ph/0012143}

\bibitem[{{Wrona} {et~al.}(2022){Wrona}, {Ratajczak},
  {Ko{\l}aczek-Szyma{\'n}ski}, {Koz{\l}owski}, {Soszy{\'n}ski}, {Iwanek},
  {Udalski}, {Szyma{\'n}ski}, {Pietrukowicz}, {Skowron}, {Skowron}, {Mr{\'o}z},
  {Poleski}, {Gromadzki}, {Ulaczyk}, \& {Rybicki}}]{2022ApJS..259...16W}
{Wrona}, M., {Ratajczak}, M., {Ko{\l}aczek-Szyma{\'n}ski}, P.~A., {et~al.}
  2022, \apjs, 259, 16, \dodoi{10.3847/1538-4365/ac4018}

\bibitem[{{Zahn}(1970)}]{1970A&A.....4..452Z}
{Zahn}, J.~P. 1970, \aap, 4, 452

\bibitem[{{Zahn}(1975)}]{1975A&A....41..329Z}
---. 1975, \aap, 41, 329

\bibitem[{{Zaritsky} {et~al.}(2004){Zaritsky}, {Harris}, {Thompson}, \&
  {Grebel}}]{2004AJ....128.1606Z}
{Zaritsky}, D., {Harris}, J., {Thompson}, I.~B., \& {Grebel}, E.~K. 2004, \aj,
  128, 1606, \dodoi{10.1086/423910}

\end{thebibliography}
	\bibliographystyle{aasjournal}
\end{document}